\documentclass[aps,prd,amsfonts,amssymb,amsmath,letterpaper,showkeys,10pt,reprint,floatfix]{revtex4-2}

\usepackage{tensor,mathtools,graphicx,enumitem} % math, graphicx and lists
\usepackage[colorlinks=true,allcolors=blue]{hyperref} % hyperlinks within paper

\pdfoutput=1 % ensures PDFLaTeX compilation

\newcommand{\gkgt}[1]{\ensuremath{\langle #1 \rangle}_{\mathrm{t}}}

\DeclareMathOperator{\diag}{diag} % diagonal matrix as list
\DeclareMathOperator{\supp}{supp} % support of
\DeclareMathOperator{\sech}{sech}

\begin{document}

%%%%%%%%%%%%%%%%%%%%%%%%%%%%%%%%%%%%%%%%%%%%%%%%%%%%%%%%%%%%%%%%%%%%%
%%%%%%%%%%%%%%%%%%%%%       Title        %%%%%%%%%%%%%%%%%%%%%%%%%%%%
%%%%%%%%%%%%%%%%%%%%%%%%%%%%%%%%%%%%%%%%%%%%%%%%%%%%%%%%%%%%%%%%%%%%%

\title{Gravitational Waves Emitted by a Uniformly Accelerated Mass: The Role of Zero-Rindler-Energy Modes in the Classical and Quantum Descriptions}

%%%%%%%%%%%%%%%%%%%%%%%%%%%%%%%%%%%%%%%%%%%%%%%%%%%%%%%%%%%%%%%%%%%%%
%%%%%%%%%%%%%%%%%%%%   Authors & Addresses   %%%%%%%%%%%%%%%%%%%%%%%%
%%%%%%%%%%%%%%%%%%%%%%%%%%%%%%%%%%%%%%%%%%%%%%%%%%%%%%%%%%%%%%%%%%%%%

\author{Felipe Portales-Oliva}
\email{felipe.portales@ufabc.edu.br}
\affiliation{Centro de Ci\^encias Naturais e Humanas, Universidade Federal do
ABC,\\Avenida dos Estados, 5001, 09210-580 Santo André, S\~ao Paulo, Brazil}

\author{Andr\'e G. S. Landulfo}
\email{andre.landulfo@ufabc.edu.br} 
\affiliation{Centro de Ci\^encias Naturais e Humanas, Universidade Federal do
ABC,\\Avenida dos Estados, 5001, 09210-580 Santo André, S\~ao Paulo, Brazil}

%%%%%%%%%%%%%%%%%%%%%%%%%%%%%%%%%%%%%%%%%%%%%%%%%%%%%%%%%%%%%%%%%%%%%
%%%%%%%%%%%%%%%%%%%%%         Abstract      %%%%%%%%%%%%%%%%%%%%%%%%%
%%%%%%%%%%%%%%%%%%%%%%%%%%%%%%%%%%%%%%%%%%%%%%%%%%%%%%%%%%%%%%%%%%%%%

\begin{abstract}
    The observation of gravitational waves opens up a new window to probe the universe and the nature of the gravitational field itself.  As a result, they serve as
    a new and promising tool to not only test our current theories but to study different models that go beyond our current understanding. In this paper, inspired by recent successes in scalar and Maxwell electrodynamics, we analyze the role played by the (quantum) Unruh effect
    on the production of both classical and quantum gravitational waves by a uniformly accelerated mass. In particular, we show the fundamental role played by zero-energy (Rindler) gravitons in building up the gravitational radiation, as measured by inertial observers, emitted by the body. 
\end{abstract}


\maketitle

%%%%%%%%%%%%%%%%%%%%%%%%%%%%%%%%%%%%%%%%%%%%%%%%%%%%%%%%%%%%%%%%%%%%%
%%%%%%%%%%%%%%%%%%%%%      Introduction     %%%%%%%%%%%%%%%%%%%%%%%%%
%%%%%%%%%%%%%%%%%%%%%%%%%%%%%%%%%%%%%%%%%%%%%%%%%%%%%%%%%%%%%%%%%%%%%
\section{Introduction} \label{sec:introduction}

The prediction of the existence of gravitational waves is one of the most important theoretical predictions brought forward by General Relativity. Today, due to the advent of gravitational wave detectors such as LIGO and Virgo, it also stands as one of the core pieces of evidence that sustain this theory as our best tool to describe gravity. 

The direct observation of these waves~\cite{abbottObservationGravitationalWaves2016} has opened new windows to probe the universe by broadening the spectrum of possible phenomenological observations and catapulted us to the era of gravitational wave astronomy.
This kind of radiation provides us with an important and interesting set of astrophysical tools, and thus, it is vital to study and understand the mechanisms under which it can be generated.

One of the processes that give rise to radiation is acceleration and the relationship between these two has attracted the interest of physicists over the years (see e.g. Refs~\cite{larmorTheoryMagneticInfluence1897,
fultonClassicalRadiationUniformly1960,
unruhAccelerationRadiationGeneralized1982,
singalEquivalencePrincipleElectric1995}). In this context, uniformly accelerated charges have received special attention, mainly due to the apparent contradictions that arise when the principle of equivalence is
considered~\cite{feynmanFeynmanLecturesGravitation1995}. Such issues are resolved when one notes that radiation is not to be regarded as a covariant concept but rather, it depends on the observer measuring it~\cite{rohrlichDefinitionElectromagneticRadiation1961,
rohrlichPrincipleEquivalence1963,boulwareRadiationUniformlyAccelerated1980}. 

In the context of Quantum Field Theory in Curved Spacetimes the connection between acceleration and radiation has only been strengthened since the discovery that an accelerated observer sees the inertial vacuum as a thermal bath of particles at the Unruh
temperature~\cite{unruhNotesBlackholeEvaporation1976}
\begin{equation}
    T_{\mathrm{U}} = \frac{\hbar a }{2 \pi c k_{\mathrm{B}}},
\end{equation}
the so-called Unruh Effect. In the decades that followed, the interplay between acceleration, radiation, and the Unruh effect has been scrutinized and their consequences analyzed through several works such as 
Refs.~\cite{unruhWhatHappensWhen1984,higuchiBremsstrahlungFullingDaviesUnruhThermal1992,cozzellaProposalObservingUnruh2017}.

Even though most of the examples above are related to electrodynamics, the emission of both classical and quantized gravitational waves by point particles has been analyzed around several background geometries. In the classical realm, Bičák~\cite{bicakGravitationalRadiationUniformly1997} studies the radiative properties of the solution provided by Bonnor and Swaminarayan~\cite{bonnorExactSolutionUniformly1964}, describing 4 accelerated
particles; Hopper and Cardoso~\cite{hopperScatteringPointParticles2018} examine the scattering of test particles and their interactions with gravitational waves around a Schwarzchild black hole; and
Poisson~\cite{poissonGravitationalRadiationParticle1993} gives detailed calculations to survey the waveforms from a binary system where one of its members is much more massive than the other. As for the quantum counterpart, quantized gravitational perturbations around a classical Schwarzschild black hole were studied by Bernar, Crispino, and
Higuchi~\cite{bernarGravitationalWavesEmitted2017}, and these same authors also provide a discussion on a de Sitter background using a multipole source~\cite{bernarGibbonsHawkingRadiationGravitons2018}.  In addition to this, the construction of propagators has also been previously reported, see, for example, the gravitational two-point functions in de Sitter spacetime of
Refs.~\cite{bernarInfraredfiniteGravitonTwopoint2014,
frobModesumConstructionCovariant2016}. 

In this paper, we aim to describe the classical and quantum gravitational waves emitted by a single mass, which is uniformly accelerated in a Minkowski background. In particular, using traceless and transverse gravitational Unruh modes, we show the main role played by zero-Rindler-energy modes in building up the gravitational radiation emitted by the mass, as measured by inertial observers in the asymptotic future. 

For this purpose, the paper is structured as follows. In
Sec.~\ref{sect:basics} we present a summary of the basics of gravitational wave theory and some tools needed for our description. After this, in Sec.~\ref{sec:Modes}, we define modes for the gravitational perturbations that are based on some of the symmetry properties of Minkowski spacetime.  Then, in Sec.~\ref{sect:accelerated-particle}, we present the physical setup we use and find the corresponding stress-energy tensor that will serve as the source of these gravitational waves. In
Sec.~\ref{sect:classical-exp} we present the classical expansion of the retarded field seen from the perspective of inertial observers in the asymptotic future. We then proceed in Sec.~\ref{sect:quantum-exp} to compare
the quantum description between the asymptotic past and asymptotic future, connecting the Fock spaces of both of these constructions when the field is initially in its vacuum state using the $S$ matrix. Finally, we give some concluding remarks to summarize our results on Sec.~\ref{sec:conclusions}.

Throughout this work, we use natural units: $ \hbar = c =1 $, and Newton's constant is kept as $ G $. We use the metric with signature $ (-,+,+,+) $ and indices of the beginning of the Latin alphabet $ (a,b,c,\ldots) $ to refer to
components in the entire spacetime (the bulk).

%%%%%%%%%%%%%%%%%%%%%%%%%%%%%%%%%%%%%%%%%%%%%%%%%%%%%%%%%%%%%%%%%%%%%
%%%%   Gravitational perturbations around a fixed metric   %%%%%%%%%%
%%%%%%%%%%%%%%%%%%%%%%%%%%%%%%%%%%%%%%%%%%%%%%%%%%%%%%%%%%%%%%%%%%%%%
\section{Gravitational perturbations around a fixed metric}
\label{sect:basics}

We begin by recalling the general formalism of Gravitational waves. Let us consider a 4-dimensional globally hyperbolic spacetime $ \left( M, g_{ab}\right),$ where $ M $ is a 4-dimensional manifold with a Lorentzian metric $ g_{a b} $. Gravitational perturbations correspond with small deviations from this metric: $ h_{a b} = \delta g_{a b} $, with $h_{a b}=h_{(ab)}$ and $ |h_{a b}| \ll 1 $. The dynamics of such perturbations are encoded in the action (without cosmological
constant)~\cite{higuchiSymmetricTensorSpherical1987}
\begin{equation}
    I_{\text{pert}}
    =
    \int_M\mathrm{d}^4{x}\,
    \mathcal{L}_{\text{pert}} 
    = 
    \int_M\mathrm{d}^4{x}\ {}
    (
        \mathcal{L}_{\text{inv}} 
        + 
        \mathcal{L}_{\text{gf}}
    ),
    \label{eq:Lagrangian-density-perturbation}
\end{equation}
where
\begin{multline}
    \mathcal{L}_{\text{inv}}
    =
    -\frac{\sqrt{-g}}{4\kappa^2} (
		\nabla_c h_{a b} \nabla^c h^{a b} 
		- \nabla_c h \nabla^c h 
    \\
		+ 2 \nabla_a h^{a b} \nabla_b h 
		- 2 \nabla^a h_{a c} \nabla_b h^{b c}
	),
    \label{eq:Lagrangian-density-perturbation-invariant}
\end{multline} 
is the Lagrangian density invariant under gauge transformations defined by an arbitrary vector field $\Lambda_a$ 
\begin{equation}
    h_{a b}
    \ \to\ 
    \tilde{h}_{a b} = h_{a b} - 2\nabla_{(a} \Lambda_{b)},
    \label{eq:gauge-transformation}
\end{equation}
and 
\begin{multline}
    \mathcal{L}_{\text{gf}}
    =
    \frac{\sqrt{-g}}{2\alpha\kappa^2}\! 
		\left(
			\nabla^a h_{a b} - \frac{1+\beta}{\beta} \nabla_b h
		\right) 
        \\ \times
		\left(
			\nabla_c h^{c b} - \frac{1+\beta}{\beta} \nabla^b h
		\right)
        ,
    \label{eq:Lagrangian-density-perturbation-gauge-fixing}
\end{multline}
is a gauge fixing term depending on the quantities  $\alpha, \beta\in \mathbb{R}$. This last term is added to the theory to eliminate spurious non-normalizable
modes, whereas the Lagrangian density of Eq.~(\ref{eq:Lagrangian-density-perturbation-invariant}) arises from the second-order perturbation of the Einstein-Hilbert Lagrangian density $ \mathcal{L}_{\text{HE}} = \kappa^{-2} \sqrt{-g} R.$ We raise and lower indices using the background metric $ g_{ab} $, $ h
= g^{a b} h_{a b} $ is the trace of the perturbation, $ R $ is the (background) Ricci scalar, $ g = \det g_{a b} $ is the determinant of the background metric, $ \nabla_a $ is the (torsion-free) covariant derivative compatible with $g_{ab}$, and $ \kappa^2 \equiv 16\pi G $.  

Given two distinct gravitational perturbations, $ h_{a b}^{(1)} $ and $ h_{a b}^{(2)} $, and a Cauchy surface $ \Sigma $, we define their tensor Klein-Gordon product as
\begin{equation}
    \gkgt{h^{(1)},h^{(2)}}
    \equiv
    -\mathrm{i} 
    \int_{\Sigma}\mathrm{d}^3 \Sigma \, n_c 
        [
            \overline{h_{a b}^{(1)}} \pi_{(2)}^{c a b}
            -
            \overline{\pi_{(1)}^{c a b }} h_{a b}^{(2)}
        ],
    \label{eq:KG-inner-prod-gravp}
\end{equation}
where the overline symbolizes the complex conjugate, $ n^a $ is the future-oriented unit vector orthogonal to $ \Sigma $, and $ \pi_{(1)}^{c a b } $ and $
\pi_{(2)}^{c a b } $ are the generalized momenta 
\begin{equation}
    \pi^{c a b} \equiv 
    \frac{1}{\sqrt{- g}} 
    \frac{\partial \mathcal{L}_\textrm{pert}}{\partial (\nabla_c h_{a b})},
    \label{eq:gen-momentum-gravp}
\end{equation}
associated to $ h_{a b}^{(1)} $ and $ h_{a b}^{(2)} $, respectively.  As the 4-vector inside the square brackets in Eq.~\eqref{eq:KG-inner-prod-gravp} has
null divergence, the result of this integral is conserved (it does not depend on the choice of Cauchy surface)~\cite{friedmanGenericInstabilityRotating1978,higuchiMassiveSymmetricTensor1989}. 

Given a matter distribution in the spacetime whose dynamics is governed by the matter action $ I_{\text{mat}} $, the stress-energy tensor is defined as
\begin{equation}
    T^{a b} 
    \equiv
    \frac{2}{\sqrt{-g}}
    \frac{\delta I_{\text{mat}}}{\delta g_{a b}}
    .
    \label{eq:stress-energy-general}
\end{equation}
In the context of gravitational perturbations we are considering, we can use the tensor above to describe the coupling of matter with gravitational waves via the interaction action~\cite{weinbergGravitationCosmologyPrinciples1972}
\begin{equation}
    I_{\text{int}} [h,T]
    \equiv
    \frac{1}{2}
    \int_M \mathrm{d}^4 x\, \sqrt{-g} \,
    T^{a b} h_{a b}.
    \label{eq:interaction-action}
\end{equation}
In regions without sources, i.e., $T_{ab}=0$, we can choose the perturbations to be traceless and transverse, the TT gauge:
\begin{align}
    \tensor{h}{^a_a} &= 0,
    &
    \nabla^a h_{a b} & = 0,
    \label{eq:TT-gauge}
\end{align}
which implies that the Euler-Lagrange equations arising from the total action $I_{\text{pert}} $ is given by
\begin{equation}
    \nabla_c\nabla^c h_{a b} 
    - 2\tensor{R}{^c_a_b^d} h_{c d}
    = 0,
    \label{eq:gravitational-field-eq-TT}
\end{equation}
where $ \tensor{R}{^d_a_b_c} $ is the Riemann curvature of the background spacetime.  For perturbations in the TT gauge, the generalized 4-momentum is simply $ \pi^{c a b} = -
(2\kappa^2)^{-1} \nabla^c h^{a b} $, which implies the tensor Klein-Gordon inner
product reduces to
\begin{equation}
    \gkgt{h^{(1)},h^{(2)}}
    =
    \frac{\mathrm{i}}{2\kappa^2} \int_{\Sigma} \mathrm{d}\Sigma\, n^c \,
    W_c [h^{(1)},h^{(2)}],
    \label{eq:KG-inner-product-TT}
\end{equation} 
with the current 
\begin{equation}
    W_a [h^{(1)},h^{(2)}]
    \equiv 
    \overline{h^{c d}_{(1)}} \nabla_a h^{(2)}_{c d}
    - h^{c d}_{(2)} \nabla_a \overline{h^{(1)}_{c d}}
    ,
    \label{eq:general-current-to-compute}
\end{equation}
which we will use to normalize the modes. 
We now present the modes in the class of background spacetimes we are interested in describing.

% %%%%%%%%%%%%%%%%%%%%%%%%%%%%%%%%%%%%%%%%%%%%%%%%%%%%%%%%%%%%%%%%%%%%%
% %%%%%%%%    Modes for the gravitational perturbation    %%%%%%%%%%%%%
% %%%%%%%%%%%%%%%%%%%%%%%%%%%%%%%%%%%%%%%%%%%%%%%%%%%%%%%%%%%%%%%%%%%%% 
\section{Modes for the gravitational perturbation around Minkowski and Rindler
backgrounds}\label{sec:Modes}

% \subsection{Minkowski and Rindler spacetimes}

\begin{figure}[bht]
    \centering
    \includegraphics[width=.46\textwidth]{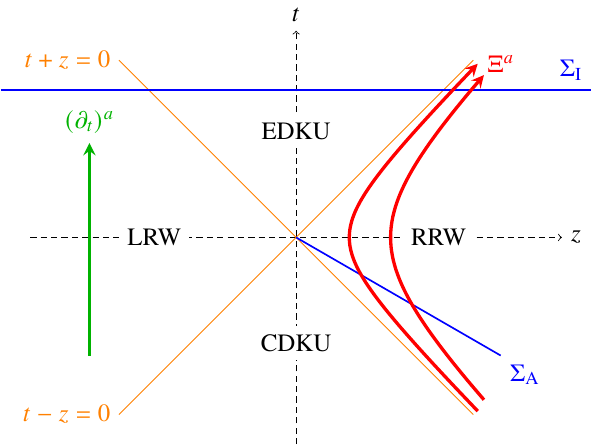}
    \caption{Schematics of Minkowski spacetime with the light-like surfaces $ t
    = \pm z $ and segments of the integral curves of each of the future-oriented
    time-like vectors. Examples of the Cauchy surfaces are in blue: $
    \Sigma_{\mathrm{I}} $ is generated by $ (\partial_t)^a $ and is associated
    with inertial observers; and $ \Sigma_{\mathrm{A}} $ is generated by $ \Xi^a
    $ and is associated with accelerated observers whose motion is constrained to
    the RRW.}\label{fig:regions}
\end{figure}

We will concentrate on gravitational perturbations around 4-dimensional
Minkowski spacetime, which is a flat and globally hyperbolic solution of
Einstein's homogeneous field equations. Mathematically, we model it as the manifold $ \mathbb{R}^4 $ endowed with the metric $ \eta_{a b} $ which, in
Cartesian (inertial) coordinates $ (t,x,y,z) $, is written as 
\begin{equation}
    \eta_{a b}
    =
    \diag(-1,1,1,1) 
    .
    \label{eq:Minkowski-metric}
\end{equation}
The condition $ t = \text{constant} $ defines a family of Cauchy surfaces in which the Killing field $ (\partial_t)^a $ is its (future-pointing) normal. By arbitrarily choosing one spatial direction, namely the $ z $ axis, we can separate this spacetime into 4 distinct regions by the light-like surfaces $ t \pm z =0 $: the Left Rindler Wedge (LRW) as the region where $ z < - |t| $, the Right Rindler Wedge (RRW) by the condition $z>|t| $, and the Expanding and Contracting Degenerate Kasner Universes (EDKU and
CDKU) by $ t > |z| $ and $ t < - |z|,$ respectively; see Fig.~\ref{fig:regions}. The Rindler wedges are globally hyperbolic static spacetimes in their own right, as they are globally hyperbolic regions where the Lorentz Boosts generators in the $ z $ direction 
\begin{equation}
    \Xi^a 
    =
    z (\partial_t)^a + t (\partial_z)^a ,
    \label{eq:Boost-generator}
\end{equation}
are (hypersurface-orthogonal) timelike Killing Fields.  Each of these wedges can be identified with a copy of the so-called Rindler spacetime, from where they receive their name. Note that the boost generators are spacelike in both the EDKU and CDKU.

The Rindler spacetime is the manifold $ \mathbb R^4 $ mapped by the coordinates $
(\lambda,\xi,x,y) $ and with metric defined by 
\begin{equation}
    \mathrm{d} {{s}}^2
    = g_{a b} \mathrm{d} x^a \mathrm{d} x^b
    = \mathrm{e}^{2 a \xi} (- \mathrm{d}\lambda^2 + \mathrm{d}\xi^2) + \mathrm{d} x^2 + \mathrm{d} y^2,
    \label{eq:Rindler-metric}
\end{equation}
where $ a $ (called the acceleration parameter) is a non-negative constant. This spacetime can be identified as the product between two subspaces: the Lorentzian orbit, given by $ \mathbb{R}^2 $ with interval $ \mathrm{d} s_{\text{orb}}^2
= \mathrm{e}^{2 a \xi}(-\mathrm{d}\lambda^2 + \mathrm{d}\xi^2) $; and the $ xy $ plane, which is another copy of $ \mathbb R^2 $, but with metric defined by $ \mathrm{d}s_{\text{ms}}^2 = \mathrm{d} x^2 + \mathrm{d} y^2 $. 

The correspondence between Rindler spacetime and the RRW becomes evident with the aid of the coordinate transformation
\begin{align}
    t &= a^{-1} \mathrm{e}^{a \xi}  \sinh(a\lambda),
    &
    z &= a^{-1} \mathrm{e}^{a \xi}  \cosh(a\lambda),
    \label{eq:RRW-coordinate-transformation}
\end{align}
that keeps $ \mathbf{x}_\perp $ invariant. From this we see that a particle that
moves along a trajectory where the spatial coordinates $ \xi $ and $
\mathbf{x}_\perp $ are kept fixed measures a proper time $ \tau = \mathrm{e}^{a
\xi} \lambda $ and a constant proper acceleration $ a
\mathrm{e}^{-a\xi} $. As a result, we associate the Rindler spacetime (and the Rindler wedges) with a family of uniformly accelerated observers moving along the $z$ direction. Note also that the only nonzero components of the Christoffel symbols are 
\begin{equation}
    \tensor{\Gamma}{^\xi_{\xi \xi}} 
    =
    \tensor{\Gamma}{^\xi_{\lambda \lambda}}
    =
    \tensor{\Gamma}{^\lambda_{\lambda \xi}}
    =
    \tensor{\Gamma}{^\lambda_{\xi \lambda}}
    =
    a,
    \label{eq:Christoffel-Rindler}
\end{equation} 
so there is no mixing between the two subspaces of Rindler spacetime when
computing the covariant derivatives. 

The treatment of the massless scalar field $ \phi $ will be useful further ahead, so it is convenient to present an abridged version here. We recall that the dynamics of the field are determined from the field equation, which in Rindler spacetime is given by
\begin{equation}
    \mathrm{e}^{- 2 a \xi}
    \left(
        - \frac{\partial^2 \phi}{\partial \lambda^2}
        + \frac{\partial^2 \phi}{\partial \xi^2}
    \right)
    + \frac{\partial^2 \phi}{\partial x^2}
    + \frac{\partial^2 \phi}{\partial y^2}
    =
    0,
    \label{eq:scalar-field-equation-Rindler}
\end{equation}
and the classical solutions can be expanded in terms of the positive-energy (with respect to Rindler time $ \lambda $) modes  given by
\begin{equation}
    v_{\omega\mathbf{k}_\perp}\!(x)
    =
        \sqrt{\frac{\sinh(\pi\omega/a)}{4\pi^4a}}
        \,
        \mathrm{e}^{\mathrm{i}(\mathbf{k}_\perp\cdot\mathbf{x}_\perp-\omega\lambda)}
        \mathrm{K}_{\mathrm{i}\omega/a}\!%\!
        \left(
            \frac{k_\perp \mathrm{e}^{a\xi}}{a} 
        \right)
    ,
    \label{eq:scalar-Rindler-Modes}
\end{equation}
where $ \omega \geq 0 $ and $ \mathbf{k}_\perp \in \mathbb{R}^2- \{0\}.$  These have been chosen so that they satisfy the normalization condition 
\begin{equation}
    \langle
        v_{\omega\mathbf{k}_\perp}
        ,
        v_{\omega'\mathbf{k}_\perp'}
    \rangle_{\mathrm{s}}
    = 
    \delta(\omega - \omega')
    \,
    \delta^2(\mathbf{k}_\perp - \mathbf{k}_\perp')
    .
    \label{eq:scalar-normalization}
\end{equation}
Here the integral is taken over a constant $ \lambda $ surface and the subscript $ \mathrm{s} $ indicates this is the scalar version of the Klein-Gordon inner product~\cite{crispinoUnruhEffectIts2008}.  On the other hand, on Minkowski spacetime with global inertial coordinate $(t, x, y,z)$, the field equation satisfies the well-known wave equation
\begin{equation}
    - \frac{\partial^2 \phi}{\partial t^2}
    + \frac{\partial^2 \phi}{\partial x^2}
    + \frac{\partial^2 \phi}{\partial y^2}
    + \frac{\partial^2 \phi}{\partial z^2}
    =
    0,
    \label{eq:scalar-field-equation-Minkowski}
\end{equation}
and its solutions can be expanded in terms of the positive-energy (with respect to the inertial time $ t $) scalar inertial modes 
\begin{equation}
    \psi_{\mathbf{k}} (t,\mathbf{x}_\perp,z)
    \equiv
    \frac{\exp[
        \mathrm{i}(
            \mathbf{k}_\perp \cdot \mathbf{x}_\perp 
            +
            k_z z 
            -
            k_0 t
        )
    ]}{\sqrt{
        2k_0 (2 \pi)^3
    }},
    \label{eq:inertial-scalar-modes}
\end{equation}
where $\mathbf{k} = (k_x,k_y,k_z)$ and $ k_0 = \sqrt{k_\perp^2+k_z^2} $ defines the energy carried by each mode according to an inertial observer.  These modes are also orthonormalized: 
\begin{equation}
    \langle
        \psi_{\mathbf{k}}
        ,
        \psi_{\mathbf{k}'}
    \rangle_{\mathrm{s}}
    = 
    \delta^3(\mathbf{k} - \mathbf{k}')
    .
    \label{eq:scalar-normalization-inertial}
\end{equation}

Since Minkowski spacetime contains two copies of Rindler spacetime, we can identify two sets of scalar Rindler modes: the left and the right ones, which can be written as linear combinations of the scalar Minkowski modes and their complex
conjugates~\cite{crispinoUnruhEffectIts2008,higuchiEntanglementVacuumLeft2017}:
\begin{align}
    v^{\mathrm{R}}_{\omega \mathbf{k}_\perp} (t,\mathbf{x}_\perp,z)
    &=
    \int_{-\infty}^\infty \mathrm{d} k_z
    [
        \alpha^{\mathrm{R}}_{\omega k_\perp k_z} 
        \psi_{\mathbf{k}}(t,\mathbf{x}_\perp,z)
    \nonumber \\
    &\qquad \qquad 
        +
        \beta^{\mathrm{R}}_{\omega k_\perp k_z} 
        \overline{\psi_{\mathbf{k}}(t,-\mathbf{x}_\perp,z)}
    ],
    \label{eq:Bogoliubov-right-Rindler}
    %%%%%%%%%%%%%%%%%%%%%%%%%%%%%%%%%%%%%%%%%%%%%%%%%%%%%%%%%%%%%%%%
    \\
    v^{\mathrm{L}}_{\omega \mathbf{k}_\perp} (t,\mathbf{x}_\perp,z)
    &=
    \int_{-\infty}^\infty \mathrm{d} k_z
    [
        \alpha^{\mathrm{L}}_{\omega k_\perp k_z} 
        \psi_{\mathbf{k}}(t,\mathbf{x}_\perp,z)
    \nonumber \\
    &\qquad \qquad 
        +
        \beta^{\mathrm{L}}_{\omega k_\perp k_z} 
        \overline{\psi_{\mathbf{k}}(t,-\mathbf{x}_\perp,z)}
    ],
    \label{eq:Bogoliubov-left-Rindler}
\end{align}
where the Bogoliubov coefficients are
\begin{align}
    \alpha^{\mathrm{R}}_{\omega k_\perp k_z}
    &=
    \alpha^{\mathrm{L}}_{\omega k_\perp -k_z}
    =
    -\mathrm{e}^{\pi\omega/a}
    \beta^{\mathrm{R}}_{\omega k_\perp k_z}
    =
    -\mathrm{e}^{\pi\omega/a}
    \beta^{\mathrm{L}}_{\omega k_\perp -k_z}
    \nonumber \\
    &=
    \frac{
        \mathrm{e}^{\pi\omega/(2 a)}
    }{
        \sqrt{4\pi a k_0 \sinh(\pi \omega / a)}
    }
    \left[
        \frac{
            k_0 + k_z
        }{
            k_0 - k_z
        }
    \right]^{
        -\mathrm{i} \omega / (2 a)
    }.
\end{align}
The integrals of Eqs.~\eqref{eq:Bogoliubov-right-Rindler}
and~\eqref{eq:Bogoliubov-left-Rindler} serve as the definition of left and right scalar Rindler modes as distributions on the entirety of Minkowski spacetime. Particularly, the
right modes converge to the form given in 
Eq.~\eqref{eq:scalar-Rindler-Modes} on the RRW and to 0 on the LRW. Similarly,  left modes are given by Eq.~\eqref{eq:scalar-Rindler-Modes} on the LRW and are null on the RRW. These expressions can be used to show that between two modes of the same wedge, the normalization condition~\eqref{eq:scalar-normalization} still holds when the integral is taken over Cauchy surfaces with constant inertial time, and that left and right scalar Rindler modes are orthogonal to each other.

We will now construct similar modes for the tensor case, applying the procedure described in Refs.~\cite{mukohyamaGaugeinvariantGravitationalPerturbations2000,
kodamaBraneWorldCosmology2000,kodamaMasterEquationGravitational2003} in Minkowski and Rindler spacetimes, taking advantage of the symmetry of the $ xy $ plane.  We will be using the scalar and vector \emph{harmonics} defined in this maximally symmetric subspace as well as \emph{gauge invariants} defined on the orbit
to build 2 independent sets of modes (or sectors), each of which arises from (and is labeled by) the type of harmonic used in its construction.  We will only show details tailored to our exposition and urge the reader to refer to the works mentioned above for a comprehensive discussion on these methods, as the
derivations and proofs are quite involved and far more general than the scope of this paper.

%%%%%%%%%%%%%%%%%%%%%%%%%%%%%%%%%%%%%%%%%%%%%%%%%%%%%%%%%%%%%%%%%%%%%
%%%%%%%%%%%%%%%      Harmonics on the plane     %%%%%%%%%%%%%%%%%%%%%
%%%%%%%%%%%%%%%%%%%%%%%%%%%%%%%%%%%%%%%%%%%%%%%%%%%%%%%%%%%%%%%%%%%%%
\subsection{Harmonics on the plane}\label{sect:plane-harmonics}

In the following, indices of the middle of the Latin alphabet $ i, j, k, \ldots,$ represent components in the $ xy $ plane.

The scalar harmonic $ \mathbb{S}(\mathbf{x}_\perp) $ is the solution to the eigenvalue problem of the Laplace-Beltrami operator, this is
\begin{equation}
    \frac{\partial^2 \mathbb{S}}{\partial x^2}
    +
    \frac{\partial^2 \mathbb{S}}{\partial y^2}
    +
    k_\perp^2 \mathbb{S}
    = 0,
    \label{eq:scalar-sector-perturbation-2scalar-harmonic-rindler}
\end{equation}
from which we find solutions of the type~\footnote{We will not worry ourselves with the normalization of the harmonics at the time being, as the final perturbations will be normalized. Calculations on the orthogonality of the harmonics are given in Appendix~\ref{sect:normalization-calculations}.}
\begin{equation}
    \mathbb{S}^{\mathbf{k}_\perp} (\mathbf{ x}_\perp) 
    = 
        \mathrm{e}^{\mathrm{i}\mathbf{k}_\perp \cdot \mathbf{x}_\perp}.
    \label{eq:scalar-harmonic-plane}
\end{equation}
The transverse momentum vector $ \mathbf{k}_\perp = (k_x,k_y) \in \mathbb{R}^2 -
\{ 0 \} $ labels the modes in such a way that the eigenvalues of the harmonic equation are identified with the modulus of this vector, i.e.,  $
|\mathbf{k}_\perp|^2 = k_x^2 + k_ y^2 = k_\perp^2 $.  Derivatives of the harmonics are used to define the auxiliary 2-vector and 2-tensor
\begin{gather}
    \mathbb{S}^{\mathbf{k}_\perp}_i 
    \equiv
    - \frac{1}{k_\perp} \partial_i \mathbb{S}^{\mathbf{k}_\perp},
    \label{eq:scalar-sector-2vector-plane}
    \\
    \mathbb{S}^{\mathbf{k}_\perp}_{i j} 
    \equiv
    \frac{1}{k_\perp^2} \partial_i\partial_j \mathbb{S}^{\mathbf{k}_\perp}
    +
    \frac{1}{2} g_{i j} \mathbb{S}^{\mathbf{k}_\perp}
    \label{eq:scalar-sector-2tensor-plane}
    ,
\end{gather}
that will represent the $ \mathbf{x}_\perp $ dependence of the scalar sector modes.

Similarly, the components of the vector harmonic $ \mathbb{V}_i (\mathbf{x}_\perp) $ are defined by the equations
\begin{align}
    \frac{\partial^2 \mathbb{V}_i}{\partial x^2}
    +
    \frac{\partial^2 \mathbb{V}_i}{\partial y^2}
    +
    k_\perp^2 \mathbb{V}_i
    &= 0,
    &
    \nabla^i \mathbb{V}_i =0.
    \label{eq:vector-harmonic-plane-equation}
\end{align}
By using the Levi-Civita tensor on the plane, defined by its Cartesian components $ \varepsilon_{x y} = -
\varepsilon_{y x} = 1 $, $ \varepsilon_{x x} = \varepsilon_{y y} = 0 $, we can write the solution in Cartesian coordinates as
\begin{equation}
    \mathbb{V}^{\mathbf{k}_\perp}_i 
    =
    - \mathrm{i} \varepsilon_{i m}
    g^{m n}
    \partial_n 
        \mathbb{S}^{\mathbf{k}_\perp} 
    .
    \label{eq:vector-harmonic-plane}
\end{equation} 
These vectors have an associated tensor defined by
\begin{align}
    \mathbb{V}^{\mathbf{k}_\perp}_{i j} 
    &\equiv
    - \frac{1}{2 k_\perp} 
    (
        \partial_i \mathbb{V}^{\mathbf{k}_\perp}_j 
        +
        \partial_j \mathbb{V}^{\mathbf{k}_\perp}_i 
    ) 
    .
    \label{eq:vector-sector-2tensor-plane}
\end{align} 
Properties of the definitions~\eqref{eq:scalar-sector-2vector-plane}, \eqref{eq:scalar-sector-2tensor-plane}, and
\eqref{eq:vector-sector-2tensor-plane} that are relevant to our study can be found on Appendix~\ref{sect:properties-of-definitions-from-harmonics}.

The tensor harmonic in the plane, $ \mathbb{T}_{i j},(\mathbf{x}_\perp) $ is defined as the symmetric solution of the system of equations  
\begin{align}
    \frac{\partial^2 \mathbb{T}_{i j}}{\partial x^2}
    +
    \frac{\partial^2 \mathbb{T}_{i j}}{\partial y^2}
    +
    k_\perp^2 \mathbb{T}_{i j}
    &= 0,
    &
    \nabla^i \mathbb{T}_{i j} &=0,
    &
    \tensor{\mathbb{T}}{^i_i} &=0.
    \label{eq:tensor-harmonic-plane-equation}
\end{align}
We find that this is only satisfied by $ \mathbb{T}_{i j}(\mathbf{x}_\perp) = 0$. Therefore, we can discard the tensor perturbations and are left with the scalar and vector sectors only.

%%%%%%%%%%%%%%%%%%%%%%%%%%%%%%%%%%%%%%%%%%%%%%%%%%%%%%%%%%%%%%%%%%%%%
%%%%%%%  Master variables and normalizable gravitational    %%%%%%%%%
%%%%%%%  perturbations in Rindler spacetime                 %%%%%%%%%
%%%%%%%%%%%%%%%%%%%%%%%%%%%%%%%%%%%%%%%%%%%%%%%%%%%%%%%%%%%%%%%%%%%%%
\subsection{Master variables and normalizable gravitational perturbations in Rindler
spacetime}\label{sect:Application-Rindler}

There have been previous reports of gravitational perturbations on Rindler spacetime, like Ref.~\cite{sugiyamaGravitationalWavesKasner2021} where the
authors chose the Regge-Wheeler gauge~\cite{reggeStabilitySchwarzschildSingularity1957,thorneIntroductionReggeWheeler2020}. However, as their even mode is not transverse, they are not suited for our purposes. For this reason, in what follows, we will build a different set of modes more suitable for our application.  We will use Greek indices of the beginning of the alphabet $ \alpha,\beta,\gamma, \ldots $ to label components in the orbit.  

The so-called master variables are governed by the equation
\begin{equation}
    \nabla^\alpha\nabla_\alpha \Omega 
    - k_\perp^2 \Omega
    = 
    0,
    \label{eq:master-variable-general}
\end{equation}
with $ \Omega\in\{\Omega_{\mathrm{s}},\Omega_{\mathrm{v}},\Omega_{\mathrm{t}}\}
$.
In Rindler coordinates, Eq.~(\ref{eq:master-variable-general}) can be cast as
\begin{equation}
    \mathrm{e}^{-2 a \xi } \left(
        \frac{\partial^2 \Omega}{\partial \xi^2} 
        - 
        \frac{\partial^2 \Omega}{\partial \lambda^2} 
    \right)
    - k_\perp^2 \Omega 
    =
    0,
    \label{eq:master-variable-Rindler}
\end{equation}
and, given an energy $ \omega \geq 0 $ measured by Rindler observers, the positive (Klein-Gordon) norm regular master variables are given by 
\begin{gather}
    \Omega_{\mathrm{s}}^{\omega {k}_\perp} (\lambda,\xi)
    =
    S_{\omega{k}_\perp} 
    \mathrm{e}^{-\mathrm{i}\omega\lambda} 
    \mathrm{K}_{\mathrm{i}\omega/a}(a^{-1} k_\perp \mathrm{e}^{a\xi}),
    \label{eq:scalar-sector-RINDLER-mastervariable}
    \\
    \Omega_{\mathrm{v}}^{\omega {k}_\perp}(\lambda,\xi)
    =
    V_{\omega{k}_\perp} \mathrm{e}^{-\mathrm{i}\omega\lambda} \mathrm{K}_{\mathrm{i}\omega/a}(a^{-1} k_\perp \mathrm{e}^{a\xi}),
    \label{eq:vector-sector-RINDLER-mastervariable}
    \\
    \Omega_{\mathrm{t}}^{\omega {k}_\perp}(\lambda,\xi)
    =
    T_{\omega{k}_\perp} \mathrm{e}^{-\mathrm{i}\omega\lambda} \mathrm{K}_{\mathrm{i}\omega/a}(a^{-1} k_\perp \mathrm{e}^{a\xi}),
    \label{eq:tensor-sector-RINDLER-mastervariable}
\end{gather}
where $ \mathrm{K}_\nu(z) $ is the modified Bessel function of the second kind.
The coefficients $ S_{\omega{k}_\perp} $ and $ V_{\omega{k}_\perp} $ are to be found through normalization.  Note we have considered the tensor master variable even though we have discarded the tensor perturbation as it is still used in the definition of the remaining sectors.

We use the master variables and the (Lorentzian)  Levi-Civita tensor of the orbit, defined in Rindler coordinates by $ \epsilon_{\lambda\xi} = - \epsilon_{\xi\lambda} = \mathrm{e}^{2 a \xi}$, $ \epsilon_{\xi\xi} = \epsilon_{\lambda\lambda} = 0 $, to construct the gauge invariants 
\begin{gather}
    F^{\omega k_\perp} = 
    ({k^2_\perp}/{4})
        \Omega_{\mathrm{s}}^{\omega {k}_\perp},
    \label{eq:scalar-gauge-invariant}
    \\
    F_{\alpha}^{\omega k_\perp}
    =
    \epsilon_{\alpha\beta}\nabla^\beta \Omega_{\mathrm{v}}^{\omega {k}_\perp}
    \label{eq:vector-gauge-invariant}
    \\
    F^{\omega k_\perp}_{\alpha\beta}
    =
        \nabla_\alpha\nabla_\beta \Omega_{\mathrm{s}}^{\omega {k}_\perp} 
        - ({k^2_\perp}/{2}) g_{\alpha\beta} \Omega_{\mathrm{s}}^{\omega {k}_\perp}
        .
        \label{eq:tensor-gauge-invariant}
\end{gather}
From these we build the gauge dependent variables
\begin{align}
    \Omega_{\mathrm{l}}^{\omega {k}_\perp}
    &\equiv
    F^{\omega k_\perp} - \Omega^{\omega {k}_\perp}_{\mathrm{t}}/2,
    \label{eq:scalar-sector-HL-Rindler}
    \\
    f_{\alpha}^{\omega {k}_\perp}
    &\equiv
    F_{\alpha}^{\omega k_\perp}
    -
    \nabla_\alpha \Omega^{\omega {k}_\perp}_{\mathrm{t}}/k_\perp,
    \label{eq:eq:vector-sector-perturbation-lorentzian-vector-RINDLER-covariant}
    \\
    f_{\alpha\beta}^{\omega {k}_\perp}
    &\equiv
    F^{\omega k_\perp}_{\alpha\beta}
        - (2/k_\perp) \nabla_{(\alpha} F_{\beta)}^{\omega k_\perp},
    \label{eq:scalar-sector-falphabeta-Rindler}
\end{align} 
which we use to write the gravitational perturbations for the scalar 
\begin{align}
    h^{(\mathrm s, \omega \mathbf{k}_\perp)}_{\alpha\beta} 
        &= 
        f^{\omega {k}_\perp}_{\alpha\beta} 
            \, 
            \mathbb{S}^{\mathbf{k}_\perp},
    \nonumber \\
    h^{(\mathrm s, \omega \mathbf{k}_\perp)}_{\alpha j} 
        &= 
        f^{\omega {k}_\perp}_\alpha 
            \, 
            \mathbb{S}^{\mathbf{k}_\perp}_j,
    \nonumber \\
    h^{(\mathrm s, \omega \mathbf{k}_\perp)}_{i j} 
        &= 
        2 g_{i j} \Omega^{\omega {k}_\perp}_{\mathrm l}  
            \, 
            \mathbb{S}^{\mathbf{k}_\perp} 
        +
        2 \Omega_{\mathrm{t}}^{\omega {k}_\perp}  
            \, 
            \mathbb{S}^{\mathbf{k}_\perp}_{i j} ,
    \label{eq:scalar-sector-perturbation-total}
\end{align}
and vector sectors 
\begin{align}
    h^{(\mathrm v, \omega \mathbf{k}_\perp)}_{\alpha\beta} 
        &= 
        0,
    \nonumber \\
    h^{(\mathrm v, \omega \mathbf{k}_\perp)}_{\alpha i}   
        &= 
        f^{\omega {k}_\perp}_\alpha 
            \,  
            \mathbb{V}_{i}^{\mathbf{k}_\perp} ,
    \nonumber \\
    h^{(\mathrm v, \omega \mathbf{k}_\perp)}_{i j}
        &= 
        2 \Omega_{\mathrm{t}}^{\omega {k}_\perp} 
            \, 
            \mathbb{V}_{i j}^{\mathbf{k}_\perp}.
    \label{eq:vector-sector-perturbation-total}
\end{align}
Given the definitions in Eqs.~\eqref{eq:scalar-gauge-invariant}-\eqref{eq:scalar-sector-falphabeta-Rindler}, we see that these modes are quite complicated, and the information between the
scalar, vector, and tensor sectors are coupled between the different modes, as $
\Omega_{\mathrm{t}}^{\omega {k}_\perp} $ and $ \Omega_{\mathrm{v}}^{\omega
{k}_\perp} $ appear explicitly in both expressions. 

Using the properties described in Appendix~\ref{sect:properties-of-definitions-from-harmonics}, we can see that the vector sector satisfies the TT gauge conditions~\eqref{eq:TT-gauge} identically. However, imposing the same requirement to the scalar sector enforces the condition 
\begin{equation}
    \Omega_{\mathrm{t}}^{\omega {k}_\perp}
    =
    \frac{k_\perp^2}{2} \Omega_{\mathrm{s}}^{\omega {k}_\perp},
    \label{eq:linked-master-variables}
\end{equation}
or, equivalently, $ \Omega_{\mathrm{l}}^{\omega {k}_\perp} = 0 $. We will now
show the appropriate gauge transformations to further simplify these modes while
still remaining in the TT gauge.

Given an arbitrary gauge transformation~\eqref{eq:gauge-transformation}, the vector $ \Lambda_a $ characterizing it can be decomposed using scalar and vector harmonics in a very similar way to what is done to the gravitational perturbation~\cite{kodamaCosmologicalPerturbationTheory1984}.  Furthermore, the
resulting tensor, $ \tilde{h}_{a b},$ will also have a unique decomposition in terms of harmonics, and the gauge transformation translates to a transformation of the modes. From this, we can show that 
\begin{align}
    \Lambda_\alpha^{(\mathrm{s})} &= \phi_\alpha(\lambda,\xi) \  \mathbb{S}^{\mathbf{k}_\perp},
    &
    \Lambda_i^{(\mathrm{s})} &= \varphi(\lambda,\xi) \  \mathbb{S}^{\mathbf{k}_\perp}_i,
    \label{eq:scalar-gauge-transformation}
\end{align}
is the most general form of vector fields that modifies the scalar sector modes, while
\begin{equation}
    \Lambda_\alpha^{(\mathrm{v})} = 0,
    \qquad\quad
    \Lambda_i^{(\mathrm{v})} = \psi(\lambda,\xi) \  \mathbb{V}^{\mathbf{k}_\perp}_i ,
    \label{eq:vector-gauge-transformation}
\end{equation}
is the most general transformation that can be applied to the vector sector
modes.  We will choose the fields $ \phi_\alpha $, $ \phi $ and $ \psi $ in such a way that the modes are decoupled while remaining in the TT gauge.

For the vector sector, we pick 
\begin{equation}
    \Lambda_\alpha^{(\mathrm{v})} = 0,
    \quad\qquad
    \Lambda_i^{(\mathrm{v})} = - \frac{1}{k_\perp}\Omega_{\mathrm{t}}^{\omega {k}_\perp} \, \mathbb{V}_i^{\mathbf{k}_\perp},
    \label{eq:vector-gauge-transformation-Rindler}
\end{equation}
to make the transformed vector perturbation (omitting the tilde as we will
discard the old form of the modes and the distinction will no longer be
necessary)
\begin{align}
    h^{(\mathrm v,\omega \mathbf{k}_\perp)}_{\alpha\beta} &= 0,
    \nonumber \\
    h^{(\mathrm v,\omega \mathbf{k}_\perp)}_{\alpha j}    &= 
        \epsilon_{\alpha\beta} \nabla^\beta \Omega_{\mathrm{v}}^{\omega {k}_\perp} 
        \mathbb{V}_{j}^{\mathbf{k}_\perp},
    \nonumber \\
    h^{(\mathrm v,\omega \mathbf{k}_\perp)}_{i j}       &= 0,
    \label{eq:vector-perturbation-Rindler}
\end{align}
which we see is written in terms of the vector master variable only. We also
note that it is proportional to the odd perturbation previously found in
Ref.~\cite{sugiyamaGravitationalWavesKasner2021}.  To transform the scalar
sector we use 
\begin{equation}
    \Lambda_\alpha^{(\mathrm{s})} = - \frac{1}{k_\perp} \epsilon_{\alpha\beta} \nabla^\beta \Omega_{\mathrm{v}}^{\omega {k}_\perp} \, \mathbb{S}^{\mathbf{k}_\perp},
    \qquad 
    \Lambda_i^{(\mathrm{s})} = 0,
    \label{eq:scalar-gauge-transformation-Rindler}
\end{equation}
and Eq.~\eqref{eq:linked-master-variables}, yielding
\begin{align}
    h^{(\mathrm s,\omega \mathbf{k}_\perp)}_{\alpha\beta} 
        &= 
        [
            \nabla_\alpha \nabla_\beta \Omega_{\mathrm{s}}^{\omega {k}_\perp} - (k_\perp^2/2)\Omega_{\mathrm{s}}^{\omega {k}_\perp} g_{\alpha\beta}
        ] 
        \mathbb{S}^{\mathbf{k}_\perp},
    \nonumber \\
    h^{(\mathrm s,\omega \mathbf{k}_\perp)}_{\alpha j} 
        &= 
        -(k_\perp/2) 
        \nabla_\alpha \Omega_{\mathrm{s}}^{\omega {k}_\perp}
        \mathbb{S}^{\mathbf{k}_\perp}_j,
    \nonumber \\
    h^{(\mathrm s,\omega \mathbf{k}_\perp)}_{i j} 
        &= 
        k_\perp^2 
        \Omega_{\mathrm{s}}^{\omega {k}_\perp}
        \mathbb{S}_{i j}^{\mathbf{k}_\perp},
    \label{eq:scalar-perturbation-Rindler}
\end{align}
which depends only on the scalar master variable.  This completely decouples the
modes while satisfying the chosen gauge condition~\eqref{eq:TT-gauge},
which is easily verified.

Normalization is a laborious but straightforward calculation (one can find some relevant identities needed in Appendix~\ref{sect:normalization-calculations}.  We find, by choosing 
\begin{equation}
    V_{\omega k_\perp} 
    =
    \frac{S_{\omega k_\perp}}{2}
    =
    \frac{\kappa}{k_\perp^2 }
    \sqrt{\frac{\sinh(\pi\omega/a)}{4 \pi^4 a}},
    \label{eq:vector-sector-normalization-constant}
\end{equation}
and labeling with $ \mathrm{p} $ either the scalar or vector sector, that the modes satisfy the normalization conditions 
\begin{gather}
    \gkgt{
        h^{ ( \mathrm{p} , \omega \mathbf{k}_\perp ) }
        ,
        h^{ ( \mathrm{p}' , \omega' \mathbf{k}_\perp' ) }
    }
    =
    \delta_{\mathrm{p} \mathrm{p}'}
    \delta ( \omega - \omega' )
    \delta^2(\mathbf{k}_\perp-\mathbf{k}_\perp')
    ,
    \label{eq:normalization-Rindler-modes}
    \\
    \gkgt{
        h^{ ( \mathrm{p} , \omega \mathbf{k}_\perp ) }
        ,
        \overline{h^{ ( \mathrm{p}' , \omega' \mathbf{k}_\perp' ) }}
    }
    = 0.
\end{gather} 
Moreover, simple identifications allow us to write these gravitational perturbations using the scalar modes of Eq.~\eqref{eq:scalar-Rindler-Modes}. As a result, we find that the vector sector is given by
\begin{align}
    h^{(\mathrm v,\omega \mathbf{k}_\perp)}_{\alpha\beta} &= 0,
    \nonumber \\
    h^{(\mathrm v,\omega \mathbf{k}_\perp)}_{\alpha j}    
        &=
        -\mathrm{i} ( \kappa / k_\perp^2) \,
        \epsilon_{\alpha\beta} 
        \varepsilon_{j l}
        \nabla^\beta 
        \nabla^l 
        v_{\omega \mathbf{k}_\perp} 
        ,
    \nonumber \\
    h^{(\mathrm v,\omega \mathbf{k}_\perp)}_{i j}       &= 0
    ,
    \label{eq:vector-perturbation-Rindler-alt}
\end{align}
while the scalar sector corresponds to
\begin{align}
    h^{(\mathrm s,\omega \mathbf{k}_\perp)}_{\alpha\beta} 
        &= 
        (2\kappa / k_\perp^{2})
        [
            \nabla_\alpha 
            \nabla_\beta 
                v_{\omega \mathbf{k}_\perp} 
            - 
            (k_\perp^2 / 2)
                g_{\alpha\beta}
                v_{\omega \mathbf{k}_\perp} 
        ] 
        ,
    \nonumber \\
    h^{(\mathrm s,\omega \mathbf{k}_\perp)}_{\alpha j} 
        &= 
        (\kappa / k_\perp^{2})
        \nabla_\alpha 
        \nabla_j
            v_{\omega \mathbf{k}_\perp}
        ,
    \nonumber \\
    h^{(\mathrm s,\omega \mathbf{k}_\perp)}_{i j} 
        &= 
        (2\kappa / k_\perp^{2})
        [
        \nabla_i 
        \nabla_j 
            v_{\omega \mathbf{k}_\perp}
        +
        (k_\perp^2 / 2)
            g_{i j}
            v_{\omega \mathbf{k}_\perp} 
        ]
        .
    \label{eq:scalar-perturbation-Rindler-alt}
\end{align}
Both representations of tensor Rindler modes, Eqs.~\eqref{eq:vector-perturbation-Rindler}
and~\eqref{eq:scalar-perturbation-Rindler} or 
Eqs.~\eqref{eq:vector-perturbation-Rindler-alt}
and~\eqref{eq:scalar-perturbation-Rindler-alt}, are equivalent and we will use either one indistinctly, depending only on what is convenient for each situation.

It is interesting to note that the form of the modes given in Eqs.~\eqref{eq:vector-perturbation-Rindler-alt} and~\eqref{eq:scalar-perturbation-Rindler-alt} enables us to write the gravitational Rindler modes as derivation operators applied over the scalar Rindler modes~\eqref{eq:scalar-Rindler-Modes}. Hence, we can write
\begin{equation}
    h^{(\mathrm{p},\omega \mathbf{k}_\perp)}_{a b}
    =
    h^{(\mathrm{p},\omega \mathbf{k}_\perp)}_{a b}
    [v_{\omega \mathbf{k}_\perp}].
    \label{eq:gravitational-modes-as-operators}
\end{equation}
As a result, we can map from Rindler spacetime to the Rindler wedges with the aid of these ``operators'' acting on the left and right scalar modes of Eqs.~\eqref{eq:Bogoliubov-right-Rindler} and~\eqref{eq:Bogoliubov-left-Rindler}, respectively, as
\begin{align}
    V^{(\mathrm{L}, \mathrm{p},\omega \mathbf{k}_\perp)}_{a b}
    &\equiv
    h^{(\mathrm{p},\omega \mathbf{k}_\perp)}_{a b}
    [v^{\mathrm{L}}_{\omega \mathbf{k}_\perp}],
    \\
    V^{(\mathrm{R}, \mathrm{p},\omega \mathbf{k}_\perp)}_{a b}
    &\equiv
    h^{(\mathrm{p},\omega \mathbf{k}_\perp)}_{a b}
    [v^{\mathrm{R}}_{\omega \mathbf{k}_\perp}].
\end{align}
The above equations give us left and right gravitational Rindler modes that can be used to describe gravitational waves from the perspective of accelerated observers.

As we are ultimately interested in describing the radiation seen from the inertial point of view (but connecting it with the physics of accelerated observers), we need to define modes that are, in fact, associated with a congruence of observers following inertial trajectories but are labeled with quantum numbers associated with uniformly accelerated observers. These are the so-called Unruh modes and they will be defined next.

%%%%%%%%%%%%%%%%%%%%%%%%%%%%%%%%%%%%%%%%%%%%%%%%%%%%%%%%%%%%%%%%%%%%%
%%%%%%%%%%%%%%%%%%%%%       Inertial modes       %%%%%%%%%%%%%%%%%%%%
%%%%%%%%%%%%%%%%%%%%%%%%%%%%%%%%%%%%%%%%%%%%%%%%%%%%%%%%%%%%%%%%%%%%%
\subsection{Inertial modes}\label{sect:inertial-modes}

In this section greek indices refer to the $ (t,z) $ coordinates.  In deriving the form of the Rindler modes given in  Eqs.~\eqref{eq:vector-perturbation-Rindler}
and~\eqref{eq:scalar-perturbation-Rindler},  we have only used the properties of the harmonics of the plane and covariant quantities in the orbit. The only time the explicit form of the master variables was used was for the normalization integrals.  Given the fact that Minkowski spacetime has the same maximally symmetric subspace and the difference lies in the orbit, we can define
gravitational (plane waves) Minkowski modes using their own master variables, $ \Psi_{\mathrm{s}}$ and $ \Psi_{\mathrm{v}} $. The vector sector of Minkowski modes is defined by
\begin{align}
    H^{(\mathrm v, \mathbf{k})}_{\alpha\beta} &= 0,
    \nonumber \\
    H^{(\mathrm v, \mathbf{k})}_{\alpha j}    &= 
        \epsilon_{\alpha\beta} \nabla^\beta \Psi_{\mathrm{v}}^{{k}_\perp k_z} 
        \mathbb{V}_{j}^{\mathbf{k}_\perp},
    \nonumber \\
    H^{(\mathrm v, \mathbf{k} )}_{i j}       &= 0,
    \label{eq:vector-perturbation-Minkowski}
\end{align}
and the scalar sector is defined by 
\begin{align}
    H^{(\mathrm s,\mathbf{k})}_{\alpha\beta} 
        &= 
        [
            \nabla_\alpha \nabla_\beta \Psi_{\mathrm{s}}^{{k}_\perp k_z} - (k_\perp^2/2)\Psi_{\mathrm{s}}^{{k}_\perp k_z} g_{\alpha\beta}
        ] 
        \mathbb{S}^{\mathbf{k}_\perp},
    \nonumber \\
    H^{(\mathrm s,\mathbf{k})}_{\alpha j} 
        &= 
        (k_\perp/2) 
        \nabla_\alpha \Psi_{\mathrm{s}}^{{k}_\perp k_z}
        \mathbb{S}^{\mathbf{k}_\perp}_j,
    \nonumber \\
    H^{(\mathrm s,\mathbf{k})}_{i j} 
        &= 
        k_\perp^2 
        \Psi_{\mathrm{s}}^{{k}_\perp k_z}
        \mathbb{S}_{i j}^{\mathbf{k}_\perp}.
    \label{eq:scalar-perturbation-Minkowski}
\end{align}
The dynamics of these modes are governed by the field equations
\begin{equation}
     \frac{\partial^2 \Psi}{\partial z^2} 
        - 
        \frac{\partial^2 \Psi}{\partial t^2} 
    - k_\perp^2 \Psi =
    0,
    \label{eq:master-variable-Minkowski}
\end{equation}
with $ \Psi \in \{ \Psi_{\mathrm{s}}, \Psi_{\mathrm{v}} \}.$ %Note that we have used $ k_z $, the third component of the wave vector $ \mathbf{k} = (k_x,k_y,k_z) $, while $ \mathbf{k}_\perp = (k_x,k_y) $ stays invariant.
Introducing the inertial energy $ k_0 \equiv  \sqrt{k_x^2 + k_y^2 + k_z^2}$,  the positive-energy master variables are chosen to be given by
\begin{equation}
    \Psi_{\mathrm{s}}^{ k_\perp k_z}
    =
    2\Psi_{\mathrm{v}}^{ k_\perp k_z}
    =
        \frac{\kappa}{2 k_\perp^2 \sqrt{\pi^3 k_0}}
        \mathrm{e}^{\mathrm{i} (-k_0 t + k_z z)}
    ,
    \label{eq:Minkowski-master-variables}
\end{equation}
in order to have the normalization relations
\begin{gather}
    \gkgt{
        H^{(\mathrm{p}, \mathbf{k} )}
        ,
        H^{(\mathrm{p}', \mathbf{k}' )}
    }
    =
    \delta_{\mathrm{p} \mathrm{p}'} \,
    \delta^3(\mathbf{k}-\mathbf{k}'),
    \label{eq:Minkowski-modes-normalization-1}
    \\
    \gkgt{
        H^{(\mathrm{p}, \mathbf{k} )}
        ,
        \overline{H^{(\mathrm{p}', \mathbf{k}' )}}
    }
    =
    0,
    \label{eq:Minkowski-modes-normalization-2}
\end{gather}
for the modes $ H^{(\mathrm{p}, \mathbf{k} )}.$ From their explicit functional dependence, it is straightforward to prove that
these gravitational inertial modes can be constructed by derivation operators applied to the scalar inertial modes, just like we did for Rindler spacetime. It is useful to note that the left and right gravitational Rindler modes are also linear combinations of positive and negative energy Minkowski tensor modes following the same structure of the scalar field seen in
Eqs.~\eqref{eq:Bogoliubov-right-Rindler} and~\eqref{eq:Bogoliubov-left-Rindler}~\footnote{After quantization, this is all that is needed to show the Unruh effect for the linearized gravitational field.}. Therefore, we can define gravitational Unruh modes from linear combinations of left- and right-Rindler modes as
\begin{align} 
    W^{(1,\mathrm{p},\omega \mathbf{k}_\perp)}_{b c} 
    &\equiv
    \frac{
        V^{(\mathrm{R}, \mathrm{p},\omega \mathbf{k}_\perp)}_{b c} 
        + \mathrm{e}^{-\pi\omega/a} 
            \,
            \overline{
                V^{(\mathrm{L}, \mathrm{p},\omega \, -\mathbf{k}_\perp)}_{b c}
            } 
    }{ 
        \sqrt{ 1-\mathrm{e}^{-2\pi\omega/a} } },
    \\ %%%%%%%%%%%%%%%%%%%%%%%%%%%%%%%%%%%%%
    W^{(2,\mathrm{p},\omega \mathbf{k}_\perp)}_{b c} 
    &\equiv
    \frac{
        V^{(\mathrm{L}, \mathrm{p},\omega \mathbf{k}_\perp)}_{b c} 
        + 
        \mathrm{e}^{-\pi\omega/a} 
            \,
            \overline{
                V^{(\mathrm{R}, \mathrm{p},\omega \, -\mathbf{k}_\perp)}_{b c}
            } 
    }{ 
        \sqrt{ 1-\mathrm{e}^{-2\pi\omega/a} } 
    }.
\end{align}    
These are purely positive-energy modes with respect to inertial time $ t $, which implies that they can be written as a linear combination of the Minkowski modes~\eqref{eq:vector-perturbation-Minkowski}
and~\eqref{eq:scalar-perturbation-Minkowski}.  However, they are intrinsically related to the physics of accelerated observers, as they are labeled using the energy and transverse momentum as seen by the fiduciary accelerated observers. 

We can write the tensor Unruh modes, in terms of the scalar Unruh modes
\begin{align}
    w^{\sigma}_{\omega \mathbf{k}_\perp} \! (x)
    &=
    \frac{
        \mathrm{e}^{ \mathrm{i} \mathbf{k}_\perp \cdot \mathbf{x}_\perp }
    }{
        4\pi^2 \sqrt{2 a}
    }
    \int_{-\infty}^{\infty} \mathrm{d}\vartheta  
    \,
    \mathrm{e}^
        {
            \mathrm{i}
            (-1)^\sigma \vartheta \omega / a
        }
    \nonumber \\ &\qquad\quad \times
    \exp[
            \mathrm{i} 
            k_\perp 
            (
                z \sinh\vartheta - t \cosh\vartheta 
            )
        ],
    \label{eq:scalar-Unruh-modes}
\end{align}
(here written as conditionally convergent distributions over the entirety Minkowski spacetime), as
\begin{align}
    W^{(\sigma,\mathrm{v},\omega \mathbf{k}_\perp)}_{\alpha\beta} &= 0,
    \nonumber \\
    W^{(\sigma,\mathrm{v},\omega \mathbf{k}_\perp)}_{\alpha j}    
        &=
        -\mathrm{i} \frac{\kappa}{k_\perp^2} \,
        \epsilon_{\alpha\beta} 
        \varepsilon_{j l}
        \nabla^\beta 
        \nabla^l 
        w^{\sigma}_{\omega \mathbf{k}_\perp} 
        ,
    \nonumber \\
    W^{(\sigma,\mathrm{v},\omega \mathbf{k}_\perp)}_{i j}       &= 0
    ,
    \label{eq:vector-perturbation-Unruh-alt}
\end{align}
for the vector sector
and
\begin{align}
    W^{(\sigma,\mathrm{s},\omega \mathbf{k}_\perp)}_{\alpha\beta} 
        &= 
        \frac{2\kappa}{k_\perp^{2}}
        [
            \nabla_\alpha 
            \nabla_\beta 
                w^{\sigma}_{\omega \mathbf{k}_\perp} 
            - 
            (k_\perp^2 / 2)
                g_{\alpha\beta}
                w^{\sigma}_{\omega \mathbf{k}_\perp} 
        ] 
        ,
    \nonumber \\
    W^{(\sigma,\mathrm{s},\omega \mathbf{k}_\perp)}_{\alpha j} 
        &= 
        \frac{\kappa}{k_\perp^{2}}
        \nabla_\alpha 
        \nabla_j
            w^{\sigma}_{\omega \mathbf{k}_\perp}
        ,
    \nonumber \\
    W^{(\sigma,\mathrm{s},\omega \mathbf{k}_\perp)}_{i j} 
        &= 
        \frac{2\kappa}{k_\perp^{2}}
        [
        \nabla_i 
        \nabla_j 
            w^{\sigma}_{\omega \mathbf{k}_\perp}
        +
        (k_\perp^2 / 2)
            g_{i j}
            w^{\sigma}_{\omega \mathbf{k}_\perp} 
        ]
        ,
    \label{eq:scalar-perturbation-Unruh-alt}
\end{align} 
for the scalar one, where $ \sigma = 1,2 $.

Both Minkowski and Unruh tensor modes form (together with their complex conjugates) two complete sets of modes expanding any linearized gravitational perturbation in the TT gauge. They will serve as our main tools to survey the radiation content of the radiation emitted by an accelerated mass in both the classical and quantum contexts.

%%%%%%%%%%%%%%%%%%%%%%%%%%%%%%%%%%%%%%%%%%%%%%%%%%%%%%%%%%%%%%%%%%%%%
%%%%%%%%%%%%%%%       Accelerated particle       %%%%%%%%%%%%%%%%%%%%
%%%%%%%%%%%%%%%%%%%%%%%%%%%%%%%%%%%%%%%%%%%%%%%%%%%%%%%%%%%%%%%%%%%%%
\section{Accelerated particle}\label{sect:accelerated-particle}

Let us now consider a particle in Minkowski spacetime $(\mathbb{R}^4,\eta_{ab})$ uniformly
accelerated by the action of an external agent. The corresponding
worldline, parameterized by the particle's proper time $ \tau \in\mathbb{R} $, is given by 
\begin{align}
    \chi^t (\tau) &= a^{-1} \sinh(a \tau),
    &
    \chi^x (\tau) &= 0
    \nonumber
    \\
    \chi^y (\tau) &= 0,
    &
    \chi^z (\tau) &= a^{-1} \cosh(a \tau).
    \label{eq:worldline}
\end{align} 
This worldline is entirely contained in the RRW, and as such, we can use Rindler coordinates \eqref{eq:RRW-coordinate-transformation} to simplify the description:
\begin{align}
    \chi^\lambda (\tau) &= \tau,
    &
    \chi^\xi (\tau) & = \chi^x (\tau) = \chi^y (\tau) = 0,
    \label{eq:worldline-Rindler}
\end{align}
i.e., the accelerated Rindler frame is the rest frame of the particle. 

In order to describe the gravitational waves emitted by such a particle, we need the corresponding stress-energy tensor~\eqref{eq:stress-energy-general} that arises from the matter action describing the particle as well as the stress-energy tensor describing the external agent accelerating the mass.  The covariant equations of motion for the particle are given by \begin{equation}
    \frac{\mathrm{d}^2 \chi^a}{\mathrm{d}\tau^2} 
        + 
        \tensor{\Gamma}{^a_b_c} 
            \frac{\mathrm{d} \chi^b}{\mathrm{d}\tau}
            \frac{\mathrm{d} \chi^c}{\mathrm{d}\tau}  
    = 
    F^a(\chi),
    \label{eq:eom-covariant} 
\end{equation}
where $ F^a  $ is a vector field representing the accelerating agent, which can be written using inertial or Rindler coordinates as
\begin{equation}
    F^b =  a^2 t (\partial_t)^b + a^2 z (\partial_z)^b
    = a (\partial_\xi)^b.
    \label{eq:4-vector-force}
\end{equation}
The equations of motion~\eqref{eq:eom-covariant} can be obtained as the Euler-Lagrange equations for the action 
\begin{equation}
I_{\text{p}}=I_k + I_{mF}
 \label{eq:matter-action-infinite-time}
\end{equation}
where 
\begin{equation}
    I_{\text{k}}
    \equiv
    - m
    \int_{-\infty}^\infty \mathrm{d}\tau
        \sqrt{ -g_{ab} \frac{\mathrm{d} \chi^a}{\mathrm{d}\tau} \frac{\mathrm{d} \chi^b}{\mathrm{d}\tau} },
    \label{eq:matter-action-infinite-time-particle-only}
\end{equation}
 is the kinetic term corresponding to the particle and 
 
\begin{equation}
    I_{mF}
    \equiv 
    \int_{-\infty}^\infty \mathrm{d}\tau 
           \frac{m}{2 a^2} F_b F^b 
    \end{equation}
gives the interaction between the external agent and the particle.

A full matter action must consider the action of the accelerating agent, this is, an action  $I_{F}$ that carries the information of the acceleration field. Therefore, the complete matter action is
\begin{equation}
    I_{\text{mat}}
    =
    I_{\text{p}}
    +
    I_{F}.
\end{equation}
From this action, we find that the stress-energy tensor can be written as 
\begin{equation}
T_{ab}={T^A}_{ab}+{T^F}_{ab}
\label{eq:total-T}
\end{equation}
where 
\begin{multline}
    {T^A}_{a b}(x) 
    =
    \frac{m}{\sqrt{-g(x)}}
    % \\ \times
    \int_{-\infty}^{\infty}\mathrm{d}\tau 
    \, 
    % (
        u_a u_b 
    %     +
    %     a^{-2}
    %     F^a F^b
    % )
    \, 
    \delta^4 {\boldsymbol(}
        x- \chi(\tau)
    {\boldsymbol)},
    \label{eq:stress-energy}
\end{multline}
comes from the variation of Eq.~(\ref{eq:matter-action-infinite-time-particle-only}) with $g_{ab}$ and gives the contribution to the stress-energy tensor coming exclusively from the accelerated motion of the mass. The tensor ${T^F}_{ab}$ comes from the variation of $I_{mF}+I_F$ with $g_{ab}$ and describes the contribution to the total stress-energy tensor coming from the accelerating agent.  Here, $ u^a  (\tau) \equiv \mathrm{d} \chi^a / \mathrm{d}\tau $ is the 4-velocity of the particle and we have introduced the Dirac delta distribution $ \delta(x) $ that ensures the stress-energy tensor is nonzero only in the events that form the trajectory. We note that $T_{ab}$ satisfies $\nabla_a T^{ab}=0,$ as it should be.
\begin{figure}[hbt]
    \centering 
    \includegraphics[width=.4\textwidth]{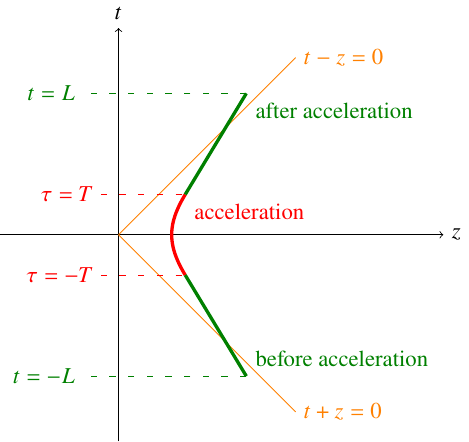}
    \caption{Support of the compactified version of the stress-energy tensor,
    showcasing the inertial (green) and accelerated (red) parts of the motion of
    the mass.}\label{fig:support}
\end{figure}

Having established the stress-energy tensors describing the uniformly accelerated particle and the external agent accelerating it, we can use Eq.~(\ref{eq:stress-energy}) in the (gauge-invariant) interaction action~(\ref{eq:interaction-action}) to write the Euler-Lagrange equations arising from the total action $$I_{\rm tot}\equiv I_{\rm pert} + I_{\rm int}$$ in transverse gauge, $\nabla^a h_{ab}=0$, as
\begin{equation}
    \nabla_c\nabla^c h_{a b} 
    = -\kappa^2 \left[{T}_{ab}-\frac{1}{2}T \eta_{ab}\right],    
    \label{eq:gravitational-field-eq-TTb}
\end{equation}
where $T\equiv \eta^{ab}T_{ab}$ and we have used the fact that ${R_{abc}}^d=0$ in Minkowski spacetime. Now, using Eq.~(\ref{eq:total-T}) in Eq.~(\ref{eq:gravitational-field-eq-TTb}) enables us to split $h_{ab}$ as 
\begin{equation}
h_{ab}\equiv h^A_{ab} + h^F_{ab}, 
\label{eq:split-h}
\end{equation}
 where $h^A_{ab}$ and $h^F_{ab}$ satisfy 
 \begin{equation}
    \nabla_c\nabla^c h^A_{a b} 
    = -\kappa^2 \left[{T^A}_{ab}-\frac{1}{2}T^A\eta_{ab}\right] 
    \label{eq:gravitational-field-eq-TThA}
\end{equation}
and
\begin{equation}
    \nabla_c\nabla^c h^F_{a b} 
    = -\kappa^2\left[{T^F}_{ab}-\frac{1}{2}T^F\eta_{ab}\right],    
    \label{eq:gravitational-field-eq-TThF}
\end{equation} 
respectively. Hence, we can see that $h^A_{ab}$ describes the gravitational waves due to the accelerated motion of the particle while $h^F_{ab}$ describes the gravitational waves coming from the external agent and its interaction with the mass. Here, we are only interested in the radiation emitted due to the accelerated motion of the particle. As a result, we will focus on $h^A_{ab}$  and its field equation, Eq.~(\ref{eq:gravitational-field-eq-TThA}).

For computational reasons, we will be interested in using a compactly supported stress-energy tensor to study the gravitational wave emission and take the physical (non-compact) limit at the end. Moreover, it is also interesting to study the case in which the acceleration time is finite (and take the infinite acceleration proper-time at the end).  In
order to include both of these conditions we introduce two positive parameters, $ L $ and $ T $, with $ L > T $, to define 
\begin{multline}
    {T^A_{L}}_{a b}(x) 
    \equiv
    \frac{m}{\sqrt{-g(x)}} \theta(L - |t|)
    \\ \times
    \int_{-\infty}^{\infty}\mathrm{d}\tau 
    % \left(
        \,
        u_a u_b
        \,
    %     +
    %     \frac{\theta(T - |\tau| )}{a^2}
    %     F^a F^b 
    % \right)
    \delta^4 {\boldsymbol(}
        x- \chi (\tau)
    {\boldsymbol)}.
    \label{eq:stress-energy-compactified}
\end{multline}
Here, $ L $ is a compactification parameter, that limits the support of the stress-energy tensor to the region of spacetime that satisfies $ -L < t < L $, while $T$ gives half of the acceleration proper-time and was introduced in such a way that the right-hand side of Eq.~\eqref{eq:eom-covariant} is turned off
outside the interval defined by $ | \tau | < T $, meaning that the particle experiences geodesic motion before and after the
acceleration. Explicitly, we can cast $\chi_L(\tau)$  in inertial coordinates as
\begin{widetext}
    \begin{equation}
        \chi^b_{L}(\tau)
        =
        \begin{cases}
            {\boldsymbol(}
                [-a^{-1} \sinh(aT) + (\tau+T) \cosh(aT)],
                0,
                0,
                [a^{-1} \cosh(aT) - (\tau+T) \sinh(aT)]
            {\boldsymbol)},
            &
            \text{if } -\Theta_{L} \leq \tau < -T,
            %%%%%%%%%%%%%%%%%%%%%%
            \\
            {\boldsymbol(}
                a^{-1} \sinh( a \tau),
                0,
                0,
                a^{-1} \cosh( a \tau ) 
            {\boldsymbol)},
            &
            \text{if } -T \leq \tau \leq T,
            %%%%%%%%%%%%%%%%%%%%%%
            \\
            {\boldsymbol(}
                [a^{-1} \sinh(aT) + (\tau-T) \cosh(aT)],
                0,
                0,
                [a^{-1} \cosh(aT) + (\tau-T) \sinh(aT)]
            {\boldsymbol)},
            &
            \text{if } T < \tau \leq \Theta_{L},
        \end{cases}
        \label{eq:worldline-compactified}
    \end{equation}
\end{widetext}
where we have defined 
\begin{equation}
    \Theta_{L} \equiv 
    T + \frac{L - a^{-1} \sinh(aT)}{\cosh(aT)} ,
    \label{eq:aux-trajectory}
\end{equation}
as an auxiliary value that describes the proper-times of ``birth'' ($\tau= - \Theta_{L} $) and ``death'' ($ \tau = \Theta_{L} $) of
the particle, as registered by an observer co-moving with the particle (which an inertial observer sees at times $ t = - L $ and $ t = L $, as explained above). The compactified trajectory can be visualized in Fig.~\ref{fig:support}

The physical setting for the particle is recovered by taking the limit $ {T^A_{\infty}}_{a b} \equiv\lim_{L\to\infty} {T^A_{L}}_{a b} $. As such, we will compute all quantities using ${T^A_{L}}_{a b} $ and then recover the true physical setup by taking the limit $ L\to\infty $ described above. We can also recover the stress-energy tensor associated with the trajectory~\eqref{eq:worldline} by taking the limit $\lim_{T\to\infty}{T^A_{\infty}}_{a b} = T^A_{a b}$.

%%%%%%%%%%%%%%%%%%%%%%%%%%%%%%%%%%%%%%%%%%%%%%%%%%%%%%%%%%%%%%%%%%%%%
%%%%   Classical expansion of the gravitational perturbation   %%%%%%
%%%%%%%%%%%%%%%%%%%%%%%%%%%%%%%%%%%%%%%%%%%%%%%%%%%%%%%%%%%%%%%%%%%%%
\section{Unruh mode expansion of the classical gravitational perturbation}
\label{sect:classical-exp}

\begin{figure}[hbt]
    \centering
    \includegraphics[width=.46\textwidth]{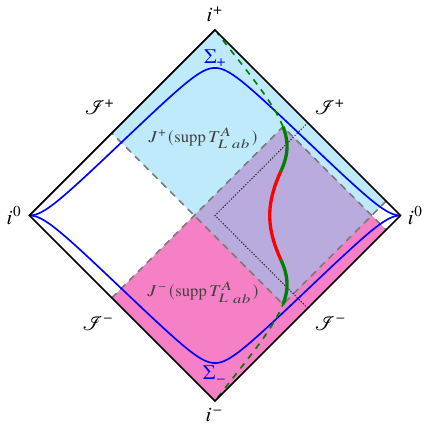}
    \caption{ 
        Conformal diagram of the setup.  
        The Cauchy surfaces are in blue.
        The green lines correspond with the inertial parts of the trajectory of
        the mass, while the accelerated portion is in red. 
        The support of the compactified version of the stress-energy tensor is the solid curve, while the limit $ L\to\infty $ is represented by the dashed extensions in green that reach the time infinities. Note the accelerated part of the motion is constrained to the RRW (delimited by the black dotted line). The causal future of the compactified trajectory is the light blue
        region, while its causal past is the light magenta one.
    }\label{fig:conformal}
\end{figure}

Let us now consider two distinct Cauchy surfaces, $ \Sigma_- $ and $ \Sigma_+ $, lying outside the causal future and past, respectively, of the support of the compactified stress-energy tensor $ {T^A_{L}}^{a b}$, see Fig.~\ref{fig:conformal} for a schematic view.

The advanced and retarded gravitational fields produced
by the compactified stress-energy tensor is obtained
from~\cite{poissonMotionPointParticles2011}

\begin{gather}
    A {\mathfrak{t}^A_L}_{a b}(x) 
    = 
    \int_{\mathbb{R}^4} \mathrm{d}^4x' \sqrt{-g(x')} 
        \tensor{{G_{\text{adv}}}}{_a_b^{c'}^{d'}}\!(x,x') \, {{\mathfrak{t}}^A_{L}}_{c' d'}(x')
    ,
    \label{eq:adv-solution}
    \\
    R {{\mathfrak{t}}^A_{L}}_{a b}(x) 
    = 
    \int_{\mathbb{R}^4} \mathrm{d}^4x' \sqrt{-g(x')} 
        \tensor{{G_{\text{ret}}}}{_a_b^{c'}^{d'}}\!(x,x') \, {{\mathfrak{t}}^A_{L}}_{c' d'}(x'),
    \label{eq:ret-solution}
\end{gather}
where we have defined ${\mathfrak{t}_L^A}_{ab}\equiv {T^A_L}_{ab}-\frac{1}{2}T^A_L\eta_{ab}$. The advanced and retarded bi-tensorial gravitational Green's functions associated with the operator $\nabla^a\nabla_a$ are reflective of the causal structure of the background spacetime. As we are working in Minkowski spacetime and we are using Cartesian coordinates, the
field equation's structure implies that 
\begin{align}
    \tensor{{G_{\text{adv}}}}{_a_b^{c'}^{d'}}(x,x') 
    &= 
    G_{\text{adv}} (x,x')
    \delta^{c'}_{(a}
    \delta^{d'}_{b)} 
    ,
    \\
    \tensor{{G_{\text{ret}}}}{_a_b^{c'}^{d'}} (x,x') 
    &= 
    G_{\text{ret}} (x,x')
    \delta^{c'}_{(a}
    \delta^{d'}_{b)} 
    ,
\end{align}
where $ G_{\text{adv}} $ and $ G_{\text{ret}} $ are the advanced and retarded
Green's functions for the scalar field. Then, we can define the regularized
particular solution by 
\begin{equation}
    E{{\mathfrak{t}}^A_{L}}_{a b}(x) 
    \equiv 
    A{\mathfrak{t}^A_{L}}_{a b}(x) 
    -
    R{\mathfrak{t}^A_{L}}_{a b}(x).
    \label{eq:regularized-particular-solution}
\end{equation}
Now, we note that for all events  $ x\in \mathbb{R}^4-J^-(\mathrm{supp}\,
\mathfrak{t}^{L}_{a b})$, the relation 
\begin{equation}
    R{\mathfrak{t}^A_{L}}_{a b}(x) 
    =
    -
    E{\mathfrak{t}^A_{L}}_{a b}(x)
    \label{eq:retarded-field-in-terms-of-regular-one}
\end{equation}
holds. In particular, this is true for all $ x\in\Sigma_+ $. 

The radiation aspects of $ R{\mathfrak{t}^A_{L}}_{a b}$ are described by its projection into the $TT$ sector. Hence, we can expand it in $\Sigma_+ $ using the Unruh modes~(\ref{eq:vector-perturbation-Unruh-alt}) and~(\ref{eq:scalar-perturbation-Unruh-alt}) as 
\begin{align}
    {^{TT\!}{R}{\mathfrak{t}^A_{L}}_{a b}}
    =
    &-
    \sum_{\sigma = 1}^2
    \sum_{\mathrm{p}=\mathrm{s},\mathrm{v}}
    \int_0^\infty\mathrm{d}\omega 
    \int_{\mathbb{R}^2}\mathrm{d}^2\mathbf{k}_\perp
    \nonumber \\ & \qquad \times
    [
        \gkgt{ W^{(\sigma,\mathrm{p},\omega\mathbf{k}_\perp)}, E\mathfrak{t}^A_{L} }
        \, 
        W^{(\sigma,\mathrm{p},\omega\mathbf{k}_\perp)}_{a b}
        \nonumber \\ & \qquad\quad +
        \overline{\gkgt{ W^{(\sigma,\mathrm{p},\omega\mathbf{k}_\perp)}, E{\mathfrak{t}^A_{L}}}}
        \, 
        \overline{W^{(\sigma,\mathrm{p},\omega\mathbf{k}_\perp)}_{a b}}
    ]
    ,
    \label{eq:classical-unruh-mode expansion}
\end{align}
where the coefficients can be obtained from the identity 
\begin{multline}
    \gkgt{ W^{(\sigma,\mathrm{p},\omega\mathbf{k}_\perp)}, E{\mathfrak{t}^A_{L}}}\\
    =
    -\frac{\mathrm{i}}{2}
    \int_{\mathbb{R}^4} 
    \mathrm{d}^4x 
    \,
    \sqrt{-g} 
    \, 
    {\mathfrak{t}^A_{L}}^{a b}
    \, 
    \overline{W^{(\sigma,\mathrm{p},\omega\mathbf{k}_\perp)}_{a b}}
    . 
    \label{eq:Unruh-expansion-coefficients}
\end{multline}
which holds for compactly supported tensors $\mathfrak{t}_{ab}$. We derive this identity in Appendix~\ref{sec:identity}. 

Now, due to the fact that the $W^{(\sigma,\mathrm{p},\omega\mathbf{k}_\perp)}$ modes are traceless, we can write 
\begin{multline}
    \int_{\mathbb{R}^4} 
    \mathrm{d}^4x 
    \,
    \sqrt{-g} 
    \, 
    {\mathfrak{t}^A_{L}}^{a b}
    \, 
    \overline{W^{(\sigma,\mathrm{p},\omega\mathbf{k}_\perp)}_{a b}}
    \\ =
    \int_{\mathbb{R}^4} 
     \mathrm{d}^4x 
    \,
    \sqrt{-g} 
    \, 
    {T^A_{L}}^{a b}
    \, 
    \overline{W^{(\sigma,\mathrm{p},\omega\mathbf{k}_\perp)}_{a b}}
    ,
\end{multline}
and thus 
\begin{equation}
    \gkgt{ W^{(\sigma,\mathrm{p},\omega\mathbf{k}_\perp)}, E{\mathfrak{t}^A_{L}}}=\gkgt{ W^{(\sigma,\mathrm{p},\omega\mathbf{k}_\perp)}, E{T^A_{L}}}.
      \label{eq:tT}
\end{equation}
Note that Eq.~(\ref{eq:tT}) implies that 
\begin{equation}
    {^{TT\!}R\mathfrak{t}^A_L}_{ab}={^{TT\!}RT^A_L}_{ab}.   
    \label{tTeq}
\end{equation}
From the setup described in Sec.~\ref{sect:accelerated-particle}, we can see from Eqs.~(\ref{eq:vector-perturbation-Unruh-alt}) and~(\ref{eq:Unruh-expansion-coefficients}) that the vector sector does not couple with the stress-energy tensor, hence
\begin{equation}
    \gkgt{ W^{(\sigma,\mathrm{v},\omega\mathbf{k}_\perp)}, ET^A_{L} }\\
    =
    0.
    \label{eq:expansion-coefficients-vector-sector}
\end{equation}

The scalar sector involves a little more work. We can first separate the contributions of the accelerated, $ \mathcal{A}^{\sigma,\omega \mathbf{k}_\perp}(T) $, and inertial,  $ \mathcal{I}^{\sigma,\omega \mathbf{k}_\perp}(T, L) $, parts of the motion as
\begin{equation}
    \gkgt{ W^{(\sigma,\mathrm{s},\omega\mathbf{k}_\perp)}, ET^A_L }
    =  
    -
    \frac{\mathrm{i}}{2}
    [
        \mathcal{A}^{\sigma,\omega \mathbf{k}_\perp}(T)
        +
        \mathcal{I}^{\sigma,\omega \mathbf{k}_\perp}(T,L)
    ].
    \label{eq:expansion-coefficients-scalar-sector-aux1}
\end{equation}
To compute $ \mathcal{A}^{\sigma,\omega \mathbf{k}_\perp}(T) $, it is more suitable to use Rindler coordinates yielding  
\begin{widetext}
    \begin{align}
        \mathcal{A}^{1,\omega \mathbf{k}_\perp}(T)
    &=
    \int_{-\infty}^\infty  \mathrm{d}\lambda 
    \int_{-\infty}^\infty  \mathrm{d}\xi 
    \int_{\mathbb{R}^2}    \mathrm{d}^2\mathbf{x}_\perp  
    \left[
        m
        \int_{-T}^{T}\mathrm{d}\tau
        \,
            u^\alpha(\tau) u^\beta(\tau)
        \,
        \delta(\lambda-\tau)
        \delta(\xi)
        \delta^2(\mathbf{x}_\perp)
    \right]
    % \nonumber \\ &\qquad\qquad\qquad
    % \times
    \overline{W^{(1,\mathrm{s},\omega\mathbf{k}_\perp)}_{\alpha\beta}(\lambda,\xi,\mathbf{x}_\perp)}
    \nonumber \\
    &=
    \frac{m}{{\sqrt{
        1- \mathrm{e}^{-2\pi \omega/a}
    }}} 
    \int_{-T}^{T}\mathrm{d}\tau
    \,
        u^\alpha(\tau) u^\beta(\tau)
    \,
    \overline{V^{(\mathrm{R},\mathrm{s},\omega\mathbf{k}_\perp)}_{\alpha\beta}}(\tau,0,0,0)
    \nonumber \\ 
    &=
    \frac{ \kappa m}{k_\perp^2} 
    \sqrt{\frac{\mathrm{e}^{\pi \omega/a}}{2\pi ^4a}}
    % \nonumber \\ &\qquad\times
    \{
        (k_\perp^2 - 2\omega^2)
        \mathrm{K}_{\mathrm{i} \omega/a}(k_\perp /a)
        + k_\perp a
        [
            \mathrm{K}_{1-\mathrm{i} \omega/a}(k_\perp /a)
            +
            \mathrm{K}_{1+\mathrm{i} \omega/a}(k_\perp /a)
        ]
    \}  
    \frac{\sin(\omega T)}{\omega},
    \label{eq:coefficient-Unruh-1-expansion}
    \end{align}
   for $\sigma=1$ and, analogously, 
    \begin{equation}
        \mathcal{A}^{2,\omega \mathbf{k}_\perp}(T)
        =
        \frac{\kappa m}{k_\perp^2} 
        \sqrt{\frac{\mathrm{e}^{-\pi \omega/a}}{2\pi ^4a}}
        \{
            (k_\perp^2 - 2\omega^2)
            \mathrm{K}_{\mathrm{i} \omega/a}(k_\perp /a)
            + k_\perp a
            [
                \mathrm{K}_{1-\mathrm{i} \omega/a}(k_\perp /a)
                +
                \mathrm{K}_{1+\mathrm{i} \omega/a}(k_\perp /a)
            ]
        \}  
        \frac{\sin(\omega T)}{\omega}
        \label{eq:coefficient-Unruh-2-expansion}
    \end{equation}
\end{widetext}
for $\sigma=2$. The inertial contributions $ \mathcal{I}^{\sigma,\omega \mathbf{k}_\perp}(T,L) $ are
not relevant to our description as we are only interested in the radiation emitted by the acceleration. They, however, must be convergent on the limit $L\to\infty $ and we prove it in  Appendix~\ref{sect:inertial}.

We can now concentrate on the case of infinite acceleration time. This corresponds to the situation of the particle following the trajectory of Eq.~\eqref{eq:worldline} with stress-energy tensor~\eqref{eq:stress-energy}.
In this case, $ \mathcal{I}^{\sigma,\omega \mathbf{k}_\perp}(\infty,\infty) = 0 $ (see Appendix~\ref{sect:inertial}), as should be the case, since there is not going to be any inertial motion. By taking the limit $T\rightarrow \infty$ we find that 
\begin{multline}
    \gkgt{ W^{(1,\mathrm{p},\omega\mathbf{k}_\perp)}, ET^A }
    =
    \gkgt{ W^{(2,\mathrm{p},\omega\mathbf{k}_\perp)}, ET^A }
    \\
    =
    -\frac{\mathrm{i} m \kappa}{ \sqrt{8\pi ^2a} } 
    \left[
        \mathrm{K}_{0}(k_\perp /a)
        + \frac{2a}{k_\perp}
            \mathrm{K}_{1}(k_\perp /a)
    \right]
    \delta(\omega).
    \label{eq:coefficient-infinite-time}
\end{multline}
where we have used the identity
\begin{equation}
    \lim_{T\to\infty}
        \omega^{-1}\sin(\omega T) 
    = 
    \pi
        \delta(\omega).
    \label{eq:delta}
\end{equation}
We can further simplify Eq.~(\ref{eq:coefficient-infinite-time})  by means of the identity~\cite{arfkenMathematicalMethodsPhysicists2013} 
$$
    \mathrm{K}_{\nu + 1} (x)
    = \mathrm{K}_{\nu - 1} (x) + (2 \nu/x) \mathrm{K}_{\nu}(x),
$$ 
which enables us to write
\begin{multline}
    \gkgt{ W^{(1,\mathrm{p},\omega\mathbf{k}_\perp)}, ET^A }
    =
    \gkgt{ W^{(2,\mathrm{p},\omega\mathbf{k}_\perp)}, ET^A}
    \\
    =
    -\frac{\mathrm{i} m \kappa}{ \sqrt{8\pi ^2a} } 
    \,
    \mathrm{K}_{2}(k_\perp /a)
    \,
    \delta(\omega).
    \label{eq:coefficient-infinite-time-reduced}
\end{multline}
From the above equations, we can see that only zero-Rindler-energy Unruh modes contribute to building the retarded field in the limit $T \to \infty$. 

Let us now explicitly compute the retarded solution~\eqref{eq:classical-unruh-mode expansion}. To this end, let us first note that, by using Eq.~(\ref{eq:coefficient-infinite-time-reduced}) together with the identity $W^{(1,\mathrm{s},\omega\mathbf{k}_\perp)}_{a b} = W^{(2,\mathrm{s},-\omega\mathbf{k}_\perp)}_{a b},$ we can  cast Eq.~\eqref{eq:classical-unruh-mode expansion} as
\begin{equation}
    {^{TT\!}}R{T}^A_{a b}
    =
    \frac{m \kappa}{ \sqrt{8\pi ^2a} }
    \int_{\mathbb{R}^2}\mathrm{d}^2\mathbf{k}_\perp
    % \\
    % \times
    \bigg[
        \mathrm{i} 
            \mathrm{K}_{2}(k_\perp /a)
        \, 
        W^{(2,\mathrm{s},0\mathbf{k}_\perp)}_{a b}
        % \nonumber \\ & \qquad\quad 
        +
        \text{c.c.}
    \bigg]
    ,
    \label{eq:classical-unruh-mode-expansion-zero-energy}
\end{equation}
where $ \text{c.c.} $ stands for complex conjugate of the expression enclosed in the bracket, and we have used Eq.~(\ref{tTeq}) to write the left-hand side in terms of ${T^A}_{ab}$. Next, by using the scalar form of the Unruh modes in the EDKU 
\begin{equation}
    w^2_{\omega \mathbf{k}_\perp} 
    =
    -\frac{
        \mathrm{i} \mathrm{e}^{\pi\omega/(2 a )}
    }{
        \sqrt{
            32 \pi^2 a 
        }
    }
    \mathrm{e}^{
        \mathrm{i} (\omega\zeta + \mathbf{k}_\perp\cdot\mathbf{x}_\perp)
    }
    \,
    \mathrm{H}^{(2)}_{\mathrm{i} \omega / a} (k_\perp \mathrm{e}^{a\eta}/a)
    ,
    \label{eq:EDKU-scalar-Unruh-mode}
\end{equation}
along with Eq.~\eqref{eq:scalar-perturbation-Unruh-alt}, we find (see Appendix~\ref{sect:bessel-integrals} for the relevant integrals needed) that the nonzero components of the perturbation are given by 
\begin{widetext}
    \begin{gather}
     {^{TT\!}}R{T}^A_{\eta \eta}
    =
     {^{TT\!}}R{T}^A_{\zeta \zeta}
    =
    \frac{m \kappa^2}{ 4 \pi a } \mathrm{e}^{2 a \eta}
    \int_{0}^{\infty} \mathrm{d} k_\perp \, k_\perp \, \mathrm{J}_0(k_\perp x_\perp)
    \,
    \mathrm{J}_{2} (k_\perp \mathrm{e}^{a\eta}/a)
    \,
    \mathrm{K}_{2}(k_\perp /a)
    ,
    \label{eq:RT-eta-eta}
    \\ %%%%%%%%%%%%%%%%%%%%%%%%%%%%%%%%%%%%%%%%%%%%%%%%%%%%%%%%%%%%%%%%%%%%%%%%%
     {^{TT\!}}R{T}^A_{\eta x}
    = 
    \frac{ m \kappa^2 }{ 4 \pi a }
    \mathrm{e}^{a\eta} 
    \cos\varphi
    \int_0^\infty \mathrm{d} k_\perp 
    \, 
    k_\perp
    \,
    \mathrm{J}_1 (k_\perp x_\perp)
    \, 
    \mathrm{J}_{1} (k_\perp \mathrm{e}^{a\eta}/a)
    \,
    \mathrm{K}_{2}(k_\perp /a)
    \label{eq:RT-eta-x}
    ,
    \\%%%%%%%%%%%%%%%%%%%%%%%%%%%%%%%%%%%%%%%%%%%%%%%%%%%%%%%%%%%%%%%%%%%%%%%%%%
    {^{TT\!}}R{T}^A_{\eta y}
    = 
    \frac{ m \kappa^2 }{ 4 \pi a }
    \mathrm{e}^{a\eta}
    \sin\varphi
    \int_0^\infty \mathrm{d} k_\perp 
    \, 
    k_\perp
    \,
    \mathrm{J}_1 (k_\perp x_\perp)
    \, 
    \mathrm{J}_{1} (k_\perp \mathrm{e}^{a\eta}/a)
    \,
    \mathrm{K}_{2}(k_\perp /a)
    ,
    \label{eq:RT-eta-y}
    \\ %%%%%%%%%%%%%%%%%%%%%%%%%%%%%%%%%%%%%%%%%%%%%%%%%%%%%%%%%%%%%%%%%%%%%%%%%
    {^{TT\!}}R{T}^A_{x y}
    =
    \frac{m \kappa^2}{ 4 \pi a }
    \sin(2\varphi) 
    \, 
    \int_0^\infty\mathrm{d}{k_\perp} 
    \, 
    k_\perp
    \,
    \mathrm{J}_{0} (k_\perp \mathrm{e}^{a\eta}/a)
    \,
    \mathrm{J}_2(k_\perp x_\perp)
    \,
    \mathrm{K}_{2}(k_\perp /a)
    ,
    \label{eq:RT-x-y}
    \\ %%%%%%%%%%%%%%%%%%%%%%%%%%%%%%%%%%%%%%%%%%%%%%%%%%%%%%%%%%%%%%%%%%%%%%%%%
     {^{TT\!}}R{T}^A_{x x}
    =
    - {^{TT\!}}R{T}^A_{y y}
    =
    \frac{m \kappa^2}{ 4 \pi a }
    \cos(2\varphi)
    \int_0^\infty\mathrm{d}{k_\perp} \, k_\perp
    \,
    \mathrm{J}_{0} (k_\perp \mathrm{e}^{a\eta}/a)
    \,
    \mathrm{J}_2(k_\perp x_\perp)
    \,
    \mathrm{K}_{2}(k_\perp /a)
    \label{eq:RT-x-x}
    .
\end{gather}
With the aid of the identities~\cite{higuchiEntanglementVacuumLeft2017}
\begin{equation}
    \int_0^\infty
    \vartheta
    \,
    \mathrm{K}_{2} (\alpha\vartheta)
    \,
    \mathrm{J}_{2} (\beta\vartheta)
    \,
    \mathrm{J}_{0} (\gamma\vartheta)
    \,
    \mathrm{d}\vartheta
    =
    -
    \frac{
        [\alpha^2+\beta^2+\gamma^2
        -
        \sqrt{
            (\alpha^2 + \beta^2 + \gamma^2)^2 - 4 \alpha^2 \beta^2
        }]^2
    }{4\alpha^2\beta^2
        \sqrt{
            (\alpha^2 + \beta^2 + \gamma^2)^2 - 4 \alpha^2 \beta^2
        }
    },
\end{equation}
valid for $ \mathrm{Re}\,\alpha > |\mathrm{Re}\, \beta| + |\mathrm{Im} \,
\gamma| ,$ and 
\begin{multline}
    \int_0^\infty
    \vartheta
    \,
    \mathrm{K}_{2} (\alpha\vartheta)
    \,
    \mathrm{J}_{1} (\beta\vartheta)
    \,
    \mathrm{J}_{1} (\gamma\vartheta)
    \,
    \mathrm{d}\vartheta
    \\
    =
    \frac{
        \alpha^2 + \beta^2 + \gamma^2
        -
        \sqrt{
            (\alpha^2
            +
            \beta^2-\gamma^2)^2
            +
            4 \alpha^2 \gamma^2
        }
    }{
        2
        \beta\gamma
    }
    \left(
        \frac{
            1
        }{
            \sqrt{
            (\alpha^2
            +
            \beta^2-\gamma^2)^2
            +
            4 \alpha^2 \gamma^2
        }
        }
        +
        \frac{
            1
        }{
            \alpha^2
        }
    \right),
\end{multline}
valid for $\gamma>0$ and $\mathrm{Re\,}\alpha > |\mathrm{Im\,}\beta|$ (see Refs.~\cite{fabrikantComputationInfiniteIntegrals2003,gradshteynTableIntegralsSeries2014}), we find that 
\begin{gather}
     {^{TT\!}}R{T}^A_{\eta \eta}
    =
     {^{TT\!}}R{T}^A_{\zeta \zeta}
    =
    \frac{m \kappa^2 a^2}{ 4 \pi }
    \,
    \left[
        \frac{
            [ a^{-2} - \mathrm{e}^{2 a\eta}/a^2 + x_\perp^2
            -
            2 a^{-1} \rho_0 (x) ]^2
        }{ 8 a^{-2} \rho_0 (x) }
    \right]
    ,
    \\ %%%%%%%%%%%%%%%%%%%%%%%%%%%%%%%%%%%%%%%%%%%%%%%%%%%%%%%%%%%%%%%%%%%%%%%%%
     {^{TT\!}}R{T}^A_{\eta x}
    = 
    \frac{ m \kappa^2 a^2 }{ 4 \pi }
    \left[
        x
        % \,
        \left(
            \frac{
                a^{-2} + x_\perp^2 + \mathrm{e}^{2 a\eta}/a^2
                -
                2 \rho_0(x) / a
            }{
                2
                x_\perp^2
            }
        \right)
        \left(
            \frac{
                1
            }{
                2 a \rho_0(x) 
            }
            +
            1
        \right)
    \right]
    ,
    \\%%%%%%%%%%%%%%%%%%%%%%%%%%%%%%%%%%%%%%%%%%%%%%%%%%%%%%%%%%%%%%%%%%%%%%%%%%
     {^{TT\!}}R{T}^A_{\eta y}
    = 
    \frac{ m \kappa^2 a^2}{ 4 \pi }
    \left[
        y
        \left(
            \frac{
                a^{-2} + x_\perp^2 + \mathrm{e}^{2 a\eta}/a^2
                -
                2 \rho_0(x) / a
            }{
                2
                x_\perp^2
            }
        \right)
        \left(
            \frac{
                1
            }{
                2 a \rho_0(x) 
            }
            +
            1
        \right)
    \right]
    ,
    \\ %%%%%%%%%%%%%%%%%%%%%%%%%%%%%%%%%%%%%%%%%%%%%%%%%%%%%%%%%%%%%%%%%%%%%%%%%
     {^{TT\!}}R{T}^A_{x y}
    =
    \frac{m \kappa^2 a^2}{ 4 \pi  }
    \left[
        x y
        \,
        \frac{
            [ a^{-2} + \mathrm{e}^{2 a\eta}/a^2 - x_\perp^2
            -
            2 a^{-1} \rho_0 (x) ]^2
        }{ 4 x_\perp^4 \rho_0 (x) }
    \right]
    ,
    \\ %%%%%%%%%%%%%%%%%%%%%%%%%%%%%%%%%%%%%%%%%%%%%%%%%%%%%%%%%%%%%%%%%%%%%%%%%
     {^{TT\!}}R{T}^A_{x x}
    =
    - {^{TT\!}}R{T}^A_{y y}
    =
    \frac{m \kappa^2 a^2 }{ 4 \pi }
    \left[
        (x^2 - y^2)
        \,
        \frac{
            [ a^{-2} + \mathrm{e}^{2 a\eta}/a^2 - x_\perp^2
            -
            2 a^{-1} \rho_0 (x) ]^2
        }{ 8 x_\perp^4 \rho_0 (x) }
    \right]
    ,
\end{gather}
\end{widetext}
where we have introduced the auxiliary distance
\begin{equation}
    \rho_0 (x)
    \equiv
    \frac{a}{2}
    \sqrt{
        \left(
            x_\perp^2 - a^{-2} \mathrm{e}^{2 a \eta} + a^{-2}
        \right)^2
        +
        \frac{4 \mathrm{e}^{2 a \eta}}{a^4}
    },
    \label{eq:covariant-distance-F}
\end{equation}
for the sake of notation.

To conclude the classical analysis of the radiation emitted by the accelerated mass, let us introduce a concept that will provide an illuminating comparison with the quantum calculations to follow. We define the so-called ``classical graviton number" radiated by the system (mass$+$accelerating agent) as seen by inertial observers) as~\cite{HMCFT}

\begin{eqnarray}
N_{\rm M}\equiv \langle KRT, KRT\rangle_{\mathfrak{t}},   
\label{ClassicalN}
\end{eqnarray}
where 
\begin{align}
    KRT_{a b}
    =
    &-
    \sum_{\sigma = 1}^2
    \sum_{\mathrm{p}=\mathrm{s},\mathrm{v}}
    \int_0^\infty\mathrm{d}\omega 
    \int_{\mathbb{R}^2}\mathrm{d}^2\mathbf{k}_\perp
    \nonumber \\ & \qquad \times
           \gkgt{ W^{(\sigma,\mathrm{p},\omega\mathbf{k}_\perp)}, ET}
        \, 
        W^{(\sigma,\mathrm{p},\omega\mathbf{k}_\perp)}_{a b},
        \nonumber \\ 
    \label{eq:KET}
\end{align}
is the (inertial) positive-energy part of the retarded solution $^{TT\!}RT$ with the limit $L\rightarrow \infty$ already been taken. As $T_{ab}=T_{ab}^A + T_{ab}^F$ we have that
\begin{eqnarray}
N_{\rm M}= N_{\rm M}^A +  N_{\rm M}^F + 2{\rm Re}\left\{
\langle KRT^A, KRT^F\rangle_{\mathfrak{t}}\right\},   
\label{ClassicalNb}
\end{eqnarray}
with 
\begin{eqnarray}
N^A_{\rm M}\equiv \langle KRT^A, KRT^A\rangle_{\mathfrak{t}},   
\label{ClassicalNA}
\end{eqnarray}
and 
\begin{eqnarray}
N_{\rm M}^F\equiv \langle KRT^F, KRT^F\rangle_{\mathfrak{t}}.  
\label{ClassicalNF}
\end{eqnarray}
In the limit of infinite acceleration proper-time we have that
\begin{equation}
 KR{T}^A_{a b}
    =
    \frac{m \kappa}{ \sqrt{8\pi ^2a} }
    \int_{\mathbb{R}^2}\mathrm{d}^2\mathbf{k}_\perp
    % \\
    % \times
        \mathrm{i} 
            \mathrm{K}_{2}(k_\perp /a)
        \, 
        W^{(2,\mathrm{s},0\mathbf{k}_\perp)}_{a b},
        % \nonumber \\ & \qquad\quad 
   \label{positiveRTA}
\end{equation}
 By using Eq.~(\ref{positiveRTA}) together with the orthonormality relations 
\begin{multline}
    \gkgt{ W^{(\sigma,\mathrm{v},\omega\mathbf{k}_\perp)},  W^{(\sigma',\mathrm{v},\omega'\mathbf{k'}_\perp)} }\\
    \\= \delta_{\sigma {\sigma'}}\delta(\omega -\omega')\delta^2(\mathbf{k}_\perp-\mathbf{k}_\perp')
\end{multline}
in Eq.~(\ref{ClassicalN}) one finds 
\begin{equation}
N_{\rm M}^A= \frac{m^2 \kappa^2}{16\pi^3 a}
    \, 
    T_{\mathrm{tot}} 
    \int_{\mathbb{R}^2}\mathrm{d}^2\mathbf{k}_\perp 
    \,
    \left[
        \mathrm{K}_{2}(k_\perp /a)
    \right]^2.
\end{equation}
which implies that the classical number of gravitons per transverse momentum, $k_\perp$, per acceleration proper-time, $T_{\rm tot}$, radiated by the mass due to its accelerated motion is given by
\begin{equation}
\frac{1}{T_{\rm out}}\frac{\mathrm{d} N^A_{k_\perp}}{\mathrm{d} k_\perp}= \frac{m^2 \kappa^2 k_\perp }{8\pi^2 a} \left[\mathrm{K}_{2}(k_\perp /a)\right]^2.
\label{classicalNgrav}
\end{equation}
%%%%%%%%%%%%%%%%%%%%%%%%%%%%%%%%%%%%%%%%%%%%%%%%%%%%%%%%%%%%%%%%%%%%%
%%%%%%   Quantum expansion of the gravitational perturbation   %%%%%%
%%%%%%%%%%%%%%%%%%%%%%%%%%%%%%%%%%%%%%%%%%%%%%%%%%%%%%%%%%%%%%%%%%%%%
\section{Unruh mode expansion of the quantum  gravitational perturbation}
\label{sect:quantum-exp}

Let us now perform the quantum analysis of the gravitational perturbations due to the accelerating mass around Minkowski spacetime. To this end, we promote the gravitational perturbations $ h_{a b} $ and its associated
generalized momentum $ \pi^{c a b} $ to operators $ \hat{h}_{a b} $ and $\hat{\pi}^{c a b} $, respectively, and impose the equal-time canonical commutation relations 
\begin{gather}
    [
        \hat{h}_{a b}(t,\mathbf{x})
        ,
        \hat{h}_{c d}(t,\mathbf{x}')
    ]
    = 0,
    \label{eq:quantization1}
    \\
    [
        \hat{h}_{a b}(t,\mathbf{x})
        ,
        n_e \hat{\pi }^{e c d}(t,\mathbf{x}')
    ]
    =
    - \mathrm{i}\delta^c_{(a} \delta^d_{b)} \delta^3(\mathbf{x}-\mathbf{x}'),
    \label{eq:quantization2}
    \\
    [
        \hat{\pi }^{f a b}(t,\mathbf{x})
        ,
        \hat{\pi }^{e c d}(t,\mathbf{x}')
    ]
    =
    0,
    \label{eq:quantization3}
\end{gather}
where $ n^a = (\partial_t)^a$ is the future-oriented normal vector orthogonal to the Cauchy surface defined by $\Sigma_{t={\rm cte.}}\equiv \{(t, {\bf x})\in \mathbb{R}^4| t={\rm cte.}\} $. 

Again, for calculational purposes, we will consider our compactified worldline which gives rise to the compactified stress-energy tensor ${T_L}_{ab}\equiv {T^A_L}_{ab} +{T^F_L}_{ab},$ where we recall that ${T^A_L}_{ab}$ is the contribution to the total stress-energy tensor coming from the accelerated mass while ${T^F_L}_{ab}$ is the contribution coming from the external agent accelerating it. In the end, we will take the limit $L\rightarrow \infty$ to recover our physical set-up. The stress-energy tensor $ {T_L}_{a b} $ will be the classical source for the quantum field $\hat{h}_{ab}$ and, in what follows, we will use two Cauchy surfaces $ \Sigma_+$ (asymptotic future),
and $ \Sigma_- $ (asymptotic past), like in Sec.~\ref{sect:classical-exp} (see Fig.~\ref{fig:conformal}). 

The influence of the classical source $ {T_L}_{a b} $ on the quantized gravitational perturbation $\hat{h}_{a b}$ is  determined by the linearized Einstein equation 

\begin{equation}
    \nabla^c\nabla_c \hat{h}_{a b} =-\kappa^2 {\mathfrak{t}_L}_{ab}\hat{\mathbb{I}}.
    \label{SourcewaveEq}
\end{equation}
One possible solution to the above equation is given by
\begin{equation}
    \hat{h}_{a b} (x)= \hat{h}^{\text{out}}_{a b}(x) + {A\mathfrak{t}_L}_{a b}(x) \hat{\mathbb{I}},
    \label{eq:quantized-field-past}
\end{equation}
where $ A{\mathfrak{t}_L}_{a b} $ is the advanced 
solution of Eq.~(\ref{SourcewaveEq}) and $ \hat{h}^{\text{out}}_{a b} $  satisfies  the homogeneous field equation in the TT gauge
\begin{equation}
    \nabla^c\nabla_c \hat{h}^{\text{out}}_{a b} = 0.
    \label{waveEq}
\end{equation}
 Given a choice of orthhornormal Minkowski positive-energy modes $ v_{ab}^{(j)}(x) $ in the TT gauge, characterized by appropriate quantum numbers $ j\in \mathfrak{J} $, we can expand the homogeneous field as 
\begin{equation}
    \hat{h}^{\text{out}}_{a b}
    =
    \sum_{j\in \mathfrak{J}} \left[
        v_{ab}^{(j)} 
        \,
        \hat{a}_{\text{out}}(\overline{v^{(j)}})
        +
        \overline{v_{ab}^{(j)}} 
        \,
        \hat{a}^\dagger_{\text{out}}(v^{(j)})
    \right].
    \label{eq:expansion-out-field}
\end{equation}
The vacuum state of such a construction is the state $ |0^{\text{M}}_{\text{out}}\rangle$ defined from the action of the annihilation operators over it by 
$$
\hat{a}_{\text{out}}(\overline{v^{(j)}})|0^{\text{M}}_{\text{out}}\rangle = 0 $$
for all $ j\in \mathfrak{J} $. We can see that, as $ A {\mathfrak{t}_L}_{a b}(x) = 0 $ for all
$ x  \in \mathbb{R}^4-J^-(\supp {\mathfrak{t}_L}_{a b})$, we can interpret the state $|0^{\text{M}}_{\text{out}}\rangle $ as the vacuum seen by inertial observers in the asymptotic future. 

Alternatively, we can also write the quantized field as the following solution to Eq.~(\ref{SourcewaveEq}): 
\begin{equation}
    \hat{h}_{a b} (x)= \hat{h}^{\text{in}}_{a b}(x) + R{\mathfrak{t}_L}_{a b}(x) \hat{\mathbb{I}},
    \label{eq:quantized-field-future}
\end{equation}
where $ R{\mathfrak{t}_L}_{a b}$ is the classical retarded field and $\hat{h}^{\text{in}}_{a b} $ is a homogeneous solution of the linearized Einstein field equations in the TT gauge [Eq.~(\ref{waveEq})]. The latter can be expanded using (another) orthonormal set of (Minkowski) positive-energy modes $ u_{a b}^{(k)}(x) $ in the TT gauge as
\begin{equation}
    \hat{h}^{\text{in}}_{a b}
    =
    \sum_{k\in \mathfrak{K}} \left[
        u_{ab}^{(k)} 
        \,
        \hat{a}_{\text{in}}(\overline{u^{(k)}})
        +
        \overline{u_{ab}^{(k)}} 
        \,
        \hat{a}^\dagger_{\text{in}}(u^{(k)})
    \right],
    \label{eq:expansion-in-field}
\end{equation}
where $\mathfrak{K}$ is a suitable set of quantum numbers. We can define the in-vacuum state $ |0^{\text{M}}_{\text{in}}\rangle $ by the action of the annihilation operator over it: $ \hat{a}_{\text{in}}(\overline{u^{(k)}})|0^{\text{M}}_{\text{in}}\rangle = 0 $ for all $ k\in\mathfrak{K} $. Following the same reasoning as above, we can see that $R{\mathfrak{t}_L}_{a b}(x) = 0 $ for all $ x
\in \mathbb{R}^4\ - J^+(\supp {\mathfrak{t}_L}_{a b}),$ which implies that $|0^{\text{M}}_{\text{in}}\rangle $ can be interpreted as the vacuum seen by inertial observers in the asymptotic past. 

The Fock space for each representation is built by the successive application of the creation operators $ \hat{a}^\dagger_{\text{in}}(u^{(k)}) $
and $ \hat{a}^\dagger_{\text{out}}(v^{(j)}) $ over their respective associated
vacua $ |0^{\text{M}}_{\text{in}}\rangle $ and $
|0^{\text{M}}_{\text{out}}\rangle $, respectively. These Fock spaces can be connected using the $S$-matrix~\cite{itzyksonQuantumFieldTheory2005}
\begin{align}
    \hat{S} 
    &\equiv
    \exp(
        \mathrm{i}
        I_{\text{int}}[ \hat{h}^{\text{out}} , T_L]
    )
    \nonumber\\
    &=
    \exp\left(
        \frac{ \mathrm{i} }{2}
        \int_{\mathbb{R}^4} \mathrm{d}^4 x \, \sqrt{-g}
        \,
        \hat{h}_{a b}^{\text{out}} 
        \,
        {\mathfrak{t}_L}^{ab}
    \right)
    \label{eq:s-matrix-definition0}
\end{align}
which relates the vacua by 
\begin{equation}
    |0^{\text{M}}_{\text{in}}\rangle 
    =
    \hat{S}
    |0^{\text{M}}_{\text{out}}\rangle   .
    \label{eq:vacua-connection-s-matrix}
\end{equation}
Note that, as $ \hat{h}_{a b}^{\text{out}} $ is in the TT gauge, we can write the $S$-matrix~(\ref{eq:s-matrix-definition0}) as 
\begin{align}
    \hat{S} 
    &=
    \exp\left(
        \frac{ \mathrm{i} }{2}
        \int_{\mathbb{R}^4} \mathrm{d}^4 x \, \sqrt{-g}
        \,
        \hat{h}_{a b}^{\text{out}} 
        \,
        {T_L}^{ab}
    \right).
    \label{eq:s-matrix-definition}
\end{align}

Now, by using that ${T_L}_{ab}={T^A_L}_{ab}+{T^F_L}_{ab}$, we can cast Eq.~(\ref{eq:s-matrix-definition}) as 
\begin{equation}
    \hat{S}=e^{{\rm i}\Theta}\hat{S}^F\hat{S}^A,
    \label{Smatrixsplit}
\end{equation}
where 
\begin{align}
    \hat{S}^F 
    &\equiv 
    \exp\left(
        \frac{ \mathrm{i} }{2}
        \int_{\mathbb{R}^4} \mathrm{d}^4 x \, \sqrt{-g}
        \,
        \hat{h}_{a b}^{\text{out}} 
        \,
        {T^F_L}^{ab}
    \right),
    \label{eq:Fs-matrix-definition}
\end{align}
\begin{align}
    \hat{S}^A 
    &\equiv 
    \exp\left(
        \frac{ \mathrm{i} }{2}
        \int_{\mathbb{R}^4} \mathrm{d}^4 x \, \sqrt{-g}
        \,
        \hat{h}_{a b}^{\text{out}} 
        \,
        {T^A_L}^{ab},
    \right).
    \label{eq:As-matrix-definition}
\end{align}
and 
\begin{align}
    \Theta  
    &\equiv 
         \int_{\mathbb{R}^4} \mathrm{d}^4 x \, \sqrt{-g}\int_{\mathbb{R}^4} \mathrm{d}^4 x' \, \sqrt{-g'} \times  \nonumber \\
        \,
      &  \Delta_{a b cd}(x,x')
        \,
        {T^F_L}^{ab}(x){T^F_L}^{cd}(x'),
    \label{eq:Fs-matrix-definition}
\end{align}
with 
\begin{equation}
\left[\hat{h}_{a b}^{\text{out}}(x),  \hat{h}_{cd}^{\text{out}} (x') \right ]\equiv {\rm i} \Delta_{a b c d}(x,x')\mathbb{I}.
\label{h-commutator}
\end{equation}
The operator $\hat{S}^A$ describes the graviton production due to the accelerating particle while $\hat{S}^F$ describes gravitons produced due to the accelerating source and its interaction with the mass (both as seen by inertial observers in the asymptotic future).  

Let us analyze first the action of $\hat{S}^A$ on $|0^{\rm M}_{\rm  out}\rangle$. To this end, we first write $\hat{h}_{a b}^{\text{out}}$ using tensor Unruh modes. In this case $ j = (\sigma,\mathrm{p},\omega,\mathbf{k}_\perp) $, $ \mathfrak{J} =
\{1,2\} \times \{\mathrm{v},\mathrm{s}\} \times [0,\infty) \times \mathbb{R}^2$, and the annihilation and creation operators are found directly from the inner product thanks to the normalization of the modes:
\begin{align}
    \hat{a}_{\text{out}}(\overline{W^{(\sigma,\mathrm{p},\omega\mathbf{k}_\perp)}}) 
    &=
    \gkgt{
        W^{(\sigma,\mathrm{p},\omega\mathbf{k}_\perp)}
        , 
        \hat{h} 
    } 
    ,
    \label{eq:unruh-annihilator}
    \\
    \hat{a}^\dagger_{\text{out}}(W^{(\sigma,\mathrm{p},\omega\mathbf{k}_\perp)}) 
    &= 
    \gkgt{
        \hat{h} 
        , 
        W^{(\sigma,\mathrm{p},\omega\mathbf{k}_\perp)}
    }.
    \label{eq:unruh-creation}
\end{align}
Then, the $S$-matrix is explicitly given by 
\begin{widetext}
    \begin{equation}
    \hat{S}^A 
    = 
    \exp\Bigg[
        \frac{ \mathrm{i} }{2}
        \int_{\mathbb{R}^4} \mathrm{d}^4 x \, \sqrt{-g}
        \,
        {T^A_L}^{a b}
        %%%%%%%%%%%
        % \nonumber \\ & \quad\times
        % \\ \times
        \Bigg(
            \sum_{\sigma,\mathrm{p}}
            \int_0^\infty\mathrm{d}\omega
            \int_{\mathbb{R}^2}\mathrm{d}^2\mathbf{k}_\perp
            %%%%%%%%%%%
            % \nonumber \\ & \quad\times
            % \\ \times
            \left[
                W^{(\sigma,\mathrm{p},\omega\mathbf{k}_\perp)}_{a b}
                \hat{a}_{\text{out}}(\overline{W^{(\sigma,\mathrm{p},\omega\mathbf{k}_\perp)}})
                +
                \text{H.c.}
            \right]
        \Bigg)
    \Bigg].
    \label{eq:s-matrix-UM-1}
\end{equation}
\end{widetext}
Here, $ \text{H.c.} $ stands for Hermitian conjugate of the expression before. This can be simplified by defining the annihilation and creation operators associated with the negative and positive energy part of the expansion, $\overline{KET^A_L} $ and $ KET^A_L $, as
\begin{align}
    \hat{a}_{\text{out}}(\overline{KET^A_L})
    &\equiv
    \sum_{\sigma,\mathrm{p}}
    \int_0^\infty\mathrm{d}\omega
    \int_{\mathbb{R}^2}\mathrm{d}^2\mathbf{k}_\perp
    \nonumber \\ &\quad\times
        \overline{\gkgt{ W^{(\sigma,\mathrm{p},\omega\mathbf{k}_\perp)}, ET^A_L }} 
        \,
        \hat{a}_{\text{out}}(\overline{W^{(\sigma,\mathrm{p},\omega\mathbf{k}_\perp)}})
    ,
    \label{eq:positive-energy-annihilation-operator}
\end{align}
\begin{align}
    \hat{a}^\dagger_{\text{out}}(KET^A_L)
    &\equiv
    \sum_{\sigma,\mathrm{p}}
        \int_0^\infty\mathrm{d}\omega
        \int_{\mathbb{R}^2}\mathrm{d}^2\mathbf{k}_\perp
        \nonumber \\ &\quad\times
            \gkgt{ W^{(\sigma,\mathrm{p},\omega\mathbf{k}_\perp)}, ET^A_L } 
            \,
            \hat{a}^\dagger_{\text{out}}(W^{(\sigma,\mathrm{p},\omega\mathbf{k}_\perp)})
    ,
    \label{eq:positive-energy-creation-operator}
\end{align}
respectively. Rearranging the integrals in Eq.~\eqref{eq:s-matrix-UM-1} and using Eq.~\eqref{eq:Unruh-expansion-coefficients} allow us to write 
\begin{equation}
    \hat{S}^A = \exp[
        \hat{a}_{\text{out}}(\overline{KET^A_L})
        -
        \hat{a}^\dagger_{\text{out}}(KET^A_L)
    ].
    \label{eq:s-matrix-UM-3}
\end{equation}

Now, we can apply the Zassenhaus formula 
$$
\mathrm{e}^{\hat{X}+\hat{Y}} = \mathrm{e}^{\hat{X}} \mathrm{e}^{\hat{Y}}
\mathrm{e}^{-[\hat{X},\hat{Y}]/2}, $$ 
valid when the operators $ \hat{X}$  and $
\hat{Y} $ satisfy $ \boldsymbol[ \hat{X} , [\hat{X},\hat{Y}] \boldsymbol] =
\boldsymbol[ \hat{Y} , [\hat{X},\hat{Y}] \boldsymbol] = 0 $, together with the identity
\begin{equation}
    [
        \hat{a}_{\text{out}}(\overline{KET^A_L})
        ,
        \hat{a}^\dagger_{\text{out}}(KET^A_L)
    ]
    =
    \| KET^A_L \|^2 \hat{\mathbb{I}}
\end{equation}
to write
\begin{equation}
    \hat{S}^A
    =
    \mathrm{e}^{-\| KET^A_L \|^2/2}
    \,
    \mathrm{e}^{ -\hat{a}^\dagger_{\text{out}}(KET^A_L) }
    \,
    \mathrm{e}^{ \hat{a}_{\text{out}}(\overline{KET^A_L}) }
    \label{eq:s-matrix-final},
    \end{equation}
where 
\begin{multline}
    \| KET^A_L \|^2
    \equiv 
    \gkgt{KET^A_L,KET^A_L}
    \\
    =
    \sum_{\sigma,p}
    \int_0^\infty\mathrm{d}\omega 
    \int_{\mathbb{R}^2}\mathrm{d}^2\mathbf{k}_\perp
    \,
    |\gkgt{ W^{(\sigma,\mathrm{p},\omega\mathbf{k}_\perp)}, ET^A_L }|^2
    .
\end{multline}
    
Now, applying Eq.~(\ref{eq:s-matrix-final}) to $|0^{\rm M}_{\rm out}\rangle$ yields 
\begin{equation}
    \hat{S}^A|0^{\text{M}}_{\text{out}}\rangle 
    =
    \mathrm{e}^{-\| KET^A_L \|^2/2}
    \,
    \mathrm{e}^{ -\hat{a}^\dagger_{\text{out}}(KET^A_L) }
    |0^{\text{M}}_{\text{out}}\rangle.
    \label{eq:vacua-connection-s-matrix-2}
\end{equation}
We can further work on this expression to show that $ \hat{S}^A|0^{\text{M}}_{\text{out}}\rangle $ is a coherent state according to future inertial observers. To this end, we apply $
\hat{a}_{\text{out}}(\overline{W^{(\sigma,\mathrm{p},\omega\mathbf{k}_\perp)}}) $
to it
\begin{widetext}
    \begin{equation}
        \hat{a}_{\text{out}}(\overline{W^{(\sigma,\mathrm{p},\omega\mathbf{k}_\perp)}})
        \hat{S}^A|0^{\text{M}}_{\text{out}}\rangle 
        % &=
        =
        \mathrm{e}^{-\| KET_L \|^2/2}
        \mathrm{e}^{ -\hat{a}^\dagger_{\text{out}}(KET_L) }
        % \nonumber
        % \\ 
        % &\quad \times
        [
            \mathrm{e}^{ \hat{a}^\dagger_{\text{out}}(KET_L) }
            \hat{a}_{\text{out}}(\overline{W^{(\sigma,\mathrm{p},\omega\mathbf{k}_\perp)}})
            % \nonumber
            % \\ 
            % &\qquad \times
            \mathrm{e}^{ - \hat{a}^\dagger_{\text{out}}(KET_L) }
        ]
        |0^{\text{M}}_{\text{out}}\rangle,
        \label{eq:coherent-state-0}
    \end{equation}
\end{widetext}
use the identity 
\begin{equation}
    e^{\hat X} \hat Y e^{-\hat{X}} 
    =
    \hat Y 
    +
    [\hat X, \hat Y]
    +
    \frac{1}{2!} \boldsymbol[
        \hat X, [\hat X, \hat Y]
    \boldsymbol]
    +
    \ldots, 
    \label{eq:expBKH}
\end{equation}
with $ \hat X = \hat{a}^\dagger_{\text{out}}(KET^A_L) $ and $
\hat Y = \hat{a}_{\text{out}}(W^{(\sigma,\mathrm{p},\omega\mathbf{k}_\perp)}) $, and the commutator
\begin{multline}
    [
        \hat{a}^\dagger_{\text{out}}(KET^A_L)
        ,
        \hat{a}_{\text{out}}(\overline{W^{(\sigma,\mathrm{p},\omega\mathbf{k}_\perp)}})
    ]\\
    =
    - 
    \gkgt{ W^{(\sigma,\mathrm{p},\omega\mathbf{k}_\perp)}, ET^A_L } 
    \hat{\mathbb{I}} ,
\end{multline}
(which implies that the series of Eq.~\eqref{eq:expBKH} will be truncated after the second term) to obtain 
\begin{equation}
    \hat{a}_{\text{out}}(\overline{W^{(\sigma,\mathrm{p},\omega\mathbf{k}_\perp)}})
     \hat{S}^A|0^{\text{M}}_{\text{out}}\rangle 
    =
    -\gkgt{ W^{(\sigma,\mathrm{p},\omega\mathbf{k}_\perp)}, ET^A_L }
     \hat{S}^A|0^{\text{M}}_{\text{out}}\rangle 
    ,
    \label{eq:coherent}
\end{equation}
i.e., $ \hat{S}^A|0^{\text{M}}_{\text{out}}\rangle $ is a multimode coherent state associated with an arbitrary Unruh mode in the asymptotic future. 

We can now consider the case where the acceleration time is infinite. This is done by first taking the limit $L\rightarrow \infty$  and then taking $T\rightarrow \infty$. By doing so, we can see that the creation operator that appears in Eq.~\eqref{eq:vacua-connection-s-matrix-2} can be explicitly written using the  coefficients in Eq.~\eqref{eq:coefficient-infinite-time-reduced} as
\begin{multline}
    \hat{a}^\dagger_{\text{out}}(KET^A)
    =
        -\frac{\mathrm{i} m \kappa}{ \sqrt{8\pi ^2a} } 
        \int_{\mathbb{R}^2}\mathrm{d}^2\mathbf{k}_\perp
        \\ \times
            \mathrm{K}_{2}(k_\perp /a)
            \,
            \hat{a}^\dagger_{\text{out}}(W^{(2,\mathrm{s},0\mathbf{k}_\perp)})
    ,
    \label{eq:positive-energy-creation-operator-inifnite-time}
\end{multline}
from which we can cast $ \hat{S}^A|0^{\text{M}}_{\text{out}}\rangle $ as

\begin{widetext}
    \begin{equation}
        \hat{S}^A|0^{\text{M}}_{\text{out}}\rangle 
        =
        \bigotimes_{\mathbf{k}_\perp \in \mathbb{R}^2}
        \exp\!
            \left[
                -
                \frac{m^2 \kappa^2}{16\pi^3 a}
                \, 
                T_{\mathrm{tot}} 
                \int_{\mathbb{R}^2}\mathrm{d}^2\mathbf{k}_\perp 
                \,
                \left[
                    \mathrm{K}_{2}(k_\perp /a)
                \right]^2
            \right]
        \exp\!
            \left[
                \frac{\mathrm{i} m \kappa}{ \sqrt{8\pi ^2a} }
                    \mathrm{K}_{2}(k_\perp /a)
                    \,
                \hat{a}^\dagger_{\text{out}}(W^{(2,\mathrm{s},0\mathbf{k}_\perp)})
            \right]
        |0^{\text{M}}_{\text{out}}\rangle   
        ,
        \label{eq:vacua-connection-s-matrix-2-infinite-time-explicit}
    \end{equation}
\end{widetext}
where we have used  
\begin{align} 
 \| KET^A \|^2 = \frac{m^2 \kappa^2}{16\pi^3 a}
    \, 
    T_{\mathrm{tot}} 
    \int_{\mathbb{R}^2}\mathrm{d}^2\mathbf{k}_\perp 
    \,
    \left[
        \mathrm{K}_{2}(k_\perp /a)
    \right]^2,
    \label{norm KET}
\end{align}
and $ T_{\mathrm{tot}}=2\pi \delta(\omega)|_{\omega=0}.$
Equation~(\ref{eq:vacua-connection-s-matrix-2-infinite-time-explicit}) showcases the fact that only zero-Rindler-energy Unruh modes participate in building the vacuum.

Now, let us look at the expectation value of the field $\hat{h}^{\rm out}_{ab}$ in the in-vacuum. In order to do this, let us use Eqs.~(\ref{eq:Fs-matrix-definition}),~(\ref{h-commutator}), and~(\ref{eq:expBKH}) to write

\begin{equation}
    \hat{S}^{F^{\dagger}} \hat{h}^{\rm out}_{ab}{\hat{S}}^F =\hat{h}^{\rm out}_{ab} - {^{TT\!}E}{T^F}_{ab} \hat{\mathbb{I}}
    \label{eq:SFh}
\end{equation}
where we have used that $\Delta_{abcd}(x,x')= {G_{\rm adv}}_{abcd}(x,x')-{G_{\rm ret}}_{abcd}(x,x')$,  
\begin{equation}
    ^{TT\!}E{T^F}_{ab}(x)=\int_{\mathbb{R}^4}d x'\sqrt{-g'}\Delta_{abcd}(x,x'){T^F}^{cd}(x'),
\end{equation}
and we have taken the $L\rightarrow \infty $ limit. Note that, as $\hat{h}^{\rm out}_{ab}$ is on the TT gauge, $^{TT\!}E{T^F}_{ab}$ above is already projected in such a subspace of solutions. By using Eqs.~(\ref{eq:vacua-connection-s-matrix}),~(\ref{Smatrixsplit}), and~(\ref{eq:SFh}) one can write 

\begin{equation}
    \langle 0^{\text{M}}_{\text{in}} |
    \hat{h}^{\text{out}}_{a b}
    |0^{\text{M}}_{\text{in}}\rangle
    =    \langle 0^{\text{M}}_{\text{out}} |\hat{S}^{A^\dagger} 
    \hat{h}^{\text{out}}_{a b}
    \hat{S}^A|0^{\text{M}}_{\text{out}}\rangle
    - {^{TT\! }E}T^F_{a b},
    \label{eq:expectation-value-evolved-fielda}
\end{equation}
which, by means of Eqs.~(\ref{eq:expansion-out-field}) and~(\ref{eq:coherent}), can be cast as
\begin{eqnarray}
    &&  \!\!\!\!\langle 0^{\text{M}}_{\text{in}} |
    \hat{h}^{\text{out}}_{a b}
  |0^{\text{M}}_{\text{in}}\rangle
   \! = \!\!\sum_{\sigma,\mathrm{p}}
        \int_0^\infty \!\!\mathrm{d}\omega
        \int_{\mathbb{R}^2}\!\mathrm{d}^2\mathbf{k}_\perp
           \left[ \gkgt{ W^{(\sigma,\mathrm{p},\omega\mathbf{k}_\perp)}, ET^A_L }\right. \nonumber \\ 
           &&\times \left.  W^{(\sigma,\mathrm{p},\omega\mathbf{k}_\perp)} + {\rm H.c}\right]-{^{TT\!}}ET^F_{a b}.
    \label{eq:expectation-value-evolved-fieldb}
\end{eqnarray}
We note that the first term in the above equation is the expansion of $ET_{ab}^A$ in terms of Unruh modes and hence, it yields ${^{TT\!}}ET^A_{a b}$. By using that, in $\Sigma_+$, we have  $^{TT\!}E{T^A}_{ab}=-^{TT\!}R{T^A}_{ab}$ and $^{TT\!}E{T^F}_{ab}=-^{TT\!}R{T^F}_{ab},$ one can write

\begin{eqnarray}
    &\langle 0^{\text{M}}_{\text{in}} |
    \hat{h}^{\text{out}}_{a b}
    |0^{\text{M}}_{\text{in}}\rangle
    ={^{TT\!}R}T^A_{a b} + {^{TT\!}R}T^F_{a b}.
    \label{eq:expectation-value-evolved-field}
\end{eqnarray}
Hence, the expectation value of the field after the interaction with the source is given by the retarded fields. Furthermore, all other physical observables will correspond with their classical counterparts. In particular, when one takes the limit of infinite acceleration for the mass, one can see from Eq.~(\ref{eq:classical-unruh-mode-expansion-zero-energy}) that only zero Rindler-energy Unruh gravitons contribute to the radiation emitted by the accelerated mass.

Moreover, we can study the expectation value of the total number of Unruh gravitons 
\begin{multline}
    \langle \hat{N} \rangle \equiv
    \sum_{\sigma,\mathrm{p}} 
    \int_0^\infty \mathrm{d}\omega 
    \int_{\mathbb{R}^2}\mathrm{d}^2\mathbf{k}_\perp 
    % \\ \times
    \langle 0^{\text{M}}_{\text{in}} |
        \hat{N}^{(\sigma,\mathrm{p})}_{\omega\mathbf{k}_\perp}
    |0^{\text{M}}_{\text{in}}\rangle,
    \label{eq:Nout}
\end{multline}
where we used the definition of the number operator $
\hat{N}^{(\sigma,\mathrm{p})}_{\omega\mathbf{k}_\perp} \equiv
\hat{a}^\dagger_{\text{out}}(\overline{W^{(\sigma,\mathrm{p},\omega\mathbf{k}_\perp)}})
\, \hat{a}_{\text{out}}(W^{(\sigma,\mathrm{p},\omega\mathbf{k}_\perp)})
$ which gives us the number of particles per each Unruh mode $ \sigma $, sector
$ \mathrm{p} $, Rindler energy $ \omega $, and transverse momentum $\mathbf{k}_\perp  $. Note that we have already taken the physical limit $L\rightarrow \infty$. Using  Eqs.~(\ref{eq:expansion-out-field}),~(\ref{eq:Fs-matrix-definition}), and~(\ref{eq:expBKH}) together with identity~(\ref{eq:Unruh-expansion-coefficients}) one can write 
\begin{multline}
    \hat{S}^{F^{\dagger}} \hat{a}_{\text{out}}(W^{(\sigma,\mathrm{p},\omega\mathbf{k}_\perp)}){\hat{S}}^F=\hat{a}_{\text{out}}(W^{(\sigma,\mathrm{p},\omega\mathbf{k}_\perp)})  \\
    + \gkgt{ W^{(\sigma,\mathrm{p},\omega\mathbf{k}_\perp)}, E{T^F_{L}}}\hat{\mathbb{I}}.
    \label{SFaout}
\end{multline}
Now, using Eqs.~(\ref{eq:vacua-connection-s-matrix}),~(\ref{Smatrixsplit}),~(\ref{eq:coherent}),~and (\ref{SFaout}) in Eq.~(\ref{eq:Nout}), one can write
\begin{eqnarray}
   && \langle \hat{N} \rangle \equiv 
    \sum_{\sigma,\mathrm{p}} 
    \int_0^\infty \mathrm{d}\omega 
    \int_{\mathbb{R}^2}\mathrm{d}^2\mathbf{k}_\perp\left[ \left|\gkgt{ W^{(\sigma,\mathrm{p},\omega\mathbf{k}_\perp)}, ET^A }\right|^2 \right. \nonumber \\
    &+&\left.\left|\gkgt{ W^{(\sigma,\mathrm{p},\omega\mathbf{k}_\perp)}, ET^F }\right|^2 \right. \nonumber \\ 
      &+& \left. \gkgt{ET^A, W^{(\sigma,\mathrm{p},\omega\mathbf{k}_\perp)} }\gkgt{ W^{(\sigma,\mathrm{p},\omega\mathbf{k}_\perp)}, ET^F }+ {\rm c.c.}\right] \nonumber \\
    &=& \|KET^A\|^2 + \|KET^F\|^2 + 2{\rm Re}\left\{\gkgt{KET^A, KET^F}\right\}, \nonumber\\
\end{eqnarray}
which coincides with the classical number of gravitons derived in Eq.~(\ref{ClassicalNb}), where we recall that, in $\Sigma_+$, $RT^X_{ab}=-ET^X_{ab},$ $X=A, F$.

In the limit of infinite acceleration proper-time (where only zero-Rindler-energy Unruh modes contribute to the radiation), we can use Eq.~(\ref{norm KET}) to compute the total number of gravitons produced in the asymptotic future yielding 
\begin{multline}
\langle \hat{N} \rangle = \frac{m^2 \kappa^2}{16\pi^3 a}
    \, 
    T_{\mathrm{tot}} 
    \int_{\mathbb{R}^2}\mathrm{d}^2\mathbf{k}_\perp 
    \,
    \left[
        \mathrm{K}_{2}(k_\perp /a) 
    \right]^2    + \|KET^F\|^2 \\
   + 2{\rm Re}\left\{\gkgt{KET^A, KET^F}\right\}.
\end{multline}
The above equation implies that the number of (zero-Rindler-energy) gravitons per transverse  momentum $k_\perp$ per proper time emitted solely due to the accelerated motion of the mass can be written as 
\begin{equation}
\frac{1}{T_{\rm out}}\frac{\mathrm{d} N^A_{k_\perp}}{\mathrm{d} k_\perp}= \frac{m^2 \kappa^2 k_\perp }{8\pi^2 a} \left[\mathrm{K}_{2}(k_\perp /a)\right]^2,
\end{equation}
which also agrees with its classical counterpart given in Eq.~(\ref{classicalNgrav}).

%%%%%%%%%%%%%%%%%%%%%%%%%%%%%%%%%%%%%%%%%%%%%%%%%%%%%%%%%%%%%%%%%%%%%
%%%%%%%%%%%%%%%           Final discussion            %%%%%%%%%%%%%%%
%%%%%%%%%%%%%%%%%%%%%%%%%%%%%%%%%%%%%%%%%%%%%%%%%%%%%%%%%%%%%%%%%%%%%

\section{Final discussion} \label{sec:conclusions}

Here we have analyzed the classical and quantum emission of gravitational
radiation by a uniformly accelerated particle. We were successful in showing
that only the zero-Rindler-energy Unruh modes, whose definition we have extended
from the scalar and vector electrodynamics to be tensor valued, contribute on
the description of the classical retarded gravitational wave solution where the
mass accelerates for infinite time. 

From the quantum analysis we see that the interaction with the accelerated mass,
codified within the $ S $ matrix we explicitly constructed, evolves the past
vacuum for it to be seen as a multimode coherent superposition of particles
according to the future observer's perspective. If the acceleration occurs
forever, this process involves only zero-Rindler-energy particles.  The
coherence of the in vacuum reflects itself in the fact that the expectation value of the evolved field corresponds with the classical result for the gravitational perturbation.

This extends the results of the authors~\cite{landulfoClassicalQuantumAspects2019,portales-olivaClassicalQuantumReconciliation2022} to spin-2 fields, and clarifies the fundamental role played by zero-Rindler-energy gravitons, recognizing they are no simple mathematical artifact and do in fact contribute to the measurable radiation content. In particular, it vindicates the claim that each graviton emitted in the inertial frame must correspond to either the absorption or emission of a zero-energy Rindler particle in the accelerated one.

%%%%%%%%%%%%%%%%%%%%%%%%%%%%%%%%%%%%%%%%%%%%%%%%%%%%%%%%%%%%%%%%%%%%%
%%%%%%%%%%%%%%%           Acknowledgments             %%%%%%%%%%%%%%%
%%%%%%%%%%%%%%%%%%%%%%%%%%%%%%%%%%%%%%%%%%%%%%%%%%%%%%%%%%%%%%%%%%%%%
\acknowledgments
The authors want to thank George Matsas for the valuable and insightful discussions on our model. This research was funded by Grant \#2019/09401-4, S\~{a}o Paulo Research
Foundation (FAPESP).

%%%%%%%%%%%%%%%%%%%%%%%%%%%%%%%%%%%%%%%%%%%%%%%%%%%%%%%%%%%%%%%%%%%%%
%%%%%%%%%%%%%%%           Appendices             %%%%%%%%%%%%%%%%%%%%
%%%%%%%%%%%%%%%%%%%%%%%%%%%%%%%%%%%%%%%%%%%%%%%%%%%%%%%%%%%%%%%%%%%%%
\appendix

%%%%  Properties of the vector and tensor fields derived from the scalar and vector harmonic
\section{Properties of the vector and tensor fields derived from the scalar
and vector harmonic}\label{sect:properties-of-definitions-from-harmonics}

For the scalar sector, the vector defined in
Eq.~\eqref{eq:scalar-sector-2vector-plane} has a divergence proportional to the
scalar harmonic 
\begin{equation}
    \nabla^i\mathbb{S}^{\mathbf{k}_\perp}_i = k_\perp \mathbb{S}^{\mathbf{k}_\perp}.
    \label{eq:app-A.1}
\end{equation}
On the other hand, the tensors of Eqs.~\eqref{eq:scalar-sector-2tensor-plane}
and \eqref{eq:vector-sector-2tensor-plane}  
are traceless 
\begin{equation}
    g^{i j} \mathbb{S}^{\mathbf{k}_\perp}_{i j} = 0,
    \qquad\quad 
    g^{i j} \mathbb{V}^{\mathbf{k}_\perp}_{i j} = 0,
    \label{eq:app-A.2}
\end{equation}
and both their divergences are proportional to the corresponding vectors
\begin{equation}
    \nabla^j\mathbb{S}_{i j}^{\mathbf{k}_\perp} = 
    \frac{k_\perp}{2}
    \mathbb{S}_i^{\mathbf{k}_\perp},
    \qquad 
    \nabla^j\mathbb{V}_{i j}^{\mathbf{k}_\perp} = 
    \frac{k_\perp}{2}
    \mathbb{V}_i^{\mathbf{k}_\perp}.
    \label{eq:app-A.3}
\end{equation}

%%%%  Explicit calculations for the normalization of the modes
\section{Explicit calculations for the normalization of the
modes}\label{sect:normalization-calculations}

When normalizing the modes of Eq.~\eqref{eq:vector-perturbation-Rindler} and/or
Eq.~\eqref{eq:scalar-perturbation-Rindler} we use the following integrals (which can be checked straightforwardly by using the definitions of $\mathbb{S}^{\mathbf{k}_\perp}$ and $\mathbb{V}_{\mathbf{k}_\perp}^{i}$):
\begin{equation}
    \int_{\mathbb{R}^2} \mathrm{d}^2\mathbf{x}_\perp  \overline{\mathbb{V}_{\mathbf{k}_\perp}^{i}} \,  \mathbb{S}^{\mathbf{k}'_\perp}_{i} 
    =0,
\end{equation}
\begin{equation}
    \int_{\mathbb{R}^2} \mathrm{d}^2\mathbf{x}_\perp  \overline{\mathbb{S}^{\mathbf{k}_\perp}} \,  \mathbb{S}^{\mathbf{k}'_\perp}
    =
    4 \pi^2 \delta^2(\mathbf{k}_\perp - \mathbf{k}'_\perp),
\end{equation}
\begin{equation}
    \int_{\mathbb{R}^2} \mathrm{d}^2\mathbf{x}_\perp  
    \overline{\mathbb{V}_{\mathbf{k}_\perp}^{i}} 
    \,  
    \mathbb{V}^{\mathbf{k}'_\perp}_{i} 
    =
    4 \pi^2 k_\perp^2 \delta^2(\mathbf{k}_\perp - \mathbf{k}'_\perp),
\end{equation} 
\begin{equation}
    \int_{\mathbb{R}^2} \mathrm{d}^2\mathbf{x}_\perp  
    \overline{\mathbb{S}_{\mathbf{k}_\perp}^{i}} 
    \,  
    \mathbb{S}^{\mathbf{k}'_\perp}_{i} 
    =
    4 \pi^2 \delta^2(\mathbf{k}_\perp - \mathbf{k}'_\perp),
\end{equation} 
\begin{equation}
    \int_{\mathbb{R}^2} \mathrm{d}^2\mathbf{x}_\perp  
    \overline{\mathbb{S}_{\mathbf{k}_\perp}^{i j}} 
    \,  
    \mathbb{S}^{\mathbf{k}'_\perp}_{i j} 
    =
    2 \pi^2 \delta^2(\mathbf{k}_\perp - \mathbf{k}'_\perp).
\end{equation} 
We also use the normalization
integral~\cite{higuchiBremsstrahlungFullingDaviesUnruhThermal1992}
\begin{multline}
    \int_{-\infty}^{\infty} 
    \mathrm{K}_{\mathrm{i}\omega/a} (k_\perp \mathrm{e}^{a\xi}/a)
    \,
    \mathrm{K}_{\mathrm{i}\omega'/a} (k_\perp \mathrm{e}^{a\xi}/a)
    \, 
    \mathrm{d}\xi 
    \\ 
    = \frac{\pi^2 a}{2\omega \sinh(\pi\omega/a)}\delta(\omega-\omega').
    \label{eq:normalization-modified-Bessel-function-2nd-kind}
\end{multline}
\begin{widetext}
    \noindent Scalar functions of the orbit that satisfy
    Eq.~\eqref{eq:master-variable-Rindler} like the
    scalar~\eqref{eq:scalar-sector-RINDLER-mastervariable},
    vector~\eqref{eq:vector-sector-RINDLER-mastervariable}, or
    tensor~\eqref{eq:tensor-sector-RINDLER-mastervariable} master variables satisfy the equations
    \begin{align}
        \nabla_\alpha 
                \overline{\Omega^{\omega {k}_\perp}}
            \nabla_\lambda
                \nabla^\alpha 
                    \Omega^{\omega' k_\perp}
            &- 
                \nabla_\alpha  
                \Omega^{\omega' k_\perp}
            \nabla_\lambda
                \nabla^\alpha \overline{\Omega^{\omega {k}_\perp}}
        \nonumber \\
        &= 
        k_\perp^2
        \left(
            \Omega^{\omega' k_\perp}
            \frac{\partial\overline{\Omega^{\omega {k}_\perp}}}{\partial\lambda} 
            -
            \frac{\partial\Omega^{\omega' k_\perp}}{\partial\lambda}  
            \overline{\Omega^{\omega k_\perp}}
        \right)
        +\frac{\partial}{\partial\xi}
        (
            \nabla_\lambda  
                \Omega^{\omega' k_\perp}
            \nabla^\xi 
                \overline{\Omega^{\omega k_\perp}}
            -
            \nabla_\lambda
                \overline{\Omega^{\omega {k}_\perp}}
            \nabla^\xi  
                \Omega^{\omega' k_\perp}
        ),
        \label{eq:current-between-covariant-derivatives-identity}
    \end{align}
    \begin{align}
    \nabla^\alpha\nabla^\beta 
        \overline{\Omega^{\omega {k}_\perp}}
    \nabla_\lambda 
    \nabla_\alpha\nabla_\beta 
        \Omega^{\omega' k_\perp}
    &-
    \nabla^\alpha\nabla^\beta 
        \Omega^{\omega' k_\perp}
    \nabla_\lambda 
    \nabla_\alpha\nabla_\beta 
        \overline{\Omega^{\omega {k}_\perp}}
    \nonumber \\
    &=
    - k_\perp^2
    (
        \nabla^\beta 
            \overline{\Omega^{\omega {k}_\perp}}
        \nabla_\lambda 
        \nabla_\beta
            \Omega^{\omega' k_\perp}
        +
        \nabla^\beta 
            \Omega^{\omega' k_\perp}
        \nabla_\lambda 
        \nabla_\beta
            \overline{\Omega^{\omega {k}_\perp}}
    )
    \nonumber \\
    & \qquad +
    \frac{\partial}{\partial\xi} 
    (
        \nabla^\xi
        \nabla^\beta 
            \overline{\Omega^{\omega {k}_\perp}}
        \nabla_\lambda 
        \nabla_\beta 
            \Omega^{\omega' k_\perp}
        -
        \nabla^\xi
        \nabla^\beta 
            \Omega^{\omega' k_\perp}
        \nabla_\lambda 
        \nabla_\beta 
            \overline{\Omega^{\omega {k}_\perp}}
    )
    ,
    \end{align}
    which are useful for the normalization of the modes, as border terms will not contribute to the result of the integrals. These translate to the orbit of Minkowski spacetime using the direct replacements $ \omega \mapsto
    k_z $, $ \omega' \mapsto k'_z $, $ \lambda\mapsto t $, and $ \xi\mapsto z $.
\end{widetext}

\section{Derivation of
Eq.~\eqref{eq:Unruh-expansion-coefficients}}\label{sec:identity}

Consider the traceless and transverse gravitational perturbation $ h_{a b} $ around Minkowski spacetime that solves 
\begin{equation}
    \nabla_c\nabla^c h_{a b}  
    = 0.
    \label{eq:homogeneous-gravitational-field-eq-TT}
\end{equation}
Let $ T_{a b} $  be any (compactly supported) symmetric tensor. Then, by using $h_{ab}$ and $T_{ab}$, we define the functional 
\begin{equation}
    I[h, T]
    \equiv 
    -\kappa^2 \int_{\mathbb{R}^4} \mathrm{d}^4x \sqrt{-g} \, T_{ab} \overline{{h}^{a b}}
    .
    \label{eq:functional-grav.pert-0}
\end{equation}
If we take a Cauchy surface outside the causal future of the support of the stress-energy tensor, i.e., $ \Sigma \subset \mathbb{R}^4-J^+(\mathrm{supp}\,T_{ab}) $, we
note that the causal past of the Cauchy surface does not contribute to the integral and thus 
\begin{equation}
    I[ h, T]
    = 
    -\kappa^2  \int_{J^+ (\Sigma)} \mathrm{d}^4x \sqrt{-g} \, T^{ab} \, \overline{ {h}_{ab}}
    .
    \label{eq:functional-grav.pert-1}
\end{equation}
By taking $ AT_{a b} $ to be the advanced particular solution  of the equation 
\begin{equation}
    \nabla_c\nabla^c AT^{a b}
    = - \kappa^2 \, T^{a b},
    \label{eq:advanced-gravitational-field-eq-TT-flat-background}
\end{equation}
we can use use Eq.~(\ref{eq:advanced-gravitational-field-eq-TT-flat-background}) in Eq.~(\ref{eq:functional-grav.pert-1}) to write
\begin{equation}
    I[h, T]
    = 
    \int_{J^+ (\Sigma)} \! \! \mathrm{d}^4x \sqrt{-g} \, 
    (\nabla_c\nabla^c \! AT^{a b}) 
    \, \overline{{h}_{ab}}.
    \label{eq:functional-grav.pert-2}
\end{equation}
A small algebraic manipulation leads us to 
\begin{equation}
    \overline{{h}_{ab}} \nabla_c\nabla^c AT^{a b}
    =
    \nabla_c W^c[h, AT]
    +
    \nabla_c\nabla^c \overline{ {h}_{ab}} AT^{ab}
    ,
    \label{eq:integrand-functional-current}
\end{equation}
where the current $ W^c[h, AT] $ is defined as in
Eq.~\eqref{eq:general-current-to-compute}. Using Gauss theorem and the fact that $h_{ab}$ satisfies Eq.~(\ref{eq:homogeneous-gravitational-field-eq-TT}), the integral reduces to 
\begin{eqnarray}
    I [ h, T] &=& \int_{ J^+(\Sigma) } 
   \mathrm{d}^4 x \sqrt{-g}  \,
    \Big[\nabla_cW^c[ h, AT]   \nonumber \\
    &+&  AT^{ab} (\nabla_c\nabla^c \overline{ {h}_{ab}}) \Big] \nonumber \\
    &=& - 2\mathrm{i}\kappa^2 \gkgt{h, AT},
\end{eqnarray}
where the inner product is taken over $ \Sigma $. This can be simplified to 
\begin{equation}
    \gkgt{ h, ET }
    =
    -\frac{\mathrm{i}}{2}
    \int_{\mathbb{R}^4} 
    \mathrm{d}^4x 
    \sqrt{-g} 
    \, 
    T^{ab} 
    \, 
    \overline{h_{ab}}
    , 
    \label{eq:gravitational-perturbation-identity}
\end{equation}
by realizing that $ ET_{a b} = AT_{a b} $ for all events in the Cauchy surface $
\Sigma $.

%%%%%%%%%%%%%%%%%%%%%%%%%%%%%%%%%%%%%%%%%%%%%%%%%%%%%%%%%%%%%%%%%%%%%%
%%%%%%%%%%%%%%%%%%%%%%%%%%%%%%%%%%%%%%%%%%%%%%%%%%%%%%%%%%%%%%%%%%%%%%
\section{Contributions due to the inertial parts of the motion}\label{sect:inertial}

The total inertial contribution of Eq.~\eqref{eq:expansion-coefficients-scalar-sector-aux1} can be further separated into two parts, one corresponding to before (represented using a $-$ sign) and the other to after acceleration (denoted by $+$) as
\begin{equation}
    \mathcal{I}^{\sigma,\omega \mathbf{k}_\perp}(T,L)
 =
    \mathcal{I}_{+}^{\sigma,\omega \mathbf{k}_\perp}(T,L)
    +
    \mathcal{I}_{-}^{\sigma,\omega \mathbf{k}_\perp}(T,L).
\end{equation}
Defining the auxiliary function 
\begin{multline}
    \Delta_\pm (t,\mathbf{x}_\perp,z)
    \\ 
    \equiv 
    \delta^2(\mathbf{x}_\perp)
    \,
    \delta {\boldsymbol(}
        z - a^{-1} \sech(aT)
        \pm
        t \tanh(aT)
    {\boldsymbol)}
    ,
\end{multline}
we can write the \emph{effective} stress-energy tensor that appears inside the integrals above as 
\begin{gather}
    {T^A_{L,\pm}}^{t t} 
    = 
    m 
    \cosh(aT)
    \,
    \theta(L-|t|) 
    \,
    \Delta_\mp (t,\mathbf{x}_\perp,z)
    ,
    \\
    {T^A_{L,\pm}}^{t z} 
    =
    {T^A_{L,\pm}}^{z t} 
    =
    \pm
    m 
    \sinh(aT)
    \,
    \theta(L-|t|) 
    \,
    \Delta_\mp (t,\mathbf{x}_\perp,z),
    \\
     {T^A_{L,\pm}}^{z z} 
    =
    m 
    \sinh^2(aT) \sech(aT)
    \,
    \theta(L-|t|) 
    \,
    \Delta_\mp (t,\mathbf{x}_\perp,z).
\end{gather}
We can now use the form of the tensor modes of Eq.~\eqref{eq:scalar-perturbation-Unruh-alt} and the distribution version of the scalar Unruh
modes~\eqref{eq:scalar-Unruh-modes} to write explicitly the components we need, obtaining 
\begin{widetext}
    \begin{multline}
        \mathcal{I}^{\sigma,\omega \mathbf{k}_\perp}_{+}
        =
        \frac{m \kappa \sech(aT)}{k_\perp^{2} 2\pi^2 \sqrt{2 a}}
        \int_{-\infty}^{\infty} \mathrm{d}\vartheta
        \,
        \mathrm{e}^{
            -
            \mathrm{i}
            (-1)^\sigma \vartheta \omega / a
        }
        \left(
            \cosh^2(\vartheta - aT)
            +
            \frac{k_\perp^2}{2} 
        \right)
        \exp[
            - \mathrm{i} k_\perp a^{-1} \sech(aT) \sinh\vartheta
        ]
        % \nonumber \\ &\qquad 
        \\
        \times
        \int^{L}_{a^{-1}\sinh(aT)}\mathrm{d} t 
        \exp\{
            -
            \mathrm{i} 
            k_\perp t
            [
                \tanh(aT) \sinh\vartheta 
                - 
                \cosh\vartheta 
            ]
        \},
    \end{multline}
    \begin{multline}
        \mathcal{I}^{\sigma,\omega \mathbf{k}_\perp}_{-}
        =
        \frac{m \kappa \sech(aT)}{k_\perp^{2} 2\pi^2 \sqrt{2 a}}
        \int_{-\infty}^{\infty} \mathrm{d}\vartheta
        \,
        \mathrm{e}^{
            -
            \mathrm{i}
            (-1)^\sigma \vartheta \omega / a
        }
        \left(
            \cosh^2(\vartheta + aT)
            +
            \frac{k_\perp^2}{2} 
        \right)
        \exp[
            - \mathrm{i} k_\perp a^{-1} \sech(aT) \sinh\vartheta
        ]
        \\  \times
        \int_{-L}^{-a^{-1}\sinh(aT)}\mathrm{d} t 
        \exp\{
            \mathrm{i} 
            k_\perp 
            t
            [
                \tanh(aT) \sinh\vartheta 
                + 
                \cosh\vartheta 
            ]
        \}.
    \end{multline}
    These depend on the integrals
    \begin{multline}
        f_+ (\vartheta,T.L)
        \equiv
        \int^{L}_{a^{-1}\sinh(aT)}\mathrm{d} t 
        \exp\{
            -
            \mathrm{i} 
            k_\perp t
            [
                \tanh(aT) \sinh\vartheta 
                - 
                \cosh\vartheta 
            ]
        \}
        \\
        = 
        - \frac{
            \mathrm{i}
        }{
            k_\perp
            [
                \tanh(aT) \sinh\vartheta 
                - 
                \cosh\vartheta 
            ]
        }
        \Big(
        \exp\{
            -
            \mathrm{i} 
            k_\perp a \sinh(aT)
            [
                \tanh(aT) \sinh\vartheta 
                - 
                \cosh\vartheta 
            ]
        \}
        % \nonumber  \\ & \qquad \qquad \qquad
        \\
        -
        \exp\{
            -
            \mathrm{i} 
            k_\perp L
            [
                \tanh(aT) \sinh\vartheta 
                - 
                \cosh\vartheta 
            ]
        \}
        \Big),
    \end{multline}
    and 
    \begin{multline}
        f_-(\vartheta,T,L)
        \equiv
        \int_{-L}^{-a^{-1}\sinh(aT)}\mathrm{d} t 
        \exp\{
            \mathrm{i} 
            k_\perp 
            t
            [
                \tanh(aT) \sinh\vartheta 
                + 
                \cosh\vartheta 
            ]
        \}
        \\
        =
        - \frac{
            \mathrm{i}
        }{
            k_\perp
            [
                \tanh(aT) \sinh\vartheta 
                + 
                \cosh\vartheta 
            ]
        }
        \Big(
            \exp\{
                -
                \mathrm{i} 
                k_\perp 
                a\sinh(aT)
                [
                    \tanh(aT) \sinh\vartheta 
                    + 
                    \cosh\vartheta 
                ]
            \}
            \\
            % \nonumber  \\ & \qquad \qquad \qquad
            -\exp\{
                -
                \mathrm{i} 
                k_\perp 
                L
                [
                    \tanh(aT) \sinh\vartheta 
                    + 
                    \cosh\vartheta 
                ]
            \}
        \Big).
    \end{multline}
\end{widetext}
We can see from the above equations that the $L$-dependent terms behave as oscillatory distributions around zero. Thus,
on the limit $ L\to\infty $, we find that $ \exp\{ - \mathrm{i} k_\perp L [
\tanh(aT) \sinh\vartheta  \pm   \cosh\vartheta ] \} $ averages out to zero. Explicitly:
\begin{multline}
    \lim_{L\to\infty}f_\pm(\vartheta,T,L) 
    \\=
    - \frac{
        \mathrm{i}
        \mathrm{e}^{
            -
            \mathrm{i} 
            k_\perp 
            a\sinh(aT)
            [
                \tanh(aT) \sinh\vartheta 
                \mp 
                \cosh\vartheta 
            ]
        }
    }{
        k_\perp
        [
            \tanh(aT) \sinh\vartheta 
            \mp 
            \cosh\vartheta 
        ]
    }.
\end{multline}
The same arguments can be used in the limit $ T\to\infty $, as the hyperbolic sine is a strictly increasing function. As a result
\begin{equation}
    \lim_{T\to\infty}
    \left(
        \lim_{L\to\infty}
        \mathcal{I}_{\pm}^{\sigma,\omega \mathbf{k}_\perp}(T,L)
    \right)
    =
    0,
\end{equation}
meaning there is no contribution from the inertial parts of the motion in the case the particle is accelerated for an infinite amount of its proper-time, as we expected.

\section{Useful integrals to compute the final form of the expansion}\label{sect:bessel-integrals}
In order to arrive at Eqs.~\eqref{eq:RT-eta-eta} to \eqref{eq:RT-x-x}, we need to use some of the generating functions of the Bessel functions. In particular, we will need the expression~\cite{arfkenMathematicalMethodsPhysicists2013,gradshteynTableIntegralsSeries2014}
\begin{equation}
     \int_{0}^{2\pi} 
        \mathrm{e}^{ \mathrm{i} u \cos(\vartheta-\varphi) }
        \, 
        \mathrm{d}\vartheta
    =
    2 \pi \mathrm{J}_0(u),
    \label{eq:e1}
\end{equation}
With this in hand, we can also find the integrals 
\begin{gather}
    \int_{0}^{2\pi} 
        \mathrm{e}^{ \mathrm{i} u \cos(\vartheta-\varphi) } 
        \sin\vartheta
        \,
        \mathrm{d}\vartheta
    =
    2 \pi \mathrm{i} \sin\varphi \, \mathrm{J}_1 (u),
    \label{eq:e2}
    \\
    \int_{0}^{2\pi}
        \mathrm{e}^{ \mathrm{i} u \cos(\vartheta-\varphi) } 
        \cos\vartheta
        \,
        \mathrm{d}\vartheta
    =
    2 \pi \mathrm{i} \cos\varphi  \, \mathrm{J}_1 (u),
    \label{eq:e3}
\end{gather}
by taking partial derivatives of \eqref{eq:e1} with respect to $u$ and $\varphi$.
Taking the partial derivatives of Eqs.~\eqref{eq:e2} and \eqref{eq:e3} with respect to $u$ and combining them, we can also show that
\begin{gather}
    \int_{0}^{2\pi}
        \mathrm{e}^{ \mathrm{i} u  \cos(\vartheta-\varphi) }
        \sin(2\vartheta) 
        \, 
        \mathrm{d}\vartheta
    =
    -2\pi\sin(2\varphi) \,  \mathrm{J}_2(u) 
    ,
    \\
        \int_{0}^{2\pi} 
        \mathrm{e}^{ \mathrm{i} u  \cos(\vartheta-\varphi) }
            \cos(2\vartheta)
        \,
        \mathrm{d}\vartheta
    = 
    -2\pi\cos(2\varphi) \,  \mathrm{J}_2(u).
\end{gather}
These expressions  are related to the generating integral for Bessel functions 
\begin{equation}
    \mathrm{J}_n(u)
    =
    \frac{1}{\pi}
    \int_0^\pi
    \cos(n \vartheta - u \sin\vartheta)
    \, \mathrm{d}\vartheta,
\end{equation}
valid for $n=0,1,2,3,\ldots$.

\bibliography{umgrav}

\begin{thebibliography}{44}%
\makeatletter
\providecommand \@ifxundefined [1]{%
 \@ifx{#1\undefined}
}%
\providecommand \@ifnum [1]{%
 \ifnum #1\expandafter \@firstoftwo
 \else \expandafter \@secondoftwo
 \fi
}%
\providecommand \@ifx [1]{%
 \ifx #1\expandafter \@firstoftwo
 \else \expandafter \@secondoftwo
 \fi
}%
\providecommand \natexlab [1]{#1}%
\providecommand \enquote  [1]{``#1''}%
\providecommand \bibnamefont  [1]{#1}%
\providecommand \bibfnamefont [1]{#1}%
\providecommand \citenamefont [1]{#1}%
\providecommand \href@noop [0]{\@secondoftwo}%
\providecommand \href [0]{\begingroup \@sanitize@url \@href}%
\providecommand \@href[1]{\@@startlink{#1}\@@href}%
\providecommand \@@href[1]{\endgroup#1\@@endlink}%
\providecommand \@sanitize@url [0]{\catcode `\\12\catcode `\$12\catcode
  `\&12\catcode `\#12\catcode `\^12\catcode `\_12\catcode `\%12\relax}%
\providecommand \@@startlink[1]{}%
\providecommand \@@endlink[0]{}%
\providecommand \url  [0]{\begingroup\@sanitize@url \@url }%
\providecommand \@url [1]{\endgroup\@href {#1}{\urlprefix }}%
\providecommand \urlprefix  [0]{URL }%
\providecommand \Eprint [0]{\href }%
\providecommand \doibase [0]{https://doi.org/}%
\providecommand \selectlanguage [0]{\@gobble}%
\providecommand \bibinfo  [0]{\@secondoftwo}%
\providecommand \bibfield  [0]{\@secondoftwo}%
\providecommand \translation [1]{[#1]}%
\providecommand \BibitemOpen [0]{}%
\providecommand \bibitemStop [0]{}%
\providecommand \bibitemNoStop [0]{.\EOS\space}%
\providecommand \EOS [0]{\spacefactor3000\relax}%
\providecommand \BibitemShut  [1]{\csname bibitem#1\endcsname}%
\let\auto@bib@innerbib\@empty
%</preamble>
\bibitem [{\citenamefont {Abbott}\ \emph {et~al.}(2016)\citenamefont {Abbott}
  \emph {et~al.}}]{abbottObservationGravitationalWaves2016}%
  \BibitemOpen
  \bibfield  {author} {\bibinfo {author} {\bibfnamefont {B.~P.}\ \bibnamefont
  {Abbott}} \emph {et~al.} (\bibinfo {collaboration} {LIGO Scientific
  Collaboration and Virgo Collaboration}),\ }\bibfield  {title} {\bibinfo
  {title} {Observation of {{Gravitational Waves}} from a {{Binary Black Hole
  Merger}}},\ }\href {https://doi.org/10.1103/PhysRevLett.116.061102}
  {\bibfield  {journal} {\bibinfo  {journal} {Phys. Rev. Lett.}\ }\textbf
  {\bibinfo {volume} {116}},\ \bibinfo {pages} {061102} (\bibinfo {year}
  {2016})},\ \Eprint {https://arxiv.org/abs/1602.03837} {arxiv:1602.03837
  [gr-qc]} \BibitemShut {NoStop}%
\bibitem [{\citenamefont {Larmor}(1897)}]{larmorTheoryMagneticInfluence1897}%
  \BibitemOpen
  \bibfield  {author} {\bibinfo {author} {\bibfnamefont {J.}~\bibnamefont
  {Larmor}},\ }\bibfield  {title} {\bibinfo {title} {On the theory of the
  magnetic influence on spectra; and on the radiation from moving ions},\
  }\href {https://doi.org/10.1080/14786449708621095} {\bibfield  {journal}
  {\bibinfo  {journal} {London Edinburgh Dublin Philos. Mag. J. Sci.}\ }\textbf
  {\bibinfo {volume} {44}},\ \bibinfo {pages} {503} (\bibinfo {year}
  {1897})}\BibitemShut {NoStop}%
\bibitem [{\citenamefont {Fulton}\ and\ \citenamefont
  {Rohrlich}(1960)}]{fultonClassicalRadiationUniformly1960}%
  \BibitemOpen
  \bibfield  {author} {\bibinfo {author} {\bibfnamefont {T.}~\bibnamefont
  {Fulton}}\ and\ \bibinfo {author} {\bibfnamefont {F.}~\bibnamefont
  {Rohrlich}},\ }\bibfield  {title} {\bibinfo {title} {Classical radiation from
  a uniformly accelerated charge},\ }\href
  {https://doi.org/10.1016/0003-4916(60)90105-6} {\bibfield  {journal}
  {\bibinfo  {journal} {Ann. Phys. (N.Y.)}\ }\textbf {\bibinfo {volume} {9}},\
  \bibinfo {pages} {499} (\bibinfo {year} {1960})}\BibitemShut {NoStop}%
\bibitem [{\citenamefont {Unruh}\ and\ \citenamefont
  {Wald}(1982)}]{unruhAccelerationRadiationGeneralized1982}%
  \BibitemOpen
  \bibfield  {author} {\bibinfo {author} {\bibfnamefont {W.~G.}\ \bibnamefont
  {Unruh}}\ and\ \bibinfo {author} {\bibfnamefont {R.~M.}\ \bibnamefont
  {Wald}},\ }\bibfield  {title} {\bibinfo {title} {Acceleration radiation and
  the generalized second law of thermodynamics},\ }\href
  {https://doi.org/10.1103/PhysRevD.25.942} {\bibfield  {journal} {\bibinfo
  {journal} {Phys. Rev. D}\ }\textbf {\bibinfo {volume} {25}},\ \bibinfo
  {pages} {942} (\bibinfo {year} {1982})}\BibitemShut {NoStop}%
\bibitem [{\citenamefont
  {Singal}(1995)}]{singalEquivalencePrincipleElectric1995}%
  \BibitemOpen
  \bibfield  {author} {\bibinfo {author} {\bibfnamefont {A.~K.}\ \bibnamefont
  {Singal}},\ }\bibfield  {title} {\bibinfo {title} {The equivalence principle
  and an electric charge in a gravitational field},\ }\href
  {https://doi.org/10.1007/BF02113077} {\bibfield  {journal} {\bibinfo
  {journal} {Gen. Relat. Gravit.}\ }\textbf {\bibinfo {volume} {27}},\ \bibinfo
  {pages} {953} (\bibinfo {year} {1995})}\BibitemShut {NoStop}%
\bibitem [{\citenamefont {Feynman}\ \emph {et~al.}(1995)\citenamefont
  {Feynman}, \citenamefont {Morinigo},\ and\ \citenamefont
  {Wagner}}]{feynmanFeynmanLecturesGravitation1995}%
  \BibitemOpen
  \bibfield  {author} {\bibinfo {author} {\bibfnamefont {R.~P.}\ \bibnamefont
  {Feynman}}, \bibinfo {author} {\bibfnamefont {F.~B.}\ \bibnamefont
  {Morinigo}},\ and\ \bibinfo {author} {\bibfnamefont {W.~G.}\ \bibnamefont
  {Wagner}},\ }\href@noop {} {\emph {\bibinfo {title} {Feynman Lectures on
  Gravitation}}},\ edited by\ \bibinfo {editor} {\bibfnamefont
  {B.}~\bibnamefont {Hatfield}}\ (\bibinfo  {publisher} {{Addison-Wesley}},\
  \bibinfo {address} {{Reading, MA, U.S.A.}},\ \bibinfo {year}
  {1995})\BibitemShut {NoStop}%
\bibitem [{\citenamefont
  {Rohrlich}(1961)}]{rohrlichDefinitionElectromagneticRadiation1961}%
  \BibitemOpen
  \bibfield  {author} {\bibinfo {author} {\bibfnamefont {F.}~\bibnamefont
  {Rohrlich}},\ }\bibfield  {title} {\bibinfo {title} {The definition of
  electromagnetic radiation},\ }\href {https://doi.org/10.1007/BF02785607}
  {\bibfield  {journal} {\bibinfo  {journal} {Nuovo Cimento}\ }\textbf
  {\bibinfo {volume} {21}},\ \bibinfo {pages} {811} (\bibinfo {year}
  {1961})}\BibitemShut {NoStop}%
\bibitem [{\citenamefont {Rohrlich}(1963)}]{rohrlichPrincipleEquivalence1963}%
  \BibitemOpen
  \bibfield  {author} {\bibinfo {author} {\bibfnamefont {F.}~\bibnamefont
  {Rohrlich}},\ }\bibfield  {title} {\bibinfo {title} {The principle of
  equivalence},\ }\href {https://doi.org/10.1016/0003-4916(63)90051-4}
  {\bibfield  {journal} {\bibinfo  {journal} {Ann. Phys. (N.Y.)}\ }\textbf
  {\bibinfo {volume} {22}},\ \bibinfo {pages} {169} (\bibinfo {year}
  {1963})}\BibitemShut {NoStop}%
\bibitem [{\citenamefont
  {Boulware}(1980)}]{boulwareRadiationUniformlyAccelerated1980}%
  \BibitemOpen
  \bibfield  {author} {\bibinfo {author} {\bibfnamefont {D.~G.}\ \bibnamefont
  {Boulware}},\ }\bibfield  {title} {\bibinfo {title} {Radiation from a
  uniformly accelerated charge},\ }\href
  {https://doi.org/10.1016/0003-4916(80)90360-7} {\bibfield  {journal}
  {\bibinfo  {journal} {Ann. Phys. (N.Y.)}\ }\textbf {\bibinfo {volume}
  {124}},\ \bibinfo {pages} {169} (\bibinfo {year} {1980})}\BibitemShut
  {NoStop}%
\bibitem [{\citenamefont {Unruh}(1976)}]{unruhNotesBlackholeEvaporation1976}%
  \BibitemOpen
  \bibfield  {author} {\bibinfo {author} {\bibfnamefont {W.~G.}\ \bibnamefont
  {Unruh}},\ }\bibfield  {title} {\bibinfo {title} {Notes on black-hole
  evaporation},\ }\href {https://doi.org/10.1103/PhysRevD.14.870} {\bibfield
  {journal} {\bibinfo  {journal} {Phys. Rev. D}\ }\textbf {\bibinfo {volume}
  {14}},\ \bibinfo {pages} {870} (\bibinfo {year} {1976})}\BibitemShut
  {NoStop}%
\bibitem [{\citenamefont {Unruh}\ and\ \citenamefont
  {Wald}(1984)}]{unruhWhatHappensWhen1984}%
  \BibitemOpen
  \bibfield  {author} {\bibinfo {author} {\bibfnamefont {W.~G.}\ \bibnamefont
  {Unruh}}\ and\ \bibinfo {author} {\bibfnamefont {R.~M.}\ \bibnamefont
  {Wald}},\ }\bibfield  {title} {\bibinfo {title} {What happens when an
  accelerating observer detects a {{Rindler}} particle},\ }\href
  {https://doi.org/10.1103/PhysRevD.29.1047} {\bibfield  {journal} {\bibinfo
  {journal} {Phys. Rev. D}\ }\textbf {\bibinfo {volume} {29}},\ \bibinfo
  {pages} {1047} (\bibinfo {year} {1984})}\BibitemShut {NoStop}%
\bibitem [{\citenamefont {Higuchi}\ \emph {et~al.}(1992)\citenamefont
  {Higuchi}, \citenamefont {Matsas},\ and\ \citenamefont
  {Sudarsky}}]{higuchiBremsstrahlungFullingDaviesUnruhThermal1992}%
  \BibitemOpen
  \bibfield  {author} {\bibinfo {author} {\bibfnamefont {A.}~\bibnamefont
  {Higuchi}}, \bibinfo {author} {\bibfnamefont {G.~E.~A.}\ \bibnamefont
  {Matsas}},\ and\ \bibinfo {author} {\bibfnamefont {D.}~\bibnamefont
  {Sudarsky}},\ }\bibfield  {title} {\bibinfo {title} {Bremsstrahlung and
  {{Fulling-Davies-Unruh}} thermal bath},\ }\href
  {https://doi.org/10.1103/PhysRevD.46.3450} {\bibfield  {journal} {\bibinfo
  {journal} {Phys. Rev. D}\ }\textbf {\bibinfo {volume} {46}},\ \bibinfo
  {pages} {3450} (\bibinfo {year} {1992})}\BibitemShut {NoStop}%
\bibitem [{\citenamefont {Cozzella}\ \emph {et~al.}(2017)\citenamefont
  {Cozzella}, \citenamefont {Landulfo}, \citenamefont {Matsas},\ and\
  \citenamefont {Vanzella}}]{cozzellaProposalObservingUnruh2017}%
  \BibitemOpen
  \bibfield  {author} {\bibinfo {author} {\bibfnamefont {G.}~\bibnamefont
  {Cozzella}}, \bibinfo {author} {\bibfnamefont {A.~G.~S.}\ \bibnamefont
  {Landulfo}}, \bibinfo {author} {\bibfnamefont {G.~E.~A.}\ \bibnamefont
  {Matsas}},\ and\ \bibinfo {author} {\bibfnamefont {D.~A.~T.}\ \bibnamefont
  {Vanzella}},\ }\bibfield  {title} {\bibinfo {title} {Proposal for
  {{Observing}} the {{Unruh Effect}} using {{Classical Electrodynamics}}},\
  }\href {https://doi.org/10.1103/PhysRevLett.118.161102} {\bibfield  {journal}
  {\bibinfo  {journal} {Phys. Rev. Lett.}\ }\textbf {\bibinfo {volume} {118}},\
  \bibinfo {pages} {161102} (\bibinfo {year} {2017})},\ \Eprint
  {https://arxiv.org/abs/1701.03446} {arxiv:1701.03446 [gr-qc]} \BibitemShut
  {NoStop}%
\bibitem [{\citenamefont {Bi{\v
  c}{\'a}k}(1968)}]{bicakGravitationalRadiationUniformly1997}%
  \BibitemOpen
  \bibfield  {author} {\bibinfo {author} {\bibfnamefont {J.}~\bibnamefont
  {Bi{\v c}{\'a}k}},\ }\bibfield  {title} {\bibinfo {title} {Gravitational
  radiation from uniformly accelerated particles in general relativity},\
  }\href {https://doi.org/10.1098/rspa.1968.0004} {\bibfield  {journal}
  {\bibinfo  {journal} {Proc. R. Soc. London A}\ }\textbf {\bibinfo {volume}
  {302}},\ \bibinfo {pages} {201} (\bibinfo {year} {1968})}\BibitemShut
  {NoStop}%
\bibitem [{\citenamefont {Bonnor}\ and\ \citenamefont
  {Swaminarayan}(1964)}]{bonnorExactSolutionUniformly1964}%
  \BibitemOpen
  \bibfield  {author} {\bibinfo {author} {\bibfnamefont {W.~B.}\ \bibnamefont
  {Bonnor}}\ and\ \bibinfo {author} {\bibfnamefont {N.~S.}\ \bibnamefont
  {Swaminarayan}},\ }\bibfield  {title} {\bibinfo {title} {An exact solution
  for uniformly accelerated particles in general relativity},\ }\href
  {https://doi.org/10.1007/BF01375497} {\bibfield  {journal} {\bibinfo
  {journal} {Z. Physik}\ }\textbf {\bibinfo {volume} {177}},\ \bibinfo {pages}
  {240} (\bibinfo {year} {1964})}\BibitemShut {NoStop}%
\bibitem [{\citenamefont {Hopper}\ and\ \citenamefont
  {Cardoso}(2018)}]{hopperScatteringPointParticles2018}%
  \BibitemOpen
  \bibfield  {author} {\bibinfo {author} {\bibfnamefont {S.}~\bibnamefont
  {Hopper}}\ and\ \bibinfo {author} {\bibfnamefont {V.}~\bibnamefont
  {Cardoso}},\ }\bibfield  {title} {\bibinfo {title} {Scattering of point
  particles by black holes: {{Gravitational}} radiation},\ }\href
  {https://doi.org/10.1103/PhysRevD.97.044031} {\bibfield  {journal} {\bibinfo
  {journal} {Phys. Rev. D}\ }\textbf {\bibinfo {volume} {97}},\ \bibinfo
  {pages} {044031} (\bibinfo {year} {2018})},\ \Eprint
  {https://arxiv.org/abs/1706.02791} {arxiv:1706.02791 [gr-qc]} \BibitemShut
  {NoStop}%
\bibitem [{\citenamefont
  {Poisson}(1993)}]{poissonGravitationalRadiationParticle1993}%
  \BibitemOpen
  \bibfield  {author} {\bibinfo {author} {\bibfnamefont {E.}~\bibnamefont
  {Poisson}},\ }\bibfield  {title} {\bibinfo {title} {Gravitational radiation
  from a particle in circular orbit around a black hole. {{I}}. {{Analytical}}
  results for the nonrotating case},\ }\href
  {https://doi.org/10.1103/PhysRevD.47.1497} {\bibfield  {journal} {\bibinfo
  {journal} {Phys. Rev. D}\ }\textbf {\bibinfo {volume} {47}},\ \bibinfo
  {pages} {1497} (\bibinfo {year} {1993})}\BibitemShut {NoStop}%
\bibitem [{\citenamefont {Bernar}\ \emph {et~al.}(2017)\citenamefont {Bernar},
  \citenamefont {Crispino},\ and\ \citenamefont
  {Higuchi}}]{bernarGravitationalWavesEmitted2017}%
  \BibitemOpen
  \bibfield  {author} {\bibinfo {author} {\bibfnamefont {R.~P.}\ \bibnamefont
  {Bernar}}, \bibinfo {author} {\bibfnamefont {L.~C.~B.}\ \bibnamefont
  {Crispino}},\ and\ \bibinfo {author} {\bibfnamefont {A.}~\bibnamefont
  {Higuchi}},\ }\bibfield  {title} {\bibinfo {title} {Gravitational waves
  emitted by a particle rotating around a {{Schwarzschild}} black hole: {{A}}
  semiclassical approach},\ }\href {https://doi.org/10.1103/PhysRevD.95.064042}
  {\bibfield  {journal} {\bibinfo  {journal} {Phys. Rev. D}\ }\textbf {\bibinfo
  {volume} {95}},\ \bibinfo {pages} {064042} (\bibinfo {year} {2017})},\
  \Eprint {https://arxiv.org/abs/1703.10648} {arxiv:1703.10648 [gr-qc]}
  \BibitemShut {NoStop}%
\bibitem [{\citenamefont {Bernar}\ \emph {et~al.}(2018)\citenamefont {Bernar},
  \citenamefont {Crispino},\ and\ \citenamefont
  {Higuchi}}]{bernarGibbonsHawkingRadiationGravitons2018}%
  \BibitemOpen
  \bibfield  {author} {\bibinfo {author} {\bibfnamefont {R.~P.}\ \bibnamefont
  {Bernar}}, \bibinfo {author} {\bibfnamefont {L.~C.~B.}\ \bibnamefont
  {Crispino}},\ and\ \bibinfo {author} {\bibfnamefont {A.}~\bibnamefont
  {Higuchi}},\ }\bibfield  {title} {\bibinfo {title} {Gibbons-{{Hawking}}
  radiation of gravitons in the {{Poincar\'e}} and static patches of de
  {{Sitter}} spacetime},\ }\href {https://doi.org/10.1103/PhysRevD.97.085005}
  {\bibfield  {journal} {\bibinfo  {journal} {Phys. Rev. D}\ }\textbf {\bibinfo
  {volume} {97}},\ \bibinfo {pages} {085005} (\bibinfo {year} {2018})},\
  \Eprint {https://arxiv.org/abs/1803.01204} {arxiv:1803.01204 [gr-qc]}
  \BibitemShut {NoStop}%
\bibitem [{\citenamefont {Bernar}\ \emph {et~al.}(2014)\citenamefont {Bernar},
  \citenamefont {Crispino},\ and\ \citenamefont
  {Higuchi}}]{bernarInfraredfiniteGravitonTwopoint2014}%
  \BibitemOpen
  \bibfield  {author} {\bibinfo {author} {\bibfnamefont {R.~P.}\ \bibnamefont
  {Bernar}}, \bibinfo {author} {\bibfnamefont {L.~C.~B.}\ \bibnamefont
  {Crispino}},\ and\ \bibinfo {author} {\bibfnamefont {A.}~\bibnamefont
  {Higuchi}},\ }\bibfield  {title} {\bibinfo {title} {Infrared-finite graviton
  two-point function in static de {{Sitter}} space},\ }\href
  {https://doi.org/10.1103/PhysRevD.90.024045} {\bibfield  {journal} {\bibinfo
  {journal} {Phys. Rev. D}\ }\textbf {\bibinfo {volume} {90}},\ \bibinfo
  {pages} {024045} (\bibinfo {year} {2014})},\ \Eprint
  {https://arxiv.org/abs/1405.3827} {arxiv:1405.3827 [gr-qc]} \BibitemShut
  {NoStop}%
\bibitem [{\citenamefont {Fr{\"o}b}\ \emph {et~al.}(2016)\citenamefont
  {Fr{\"o}b}, \citenamefont {Higuchi},\ and\ \citenamefont
  {Lima}}]{frobModesumConstructionCovariant2016}%
  \BibitemOpen
  \bibfield  {author} {\bibinfo {author} {\bibfnamefont {M.~B.}\ \bibnamefont
  {Fr{\"o}b}}, \bibinfo {author} {\bibfnamefont {A.}~\bibnamefont {Higuchi}},\
  and\ \bibinfo {author} {\bibfnamefont {W.~C.~C.}\ \bibnamefont {Lima}},\
  }\bibfield  {title} {\bibinfo {title} {Mode-sum construction of the covariant
  graviton two-point function in the {{Poincar\'e}} patch of de {{Sitter}}
  space},\ }\href {https://doi.org/10.1103/PhysRevD.93.124006} {\bibfield
  {journal} {\bibinfo  {journal} {Phys. Rev. D}\ }\textbf {\bibinfo {volume}
  {93}},\ \bibinfo {pages} {124006} (\bibinfo {year} {2016})},\ \Eprint
  {https://arxiv.org/abs/1603.07338v2} {arxiv:1603.07338v2 [gr-qc]}
  \BibitemShut {NoStop}%
\bibitem [{\citenamefont
  {Higuchi}(1987)}]{higuchiSymmetricTensorSpherical1987}%
  \BibitemOpen
  \bibfield  {author} {\bibinfo {author} {\bibfnamefont {A.}~\bibnamefont
  {Higuchi}},\ }\bibfield  {title} {\bibinfo {title} {Symmetric tensor
  spherical harmonics on the {{N}}-sphere and their application to the de
  {{Sitter}} group {{SO}}({{N}},1)},\ }\href {https://doi.org/10.1063/1.527513}
  {\bibfield  {journal} {\bibinfo  {journal} {J. Math. Phys.}\ }\textbf
  {\bibinfo {volume} {28}},\ \bibinfo {pages} {1553} (\bibinfo {year}
  {1987})}\BibitemShut {NoStop}%
\bibitem [{\citenamefont
  {Friedman}(1978)}]{friedmanGenericInstabilityRotating1978}%
  \BibitemOpen
  \bibfield  {author} {\bibinfo {author} {\bibfnamefont {J.~L.}\ \bibnamefont
  {Friedman}},\ }\bibfield  {title} {\bibinfo {title} {Generic instability of
  rotating relativistic stars},\ }\href {https://doi.org/10.1007/BF01202527}
  {\bibfield  {journal} {\bibinfo  {journal} {Commun. Math. Phys.}\ }\textbf
  {\bibinfo {volume} {62}},\ \bibinfo {pages} {247} (\bibinfo {year}
  {1978})}\BibitemShut {NoStop}%
\bibitem [{\citenamefont {Higuchi}(1989)}]{higuchiMassiveSymmetricTensor1989}%
  \BibitemOpen
  \bibfield  {author} {\bibinfo {author} {\bibfnamefont {A.}~\bibnamefont
  {Higuchi}},\ }\bibfield  {title} {\bibinfo {title} {Massive symmetric tensor
  field in spacetimes with a positive cosmological constant},\ }\href
  {https://doi.org/10.1016/0550-3213(89)90507-5} {\bibfield  {journal}
  {\bibinfo  {journal} {Nucl. Phys. B}\ }\textbf {\bibinfo {volume} {325}},\
  \bibinfo {pages} {745} (\bibinfo {year} {1989})}\BibitemShut {NoStop}%
\bibitem [{\citenamefont
  {Weinberg}(1972)}]{weinbergGravitationCosmologyPrinciples1972}%
  \BibitemOpen
  \bibfield  {author} {\bibinfo {author} {\bibfnamefont {S.}~\bibnamefont
  {Weinberg}},\ }\href@noop {} {\emph {\bibinfo {title} {Gravitation and
  {{Cosmology}}: {{Principles}} and {{Applications}} of the {{General Theory}}
  of {{Relativity}}}}}\ (\bibinfo  {publisher} {{Wiley}},\ \bibinfo {address}
  {{New York, NY, U.S.A.}},\ \bibinfo {year} {1972})\BibitemShut {NoStop}%
\bibitem [{\citenamefont {Crispino}\ \emph {et~al.}(2008)\citenamefont
  {Crispino}, \citenamefont {Higuchi},\ and\ \citenamefont
  {Matsas}}]{crispinoUnruhEffectIts2008}%
  \BibitemOpen
  \bibfield  {author} {\bibinfo {author} {\bibfnamefont {L.~C.~B.}\
  \bibnamefont {Crispino}}, \bibinfo {author} {\bibfnamefont {A.}~\bibnamefont
  {Higuchi}},\ and\ \bibinfo {author} {\bibfnamefont {G.~E.~A.}\ \bibnamefont
  {Matsas}},\ }\bibfield  {title} {\bibinfo {title} {The {{Unruh}} effect and
  its applications},\ }\href {https://doi.org/10.1103/RevModPhys.80.787}
  {\bibfield  {journal} {\bibinfo  {journal} {Rev. Mod. Phys.}\ }\textbf
  {\bibinfo {volume} {80}},\ \bibinfo {pages} {787} (\bibinfo {year} {2008})},\
  \Eprint {https://arxiv.org/abs/0710.5373v1} {arxiv:0710.5373v1 [gr-qc]}
  \BibitemShut {NoStop}%
\bibitem [{\citenamefont {Higuchi}\ \emph {et~al.}(2017)\citenamefont
  {Higuchi}, \citenamefont {Iso}, \citenamefont {Ueda},\ and\ \citenamefont
  {Yamamoto}}]{higuchiEntanglementVacuumLeft2017}%
  \BibitemOpen
  \bibfield  {author} {\bibinfo {author} {\bibfnamefont {A.}~\bibnamefont
  {Higuchi}}, \bibinfo {author} {\bibfnamefont {S.}~\bibnamefont {Iso}},
  \bibinfo {author} {\bibfnamefont {K.}~\bibnamefont {Ueda}},\ and\ \bibinfo
  {author} {\bibfnamefont {K.}~\bibnamefont {Yamamoto}},\ }\bibfield  {title}
  {\bibinfo {title} {Entanglement of the vacuum between left, right, future,
  and past: {{The}} origin of entanglement-induced quantum radiation},\ }\href
  {https://doi.org/10.1103/PhysRevD.96.083531} {\bibfield  {journal} {\bibinfo
  {journal} {Phys. Rev. D}\ }\textbf {\bibinfo {volume} {96}},\ \bibinfo
  {pages} {083531} (\bibinfo {year} {2017})},\ \Eprint
  {https://arxiv.org/abs/1709.05757} {arxiv:1709.05757 [hep-th]} \BibitemShut
  {NoStop}%
\bibitem [{\citenamefont
  {Mukohyama}(2000)}]{mukohyamaGaugeinvariantGravitationalPerturbations2000}%
  \BibitemOpen
  \bibfield  {author} {\bibinfo {author} {\bibfnamefont {S.}~\bibnamefont
  {Mukohyama}},\ }\bibfield  {title} {\bibinfo {title} {Gauge-invariant
  gravitational perturbations of maximally symmetric spacetimes},\ }\href
  {https://doi.org/10.1103/PhysRevD.62.084015} {\bibfield  {journal} {\bibinfo
  {journal} {Phys. Rev. D}\ }\textbf {\bibinfo {volume} {62}},\ \bibinfo
  {pages} {084015} (\bibinfo {year} {2000})},\ \Eprint
  {https://arxiv.org/abs/hep-th/0004067} {arxiv:hep-th/0004067} \BibitemShut
  {NoStop}%
\bibitem [{\citenamefont {Kodama}\ \emph {et~al.}(2000)\citenamefont {Kodama},
  \citenamefont {Ishibashi},\ and\ \citenamefont
  {Seto}}]{kodamaBraneWorldCosmology2000}%
  \BibitemOpen
  \bibfield  {author} {\bibinfo {author} {\bibfnamefont {H.}~\bibnamefont
  {Kodama}}, \bibinfo {author} {\bibfnamefont {A.}~\bibnamefont {Ishibashi}},\
  and\ \bibinfo {author} {\bibfnamefont {O.}~\bibnamefont {Seto}},\ }\bibfield
  {title} {\bibinfo {title} {Brane world cosmology: {{Gauge-invariant}}
  formalism for perturbation},\ }\href
  {https://doi.org/10.1103/PhysRevD.62.064022} {\bibfield  {journal} {\bibinfo
  {journal} {Phys. Rev. D}\ }\textbf {\bibinfo {volume} {62}},\ \bibinfo
  {pages} {064022} (\bibinfo {year} {2000})},\ \Eprint
  {https://arxiv.org/abs/hep-th/0004160} {arxiv:hep-th/0004160} \BibitemShut
  {NoStop}%
\bibitem [{\citenamefont {Kodama}\ and\ \citenamefont
  {Ishibashi}(2003)}]{kodamaMasterEquationGravitational2003}%
  \BibitemOpen
  \bibfield  {author} {\bibinfo {author} {\bibfnamefont {H.}~\bibnamefont
  {Kodama}}\ and\ \bibinfo {author} {\bibfnamefont {A.}~\bibnamefont
  {Ishibashi}},\ }\bibfield  {title} {\bibinfo {title} {A {{Master Equation}}
  for {{Gravitational Perturbations}} of {{Maximally Symmetric Black Holes}} in
  {{Higher Dimensions}}},\ }\href {https://doi.org/10.1143/PTP.110.701}
  {\bibfield  {journal} {\bibinfo  {journal} {Prog. Theor. Phys.}\ }\textbf
  {\bibinfo {volume} {110}},\ \bibinfo {pages} {701} (\bibinfo {year}
  {2003})},\ \Eprint {https://arxiv.org/abs/hep-th/0305147}
  {arxiv:hep-th/0305147} \BibitemShut {NoStop}%
\bibitem [{Note1()}]{Note1}%
  \BibitemOpen
  \bibinfo {note} {We will not worry ourselves with the normalization of the
  harmonics at the time being, as the final perturbations will be normalized.
  Calculations on the orthogonality of the harmonics are given in Appendix~\ref
  {sect:normalization-calculations}.}\BibitemShut {Stop}%
\bibitem [{\citenamefont {Sugiyama}\ \emph {et~al.}(2021)\citenamefont
  {Sugiyama}, \citenamefont {Yamamoto},\ and\ \citenamefont
  {Kobayashi}}]{sugiyamaGravitationalWavesKasner2021}%
  \BibitemOpen
  \bibfield  {author} {\bibinfo {author} {\bibfnamefont {Y.}~\bibnamefont
  {Sugiyama}}, \bibinfo {author} {\bibfnamefont {K.}~\bibnamefont {Yamamoto}},\
  and\ \bibinfo {author} {\bibfnamefont {T.}~\bibnamefont {Kobayashi}},\
  }\bibfield  {title} {\bibinfo {title} {Gravitational waves in {{Kasner}}
  spacetimes and {{Rindler}} wedges in {{Regge-Wheeler}} gauge: {{Formulation}}
  of {{Unruh}} effect},\ }\href {https://doi.org/10.1103/PhysRevD.103.083503}
  {\bibfield  {journal} {\bibinfo  {journal} {Phys. Rev. D}\ }\textbf {\bibinfo
  {volume} {103}},\ \bibinfo {pages} {083503} (\bibinfo {year} {2021})},\
  \Eprint {https://arxiv.org/abs/2012.15004} {arxiv:2012.15004 [gr-qc]}
  \BibitemShut {NoStop}%
\bibitem [{\citenamefont {Regge}\ and\ \citenamefont
  {Wheeler}(1957)}]{reggeStabilitySchwarzschildSingularity1957}%
  \BibitemOpen
  \bibfield  {author} {\bibinfo {author} {\bibfnamefont {T.}~\bibnamefont
  {Regge}}\ and\ \bibinfo {author} {\bibfnamefont {J.~A.}\ \bibnamefont
  {Wheeler}},\ }\bibfield  {title} {\bibinfo {title} {Stability of a
  {{Schwarzschild Singularity}}},\ }\href
  {https://doi.org/10.1103/PhysRev.108.1063} {\bibfield  {journal} {\bibinfo
  {journal} {Phys. Rev.}\ }\textbf {\bibinfo {volume} {108}},\ \bibinfo {pages}
  {1063} (\bibinfo {year} {1957})}\BibitemShut {NoStop}%
\bibitem [{\citenamefont {Thorne}(2020)}]{thorneIntroductionReggeWheeler2020}%
  \BibitemOpen
  \bibfield  {author} {\bibinfo {author} {\bibfnamefont {K.~S.}\ \bibnamefont
  {Thorne}},\ }\bibfield  {title} {\bibinfo {title} {Introduction to {{Regge}}
  and {{Wheeler}} ``{{Stability}} of a {{Schwarzschild Singularity}}''},\ }in\
  \href {https://doi.org/10.1142/11643} {\emph {\bibinfo {booktitle} {Tullio
  {{Regge}} an {{Eclectic Genius}}: {{From Quantum Gravity}} to {{Computer
  Play}}}}}\ (\bibinfo  {publisher} {{World Scientific}},\ \bibinfo {address}
  {{Singapore}},\ \bibinfo {year} {2020})\ pp.\ \bibinfo {pages}
  {3--12}\BibitemShut {NoStop}%
\bibitem [{\citenamefont {Kodama}\ and\ \citenamefont
  {Sasaki}(1984)}]{kodamaCosmologicalPerturbationTheory1984}%
  \BibitemOpen
  \bibfield  {author} {\bibinfo {author} {\bibfnamefont {H.}~\bibnamefont
  {Kodama}}\ and\ \bibinfo {author} {\bibfnamefont {M.}~\bibnamefont
  {Sasaki}},\ }\bibfield  {title} {\bibinfo {title} {Cosmological
  {{Perturbation Theory}}},\ }\href {https://doi.org/10.1143/PTPS.78.1}
  {\bibfield  {journal} {\bibinfo  {journal} {Prog. Theor. Phys. Supp.}\
  }\textbf {\bibinfo {volume} {78}},\ \bibinfo {pages} {1} (\bibinfo {year}
  {1984})}\BibitemShut {NoStop}%
\bibitem [{Note2()}]{Note2}%
  \BibitemOpen
  \bibinfo {note} {After quantization, this is all that is needed to show the
  Unruh effect for the linearized gravitational field.}\BibitemShut {Stop}%
\bibitem [{\citenamefont {Poisson}\ \emph {et~al.}(2011)\citenamefont
  {Poisson}, \citenamefont {Pound},\ and\ \citenamefont
  {Vega}}]{poissonMotionPointParticles2011}%
  \BibitemOpen
  \bibfield  {author} {\bibinfo {author} {\bibfnamefont {E.}~\bibnamefont
  {Poisson}}, \bibinfo {author} {\bibfnamefont {A.}~\bibnamefont {Pound}},\
  and\ \bibinfo {author} {\bibfnamefont {I.}~\bibnamefont {Vega}},\ }\bibfield
  {title} {\bibinfo {title} {The {{Motion}} of {{Point Particles}} in {{Curved
  Spacetime}}},\ }\href {https://doi.org/10.12942/lrr-2011-7} {\bibfield
  {journal} {\bibinfo  {journal} {Living Rev. Relativ.}\ }\textbf {\bibinfo
  {volume} {14}},\ \bibinfo {pages} {7} (\bibinfo {year} {2011})},\ \Eprint
  {https://arxiv.org/abs/1102.0529} {arxiv:1102.0529 [gr-qc]} \BibitemShut
  {NoStop}%
\bibitem [{\citenamefont {Arfken}\ \emph {et~al.}(2013)\citenamefont {Arfken},
  \citenamefont {Weber},\ and\ \citenamefont
  {Harris}}]{arfkenMathematicalMethodsPhysicists2013}%
  \BibitemOpen
  \bibfield  {author} {\bibinfo {author} {\bibfnamefont {G.~B.}\ \bibnamefont
  {Arfken}}, \bibinfo {author} {\bibfnamefont {H.~J.}\ \bibnamefont {Weber}},\
  and\ \bibinfo {author} {\bibfnamefont {F.~E.}\ \bibnamefont {Harris}},\
  }\href {https://doi.org/10.1016/C2009-0-30629-7} {\emph {\bibinfo {title}
  {Mathematical Methods for Physicists: {{A}} Comprehensive Guide}}},\ \bibinfo
  {edition} {7th}\ ed.\ (\bibinfo  {publisher} {{Academic Press}},\ \bibinfo
  {address} {{Boston, MA, U.S.A.}},\ \bibinfo {year} {2013})\BibitemShut
  {NoStop}%
\bibitem [{\citenamefont
  {Fabrikant}(2003)}]{fabrikantComputationInfiniteIntegrals2003}%
  \BibitemOpen
  \bibfield  {author} {\bibinfo {author} {\bibfnamefont {V.~I.}\ \bibnamefont
  {Fabrikant}},\ }\bibfield  {title} {\bibinfo {title} {Computation of infinite
  integrals involving three {{Bessel}} functions by introduction of new
  formalism},\ }\href {https://doi.org/10.1002/zamm.200310059} {\bibfield
  {journal} {\bibinfo  {journal} {Z. Angew. Math. Mech.}\ }\textbf {\bibinfo
  {volume} {83}},\ \bibinfo {pages} {363} (\bibinfo {year} {2003})}\BibitemShut
  {NoStop}%
\bibitem [{\citenamefont {Gradshteyn}\ and\ \citenamefont
  {Ryzhik}(2014)}]{gradshteynTableIntegralsSeries2014}%
  \BibitemOpen
  \bibfield  {author} {\bibinfo {author} {\bibfnamefont {I.~S.}\ \bibnamefont
  {Gradshteyn}}\ and\ \bibinfo {author} {\bibfnamefont {I.~M.}\ \bibnamefont
  {Ryzhik}},\ }\href {https://doi.org/10.1016/C2010-0-64839-5} {\emph {\bibinfo
  {title} {Table of {{Integrals}}, {{Series}}, and {{Products}}}}},\ \bibinfo
  {edition} {8th}\ ed.,\ edited by\ \bibinfo {editor} {\bibfnamefont
  {D.}~\bibnamefont {Zwillinger}}\ and\ \bibinfo {editor} {\bibfnamefont
  {V.}~\bibnamefont {Moll}}\ (\bibinfo  {publisher} {{Academic Press}},\
  \bibinfo {address} {{Boston, MA, U.S.A.}},\ \bibinfo {year}
  {2014})\BibitemShut {NoStop}%
\bibitem [{\citenamefont {Higuchi}\ and\ \citenamefont {Matsas}(1993)}]{HMCFT}%
  \BibitemOpen
  \bibfield  {author} {\bibinfo {author} {\bibfnamefont {A.}~\bibnamefont
  {Higuchi}}\ and\ \bibinfo {author} {\bibfnamefont {G.~E.~A.}\ \bibnamefont
  {Matsas}},\ }\bibfield  {title} {\bibinfo {title} {Fulling-davies-unruh
  effect in classical field theory},\ }\href
  {https://doi.org/https://doi.org/10.1103/PhysRevD.48.689} {\bibfield
  {journal} {\bibinfo  {journal} {Phys. Rev. D}\ }\textbf {\bibinfo {volume}
  {48}},\ \bibinfo {pages} {689} (\bibinfo {year} {1993})}\BibitemShut
  {NoStop}%
\bibitem [{\citenamefont {Itzykson}\ and\ \citenamefont
  {Zuber}(2005)}]{itzyksonQuantumFieldTheory2005}%
  \BibitemOpen
  \bibfield  {author} {\bibinfo {author} {\bibfnamefont {C.}~\bibnamefont
  {Itzykson}}\ and\ \bibinfo {author} {\bibfnamefont {J.-B.}\ \bibnamefont
  {Zuber}},\ }\href@noop {} {\emph {\bibinfo {title} {Quantum {{Field
  Theory}}}}}\ (\bibinfo  {publisher} {{Dover}},\ \bibinfo {address} {{Mineola,
  NY, U.S.A.}},\ \bibinfo {year} {2005})\BibitemShut {NoStop}%
\bibitem [{\citenamefont {Landulfo}\ \emph {et~al.}(2019)\citenamefont
  {Landulfo}, \citenamefont {Fulling},\ and\ \citenamefont
  {Matsas}}]{landulfoClassicalQuantumAspects2019}%
  \BibitemOpen
  \bibfield  {author} {\bibinfo {author} {\bibfnamefont {A.~G.~S.}\
  \bibnamefont {Landulfo}}, \bibinfo {author} {\bibfnamefont {S.~A.}\
  \bibnamefont {Fulling}},\ and\ \bibinfo {author} {\bibfnamefont {G.~E.~A.}\
  \bibnamefont {Matsas}},\ }\bibfield  {title} {\bibinfo {title} {Classical and
  quantum aspects of the radiation emitted by a uniformly accelerated charge:
  {{Larmor-Unruh}} reconciliation and zero-frequency {{Rindler}} modes},\
  }\href {https://doi.org/10.1103/PhysRevD.100.045020} {\bibfield  {journal}
  {\bibinfo  {journal} {Phys. Rev. D}\ }\textbf {\bibinfo {volume} {100}},\
  \bibinfo {pages} {045020} (\bibinfo {year} {2019})},\ \Eprint
  {https://arxiv.org/abs/1907.06665} {arxiv:1907.06665 [gr-qc]} \BibitemShut
  {NoStop}%
\bibitem [{\citenamefont {{Portales-Oliva}}\ and\ \citenamefont
  {Landulfo}(2022)}]{portales-olivaClassicalQuantumReconciliation2022}%
  \BibitemOpen
  \bibfield  {author} {\bibinfo {author} {\bibfnamefont {F.}~\bibnamefont
  {{Portales-Oliva}}}\ and\ \bibinfo {author} {\bibfnamefont {A.~G.~S.}\
  \bibnamefont {Landulfo}},\ }\bibfield  {title} {\bibinfo {title} {Classical
  and quantum reconciliation of electromagnetic radiation: {{Vector Unruh}}
  modes and zero-{{Rindler-energy}} photons},\ }\href
  {https://doi.org/10.1103/PhysRevD.106.065002} {\bibfield  {journal} {\bibinfo
   {journal} {Phys. Rev. D}\ }\textbf {\bibinfo {volume} {106}},\ \bibinfo
  {pages} {065002} (\bibinfo {year} {2022})},\ \Eprint
  {https://arxiv.org/abs/2205.15183} {arxiv:2205.15183 [gr-qc]} \BibitemShut
  {NoStop}%
\end{thebibliography}%
\end{document}